\documentclass[authoryear,final,3p,times]{elsarticle} 

\usepackage{amssymb}

\usepackage{rotating,longtable,boldline,amsmath}
\usepackage[header]{appendix}
\usepackage{comment}
\usepackage[normalem]{ulem}
\usepackage{color}
\usepackage{soul}
\usepackage{mathrsfs}
\usepackage{amsbsy}

\newenvironment{DIFnomarkup}{}{}

\DeclareMathOperator{\sgn}{sgn}
\DeclareMathOperator{\ord}{ord}
\DeclareMathOperator{\diag}{diag}
\newcommand{\id}{\mathrm{d}} 
\newcommand{\tast}{t^*} 
\newcommand{\skewmat}[1]{\left[#1\right]_{\times}} 
\newcommand{\intwave}[1]{\int_{\epsilon^{-1}\hS-\frac{\Lambda}{2}}^{\epsilon^{-1}\hS+\frac{\Lambda}{2}}#1\:\mathrm{d}S} 
\newcommand{\intwaveref}[1]{\int_{\epsilon^{-1}\hS-\frac{1}{2}}^{\epsilon^{-1}\hS+\frac{1}{2}}#1\:\mathrm{d}S} 

\newcommand{\hS}{\mathcal{S}}
\newcommand{\hK}{\mathcal{K}}  
\newcommand{\hT}{\mathcal{T}}  
\newcommand{\hZ}{\mathcal{Z}}       
\newcommand{\hPsi}{\varPsi}
\newcommand{\hR}{\mathcal{R}}      
\newcommand{\hC}{\mathcal{C}}  

\newcommand{\aS}{\mathscr{S}}

\newcommand{\aref}{\alpha^{u}} 
\newcommand{\lamref}{\lambda^{u}} 
\newcommand{\rref}{r^{u}}
\newcommand{\brref}{\mathbf{r}^{u}} 
\newcommand{\bRref}{\mathbf{R}^{u}} 
\newcommand{\Psiref}{\Psi^{u}}
\newcommand{\hPsiref}{\varPsi^{u}}
\newcommand{\hZref}{\mathcal{Z}^{u}}

\newcommand{\bnref}{\mathbf{n}^{u}}  
\newcommand{\bbref}{\mathbf{b}^{u}}  
\newcommand{\btref}{\mathbf{t}^{u}}  
\newcommand{\bdoneref}{\mathbf{d}_1^{u}}  
\newcommand{\bdtworef}{\mathbf{d}_2^{u}}  
\newcommand{\bdthreeref}{\mathbf{d}_3^{u}}  
  
\newcommand{\kapref}{\kappa^{u}}  
\newcommand{\hKref}{\mathcal{K}^{u}}  
\newcommand{\tauref}{\tau^{u}}  
\newcommand{\hTref}{\mathcal{T}^{u}}
\newcommand{\hRref}{\mathcal{R}^{u}}  
\newcommand{\buref}{\mathbf{u}^{u}}

\newcommand{\bUref}{\mathbf{U}^{u}}

\newcommand{\Jref}{J^{u}} 

\newcommand{\bUlead}{\mathbf{U}^{(0)}}

\newcommand{\Uilead}{U_i^{(0)}} 
\newcommand{\bWlead}{\mathbf{W}^{(0)}} 
\newcommand{\Wilead}{W_i^{(0)}} 
\newcommand{\bRlead}{\mathbf{R}^{(0)}}  
\newcommand{\Zlead}{Z^{(0)}} 
\newcommand{\bRleadperp}{{\mathbf{R}^{(0)}}^{\perp}} 
\newcommand{\bMlead}{\mathbf{M}^{(0)}} 
\newcommand{\Monelead}{M_1^{(0)}}  
\newcommand{\Mtwolead}{M_2^{(0)}}  
\newcommand{\Mthreelead}{M_3^{(0)}}  
\newcommand{\Milead}{M_i^{(0)}}  
\newcommand{\bFlead}{\mathbf{F}^{(0)}}

\newcommand{\Filead}{F_i^{(0)}}  
\newcommand{\bOmegalead}{\boldsymbol{\Omega}^{(0)}}

\newcommand{\Omegailead}{\Omega_i^{(0)}}
\newcommand{\bdonelead}{\mathbf{d}_1^{(0)}}  
\newcommand{\bdtwolead}{\mathbf{d}_2^{(0)}}  
\newcommand{\bdthreelead}{\mathbf{d}_3^{(0)}}  
\newcommand{\bdilead}{\mathbf{d}_i^{(0)}}  
\newcommand{\bdjlead}{\mathbf{d}_j^{(0)}} 

\newcommand{\sse}{\mathsf{e}}
\newcommand{\ssMelead}{\mathsf{M}_{e}^{(0)}} 
\newcommand{\ssFelead}{\mathsf{F}_{e}^{(0)}} 
\newcommand{\ssK}{\mathsf{K}}
\newcommand{\ssA}{\mathsf{A}}
\newcommand{\sstA}{\mathsf{\tilde{A}}}
\newcommand{\ssV}{\mathsf{V}}
\newcommand{\ssW}{\mathsf{W}}
\newcommand{\sszero}{\mathsf{0}}
\newcommand{\ssdthreelead}{\mathsf{d}_3^{(0)}} 
\newcommand{\ssUlead}{\mathsf{U}^{(0)}}  
\newcommand{\ssMlead}{\mathsf{M}^{(0)}}  
\newcommand{\ssFlead}{\mathsf{F}^{(0)}} 
\newcommand{\ssRlead}{\mathsf{R}^{(0)}}
\newcommand{\ssRref}{\mathsf{R}^{u}}  

\newcommand{\bUfirst}{\mathbf{U}^{(1)}} 
\newcommand{\Uonefirst}{U_1^{(1)}}  
\newcommand{\Utwofirst}{U_2^{(1)}}  
\newcommand{\Uthreefirst}{U_3^{(1)}} 
\newcommand{\Uifirst}{U_i^{(1)}} 
\newcommand{\Ujfirst}{U_j^{(1)}} 
\newcommand{\bWfirst}{\mathbf{W}^{(1)}} 
\newcommand{\Wifirst}{W_i^{(1)}}  
\newcommand{\bMfirst}{\mathbf{M}^{(1)}} 
\newcommand{\Monefirst}{M_1^{(1)}}  
\newcommand{\Mtwofirst}{M_2^{(1)}}  
\newcommand{\Mthreefirst}{M_3^{(1)}}  
\newcommand{\Mifirst}{M_i^{(1)}}  
\newcommand{\bFfirst}{\mathbf{F}^{(1)}} 
\newcommand{\Fonefirst}{F_1^{(1)}}  
\newcommand{\Ftwofirst}{F_2^{(1)}}  
\newcommand{\Fthreefirst}{F_3^{(1)}} 
\newcommand{\Fifirst}{F_i^{(1)}}   

\newcommand{\Omegaifirst}{\Omega_i^{(1)}}

\newcommand{\bdthreefirst}{\mathbf{d}_3^{(1)}}  
\newcommand{\bdifirst}{\mathbf{d}_i^{(1)}}  
\newcommand{\bdjfirst}{\mathbf{d}_j^{(1)}}  

\newcommand{\sstUfirst}{\mathsf{\tilde{U}}^{(1)}}  
\newcommand{\sstMfirst}{\mathsf{\tilde{M}}^{(1)}}  
\newcommand{\sstFfirst}{\mathsf{\tilde{F}}^{(1)}}
\newcommand{\ssYfirst}{\mathsf{Y}^{(1)}}
\newcommand{\sstYfirst}{\mathsf{\tilde{Y}}^{(1)}} 
\newcommand{\ssYadj}{\mathsf{Y}^{*}}
\newcommand{\sstYadj}{\mathsf{\tilde{Y}}^{*}}
\newcommand{\ssPhiadj}{\mathsf{\Phi}^{*}} 
\newcommand{\ssMadj}{\mathsf{M}^{*}}  
\newcommand{\ssFadj}{\mathsf{F}^{*}}  
\newcommand{\ssPhi}{\mathsf{\Phi}}  
\newcommand{\ssTheta}{\mathsf{\Theta}}

\newcommand{\br}{\mathbf{r}} 
\newcommand{\bR}{\mathbf{R}} 
\newcommand{\bfo}{\mathbf{f}}  
\newcommand{\bF}{\mathbf{F}}  
\newcommand{\bmo}{\mathbf{m}}  
\newcommand{\bM}{\mathbf{M}}  
\newcommand{\bfe}{\mathbf{f}_{e}}  
\newcommand{\bFe}{\mathbf{F}_{e}}  
\newcommand{\bme}{\mathbf{m}_{e}}  
\newcommand{\bMe}{\mathbf{M}_{e}}
\newcommand{\bAe}{\mathbf{A}_{e}}
\newcommand{\bBe}{\mathbf{B}_{e}}
\newcommand{\bCe}{\mathbf{C}_{e}}
\newcommand{\bDe}{\mathbf{D}_{e}}
\newcommand{\bEe}{\mathbf{E}_{e}}
\newcommand{\bfelead}{\mathbf{f}_{e}^{(0)}}
\newcommand{\bmelead}{\mathbf{m}_{e}^{(0)}}  
\newcommand{\bFelead}{\mathbf{F}_{e}^{(0)}}
\newcommand{\bMelead}{\mathbf{M}_{e}^{(0)}}
\newcommand{\bd}{\mathbf{d}}  
\newcommand{\bu}{\mathbf{u}}  
\newcommand{\bU}{\mathbf{U}}
\newcommand{\bomega}{\boldsymbol{\omega}}  
\newcommand{\bOmega}{\boldsymbol{\Omega}}  
\newcommand{\bPhi}{\boldsymbol{\Phi}}  
\newcommand{\bTheta}{\boldsymbol{\Theta}}  
\newcommand{\be}{\mathbf{e}}  
\newcommand{\bv}{\mathbf{v}}  
\newcommand{\bV}{\mathbf{V}}
\newcommand{\bW}{\mathbf{W}}    
\newcommand{\bn}{\mathbf{n}}  
\newcommand{\bb}{\mathbf{b}}  
\newcommand{\bt}{\mathbf{t}}

\newcommand{\bY}{\mathbf{Y}}

\newcommand{\zpa}{\zeta_{\parallel}}
\newcommand{\zpe}{\zeta_{\perp}}

\newcommand{\Apa}{A_{\parallel}}
\newcommand{\Bpa}{B_{\parallel}}
\newcommand{\Cpa}{C_{\parallel}}
\newcommand{\Aparef}{A_{\parallel}^{u}}
\newcommand{\Bparef}{B_{\parallel}^{u}}
\newcommand{\Cparef}{C_{\parallel}^{u}}

\newcommand \beq{\begin{equation}}
\newcommand \eeq{\end{equation}}
\newcommand \beqn{\begin{equation*}}
\newcommand \eeqn{\end{equation*}}
\newcommand \beqa{\begin{align}}
\newcommand \eeqa{\end{align}}
\newcommand \beqan{\begin{align*}}
\newcommand \eeqan{\end{align*}}

\newcommand{\pd}[2]{\frac{\partial #1}{\partial #2}} 
\newcommand{\pdd}[2]{\frac{\partial^2 #1}{\partial #2^2}} 
 
\newcommand{\pdS}[1]{\frac{\partial #1}{\partial S}\bigg\lvert_{\hS}} 
 
\newcommand{\pdSinline}[1]{\partial #1/\partial S\lvert_{\hS}} 
\newcommand{\pdSt}[1]{\frac{\partial #1}{\partial S}\big\lvert_{\hS}}
\newcommand{\pdhS}[1]{\frac{\partial #1}{\partial \hS}\bigg\lvert_{S}}
 
\newcommand{\pdhSinline}[1]{\partial #1/\partial\hS\lvert_{S}} 
\newcommand{\pdhSt}[1]{\frac{\partial #1}{\partial \hS}\big\lvert_{S}}

\begin{document}

\begin{frontmatter}
 
\title{Effective extensional-torsional elasticity and dynamics of helical filaments under distributed loads}


\author[KCL]{Michael Gomez}
\ead{michael.gomez@kcl.ac.uk}
\author[DAMTP]{Eric Lauga}
\ead{e.lauga@damtp.cam.ac.uk}

\address[KCL]{Department of Engineering, King's College London, Strand, London WC2R 2LS, UK}
\address[DAMTP]{Department of Applied Mathematics and Theoretical Physics, University of Cambridge, Wilberforce Road, Cambridge CB3 0WA, UK}

\begin{abstract}
We study slender, helical elastic rods subject to distributed forces and moments. Focussing on the case when the helix axis remains straight, we employ the method of multiple scales to systematically derive an `equivalent-rod' theory from the Kirchhoff rod equations: the helical filament is described as a naturally-straight rod (aligned with the helix axis) for which the extensional and torsional deformations are coupled. Importantly, our analysis is asymptotically exact in the limit of a `highly-coiled' filament (i.e., when the helical wavelength is much smaller than the characteristic lengthscale over which the filament bends due to external loading) and is able to account for large, unsteady displacements. In addition, our analysis yields explicit conditions on the external loading that must be satisfied for a straight helix axis. In the small-deformation limit, we exactly recover the coupled wave equations used to describe the free vibrations of helical coil springs, thereby justifying previous equivalent-rod approximations in which linearised stiffness coefficients are assumed to apply locally and dynamically. We then illustrate our theory with two loading scenarios: (I) a heavy helical rod deforming under its own weight; and (II) the dynamics of axial rotation (twirling) in viscous fluid, which may be considered as a simple model for a bacteria flagellar filament. In both scenarios, we demonstrate excellent agreement with solutions of the full Kirchhoff rod equations, even beyond the formal limit of validity of the `highly-coiled' assumption. More broadly, our analysis provides a framework to develop reduced models of helical rods in a wide variety of physical and biological settings, and yields analytical insight into their elastic instabilities. In particular, our analysis indicates that tensile instabilities are a generic phenomenon when helical rods are subject to a combination of distributed forces and moments.
\end{abstract}


\end{frontmatter}

\section{Introduction}
\label{sec:intro}
\subsection{Background}
Slender elastic rods (also known as `filaments') with an intrinsic helical geometry are encountered in a wide range of physical and biological systems. In engineering, helical coil springs have long been used to store energy and absorb shock, with applications ranging from computer keyboards and mattresses to vehicle suspension systems~\citep{kobelev2021}. In biology, helical forms appear in a variety of settings, including the tendrils of climbing plants~\citep{mcmillen2002}, arteries and veins in the human umbilical cord~\citep{malpas1966}, the shape of some viruses~\citep{stubbs2012}, and --- perhaps most famous of all --- DNA, a biopolymer comprising two helical strands of nucleic acid that spiral around one another~\citep{watson1953}. Furthermore, the majority of bacteria are propelled by helical filaments whose rotation in a viscous fluid induces forward propulsion due to their chiral shape~\citep{lauga2020}. Recent interest in artificial swimmers has seen the design of bio-inspired devices driven by helical filaments~\citep{zhang2009,katsamba2019,huang2019,lim2023}. 

In these systems, the elasticity of the filament often plays a key role in its function. Indeed, the flexibility of a traditional coil spring is essential for its ability to absorb energy, and the linear relationship between small longitudinal displacements of a spring and the applied force serves as the paradigmatic example of Hooke's Law. Similarly, the elasticity of biofilaments often plays an important role. For example, the variety of polymeric filaments (from actin to microtubules) inside the cytoskeleton of cells have a range of bending rigidities tuned to their structural functions in the cells~\citep{howard2001}. The flexibility of DNA is known to be necessary for a variety of processes including replication, packing inside eukaryotic nuclei, and binding to proteins \citep{peters2010,marin2021}. Moreover, the run-and-tumble motion of multi-flagellated bacteria (such as \emph{E.~coli}) requires that the flagellar filaments are sufficiently flexible to form a tight bundle behind the cell body during steady swimming, yet stiff enough to unbundle once one of the rotary motors slows down or reverses direction~\citep{berg2003,berg2004,riley2018}.


To model the elastic deformations and dynamics of helical filaments, the Kirchhoff rod equations are commonly used~\citep{goriely1997c,goriely2017}. These equations are geometrically nonlinear and so can account for large, global displacements of the rod in three dimensions, while maintaining a mechanically-linear (i.e., Hookean) constitutive law; such large displacements are consistent with the assumption of small local strains, as required for linear elasticity, provided that the lengthscale of the deformation is much larger than the cross-section dimensions of the rod~\citep{audoly2010}. This geometric nonlinearity also makes a mathematical analysis of the rod equations difficult, so that the dynamic behaviour of helical rods under external loading is still generally poorly understood. Most research in the area tends to be computational in nature \citep{shum2012,jawed2015,jawed2017,park2017,park2019}, though simulations are computationally expensive due to the inherent three-dimensional geometry of helical filaments, meaning only a relatively small number of simulations may feasibly be performed. One alternative approach is to use a coarse-grained elastic model --- for example, by replacing the filament by bead-spring interactions --- to decrease the computational expense. Coarse-grained models have been applied successfully to study flagellar bundling \citep{flores2005,watari2010,nguyen2017,nguyen2018} and the stability of a Slinky toy~\citep{holmes2014}, although such models are generally still too complex to be solved analytically.


As an alternative to numerical simulations, analytical models are of fundamental importance: they clarify the dependence on (possibly many) parameters of a system, and provide a basis to guide more detailed simulations or experiments. In general, to make analytical progress with the Kirchhoff rod equations an approximation must be made, for which previous work can roughly be split into two groups. The first group is based on the assumption that the filament is relatively stiff compared to the external loads, or close to a buckling threshold, so that the deformed shape can be analysed as a small perturbation away from a known base state (such as the natural helical shape). Early work considered helices of small pitch angle and treated the natural shape as a small perturbation from a straight rod~\citep{haringx1949}. \cite{goriely1997a} and \cite{goriely1997b} developed a general perturbation scheme to study, respectively, the linear stability and weakly-nonlinear dynamics of elastic rods; this scheme was then applied to helical rods by \cite{goriely1997c}, who quantified the early-time dynamics of buckling under axial compression. Moreover, the assumption of small deformations is useful when modelling helices rotating in viscous fluid, since the fluid and elastic problems can be decoupled to a first approximation: the viscous drag is computed for an undeformed helix rigidly rotating about its axis, and this drag is then used to calculate the deformed shape~\citep{takano2003,kim2005}. This asymptotic procedure was explored in detail by \cite{katsamba2019}, who were also able to calculate the next-order correction to the drag forces in this approximation.


 The second group of analytical models consists of `equivalent-rod' approximations. These model the helical filament as a naturally-straight rod whose centreline is aligned with the helix axis; the effective elastic properties of the rod are chosen so that the deformation captures that of the full helix in simple loading situations, for example end-to-end compression or uniform bending. Such approximations have long been used in engineering to describe the lateral vibrations and buckling behaviour of helical coil springs, where they are known as `equivalent-column' approximations; see Chapter $3$ of \cite{kobelev2021} and references therein. In the absence of distributed loads, a theoretical basis for an equivalent-rod approximation is provided by the work of \cite{love1944}, who derived equilibrium equations for helical filaments based on the inextensible Kirchhoff rod equations under terminal loads. In the case where the loads form a wrench whose axis coincides with the axis of the helix, an exact solution of these equations is that of another helix with modified geometry \citep{love1944}. By linearising this solution in the limit of small deformations, stiffness coefficients for an equivalent-rod approximation can be obtained. Using an ad hoc assumption that these linearised stiffness coefficients can be applied locally (i.e., for each infinitesimal spring element) in dynamic problems, \cite{phillips1972} analysed free extensional-torsional vibrations of helical springs, obtaining good agreement with experiments even for large longitudinal displacements.  This work has since been extended, for example to address the radial expansion of impacted springs \citep{costello1975}, filament extensibility~\citep{jiang1991,jiang1998}, and the effects of shear deformations \citep{kruzelecki1990,michalczyk2019}. However, the validity of equivalent-rod approximations remains unclear when dynamic effects and distributed loads are present.
 

Equivalent-rod models have also been developed in other contexts. \cite{kehrbaum2000} analysed the mechanical properties of straight elastic rods with high intrinsic twist, as a model for DNA molecules deforming over lengthscales much larger than that of individual base pairs. Using a Hamiltonian formulation of the Kirchhoff rod equations in the absence of distributed loads, in which the twist lengthscale plays the role of a `fast' time-like variable in a dynamical system, they used standard averaging theory to obtain an effective constitutive law governing bending over relatively large or `slow' lengthscales. For linearly elastic rods, this effective constitutive law is isotropic, even if the rod is locally anisotropic. \cite{rey2000} built upon this work to consider twisted rods under general boundary conditions and buckling under compression. \cite{healey2011} studied hyperelastic rods possessing helical material symmetry, i.e., straight rods whose material parameters vary longitudinally according to a uniform circular helix. By applying both standard averaging theory and Gamma-convergence methods in the limit of zero helical pitch, an effective hemitropic constitutive law was derived that retains the chirality of the original helical symmetry \citep{healey2002}.

There are several hints that a similar averaged theory may be fruitful to describe the mechanics of helical filaments when distributed loads \emph{are} present. For example, the instabilities of a helical rod rotating in viscous fluid share many features with those of a naturally-straight rod. If a naturally-straight rod is rotated about its axis while the other end is free, the straight (twirling) state becomes unstable at a critical frequency and transitions to an overwhirling state characterised by significant bending~\citep{ryan2022}. A helical rod (with sufficiently shallow pitch angle) rotated about its axis exhibits an analogous instability, for which the critical frequency scales with the bending stiffness and filament length identically to the straight-rod case~\citep{park2017}. At large pitch angles, the translation-rotation coupling in the viscous drag becomes significant, and the filament may buckle under a combination of the hydrodynamic torque and propulsive force. In this regime, \cite{vogel2012} were able to obtain good agreement with a straight-rod model; \cite{jawed2015} found that the critical frequency scales identically to an effective beam of equal length. 


\subsection{Summary and structure of this paper}
\label{sec:papersummary}
In this paper, we develop a reduced model for a helical rod undergoing unsteady deformations in the presence of distributed forces and moments. The fundamental assumptions of our theory are that (i) the helix wavelength is much smaller than the characteristic lengthscale over which the filament bends, which we refer to as the `highly-coiled' assumption (we emphasise that our definition of `highly-coiled' here should not be confused with the limit in which the helical pitch approaches zero); (ii) the filament is sufficiently slender so that the local strains remain small and we may assume linear elasticity; (iii) the filament has uniform, circular cross-section; and (iv) the helix axis remains straight. While in general multiple types of deformation are possible, namely extensional-rotational deformations about the helix axis and bending of the helix axis itself, we ignore axis bending here as it allows us to understand how the helical geometry depends on the distributed loads without introducing additional complexity. We therefore refer to our reduced model as a `(straight) equivalent rod' theory, to distinguish it from `equivalent-column' approximations that incorporate axis bending and buckling \citep{kobelev2021}.

Inspired by the work of \cite{kehrbaum2000} and \cite{rey2000} on straight rods with high intrinsic twist, we employ a homogenisation procedure in which the helix wavelength acts as a `fast' time-like variable. The basis of our method is Love's helical solution of the Kirchhoff rod equations, in the case of a constant wrench aligned with the helix axis \citep{love1944}. Under the highly-coiled assumption, the force and moment resultants in the rod are approximately constant over each helical wavelength; provided that the helix axis remains straight, these resultants form a wrench aligned with the helix axis, so that the solution is locally a rigid transformation of a helix with modified geometric parameters. We then apply the method of multiple scales to describe `slow' variations in the geometric parameters under distributed loading. These variations in helical geometry correspond to extensional and torsional deformations about the helix axis, so that we obtain an equivalent-rod theory.

The remainder of this paper is organised as follows. In \S\ref{sec:formulation}, we present the Kirchhoff rod equations and their non-dimensionalisation, before introducing the highly-coiled assumption. We also discuss the numerical implementation of the full Kirchhoff rod equations, which we later use to generate simulation results for specific loading scenarios. In \S\ref{sec:multiscales}, we apply the method of multiple scales to derive the equivalent-rod equations, showing how these arise as solvability conditions on an appropriate first-order problem when the solution is expanded as an asymptotic series. In \S \ref{sec:equivalent-rod}, we analyse the equivalent-rod equations in detail: we first show how the equations can be written in terms of useful pairs of variables for analysis, before considering singularities and the small-deformation limit of the equations. We next apply our theory to two specific scenarios: the compression/extension of a heavy helical column in \S \ref{sec:scenario1}, and the dynamics of a helix rotating in viscous fluid in \S \ref{sec:scenario2}. In both scenarios, we compare the predictions of the equivalent-rod theory with numerical simulations. Finally, in \S\ref{sec:conclusion}, we summarise our findings and conclude. We discuss how the framework introduced here may be extended to incorporate other effects, including axis bending and different cross-section shapes, and we comment on the physical significance of singularities for instabilities of helical rods subject to distributed loads.
 
\section{Theoretical formulation}
\label{sec:formulation}
\begin{figure}
\centering
\includegraphics[width=0.79\textwidth]{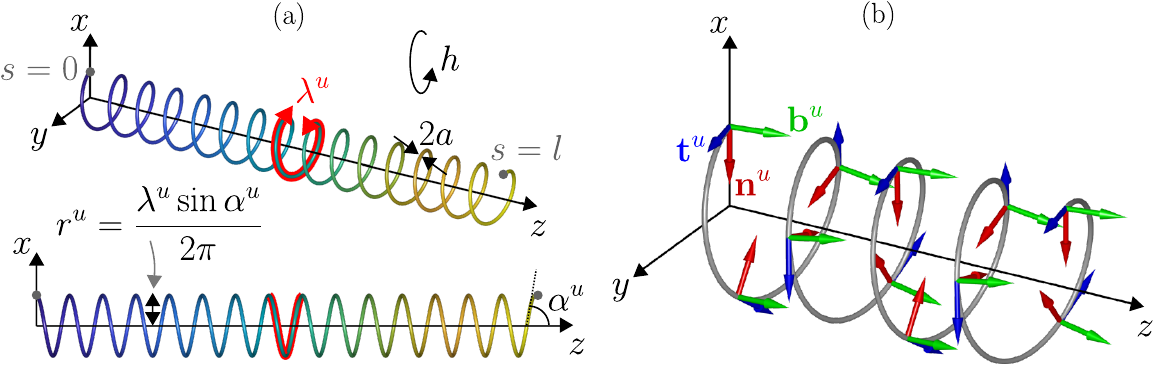} 
\caption{(a) Schematic diagram of the undeformed helical filament, showing the notation we use: total contour length $l$, pitch angle $\aref$, chirality index $h$, contour wavelength $\lamref$, helix radius $\rref$, filament radius $a$, arclength $s$ and Cartesian coordinates $(x,y,z)$ (the helix frame). (b) Corresponding Frenet-Serret basis vectors.}
\label{fig:undeformedschematic}
\end{figure}

In this section we derive the equations governing the helical filament in the framework of Kirchhoff rod theory. We present only the key ingredients here; for a detailed treatment see \cite{audoly2010} or \cite{goriely2017}. We consider an elastic filament whose undeformed centreline is a uniform helix with contour length $l$, pitch angle $\aref$ (the angle between the centreline tangent and the helix axis), contour wavelength $\lamref$, and chirality index $h = \pm 1$ ($h = -1$ or $+1$ corresponds, respectively, to a left-handed or a right-handed helix); see Fig.~\ref{fig:undeformedschematic}a. Simple geometry states that the radius of the cylinder on which the helix is wound is $\rref = \lamref\sin\aref/(2\pi)$. Here and throughout this paper, we use the superscript $^{u}$ to denote quantities of the undeformed shape that may change during deformation, and drop these for the deformed (current) shape. The filament has a circular cross-section of constant radius $a$ and we assume that it is sufficiently slender (i.e., $a \ll \lamref$) so that it is inextensible and unshearable. We also assume that the filament is composed of a uniform, isotropic material of density $\rho_s$ and that the strains remain small; we may then use a linearly elastic constitutive law with (constant) Young's modulus $E$ and Poisson ratio $\nu$.

\subsection{Kinematics}
As shown in Fig.~\ref{fig:undeformedschematic}a, we introduce Cartesian coordinates $Oxyz$ (the `helix frame') such that the undeformed helix axis lies on the $z$-axis, and the filament base (taken to be at $z = 0$) lies on the $x$-axis. The corresponding unit Cartesian vectors are denoted $\lbrace\be_x,\be_y,\be_z\rbrace$. If the filament base is not fixed in space but is allowed to move (e.g., if it is attached to a freely-moving body), the helix frame rotates and translates relative to the laboratory frame. 

\paragraph{Deformed configuration} During deformation, we write $\br(s,t)$ for the centreline position at arclength $s$ (measured from the filament base) and time $t$. Under the inextensibility assumption, the unit tangent vector, $\bd_3(s,t)$, is \citep{coleman1993}
\beq
\bd_3 = \pd{\br}{s}, \label{eqn:inextensibilitydim}
\eeq
which we refer to as the inextensibility constraint. To describe the local orientation of the rod, we introduce the additional vectors $\bd_1(s,t)$ and $\bd_2(s,t)$ such that $\lbrace\bd_1,\bd_2,\bd_3\rbrace$  (referred to as \emph{directors}) form a right-handed orthonormal basis for each $s$ and $t$ \citep{antman2005}; the vectors  $\lbrace\bd_1,\bd_2\rbrace$ span the normal plane through the cross-section, changing orientation as the rod bends and twists. We can then quantify mechanical strains via the strain vector, $\bu$, and the angular velocity vector, $\bomega$ \citep[also known as the twist vector and spin vector, respectively; see][]{goriely1997c}, which account for rotations of the directors as $s$ or $t$ varies (hence preserving their orthonormality):
\beq
\pd{\bd_i}{s} = \bu \times \bd_i, \quad \pd{\bd_i}{t} = \bomega \times \bd_i, \qquad i = 1,2,3. \label{eqn:kinematicstrainspindim}
\eeq
Expressing $\bu$ and $\bomega$ in terms of components with respect to the director basis,
\beqn
\bu =u_1 \bd_1 + u_2 \bd_2 + u_3 \bd_3, \quad \bomega = \omega_1 \bd_1 +\omega_2 \bd_2 + \omega_3 \bd_3,
\eeqn
we can interpret $u_1$ and $u_2$ as the bending (flexural) strains (the rate at which the tangent vector $\bd_3$ rotates about $\bd_1$ and $\bd_2$, respectively, as $s$ increases) and $u_3$ as the twist (torsional) strain (the rate at which $\bd_1$ and $\bd_2$ rotate about $\bd_3$, which incorporates both axial twist and centreline torsion)~\citep{antman2005,audoly2010}. The interpretation of the components $\omega_i$ is analogous with $s$ replaced by time, $t$. 


\paragraph{Undeformed configuration} The undeformed centreline can be written in cylindrical polar coordinates as
\beq
\brref(s)  = \rref\,\be_r(s) + s\cos\aref\,\be_z \qquad 0 < s < l, \label{eqn:undeformedcentrelinedim}
\eeq
where the local unit vectors evaluated on the centreline and the winding angle of the helix (i.e., the polar angle between the radius to the filament and the $x$-axis) are
\beq
\be_r(s) = \cos\Psiref(s)\,\be_x + \sin\Psiref(s)\,\be_y, \quad \be_{\theta}(s) = -\sin\Psiref(s)\,\be_x + \cos\Psiref(s)\,\be_y, \qquad \Psiref(s) = \frac{2\pi h s}{\lamref}. \label{eqn:defnundeformedpsidim}
\eeq

Because the filament has a circular cross-section, without loss of generality\footnote{In general, the undeformed directors $\lbrace\bd_1^u(s),\bd_2^u(s),\bd_3^u(s)\rbrace$ are chosen such that $\bd_3^u(s)=\bt^u(s)=\id\br^u/\id s$ is the unit tangent vector and $\lbrace\bd_1^u(s),\bd_2^u(s)\rbrace$ lie along the principal axes of inertia of the cross-section \citep{coleman1993}. Because we consider an isotropic rod with circular cross-section, we are free to choose any orientation for $\bd_1^u(s)$ and $\bd_2^u(s)$ provided that $\lbrace\bd_1^u(s),\bd_2^u(s),\bd_3^u(s)\rbrace$ are orthonormal (and twice continuously differentiable) at each $s$. In particular, we choose $\bd_1^u(s) = \bn^u(s)$ and $\bd_2^u(s)=\bb^u(s)$, where $\bn^u(s)$ and $\bb^u(s)$ are defined in Eq.~\eqref{eqn:undeformedFSframe}.} we may choose the undeformed director basis to coincide with the Frenet-Serret frame $\lbrace\bnref,\bbref,\btref\rbrace$; here $\bnref$, $\bbref$ and $\btref$  correspond to the unit normal, binormal and tangent vectors along $\brref(s)$, respectively:
\beq
\bnref(s) = -\be_r(s), \quad \bbref(s) = -\cos\aref\,\be_{\theta}(s)+ h\sin\aref\,\be_z, \quad \btref(s) = h\sin\aref\,\be_{\theta}(s) + \cos\aref\,\be_z. \label{eqn:undeformedFSframe}
\eeq
These are illustrated in Fig.~\ref{fig:undeformedschematic}b. Setting $\lbrace\bd_1,\bd_2,\bd_3\rbrace = \lbrace\bnref,\bbref,\btref\rbrace$ in Eq.~\eqref{eqn:kinematicstrainspindim} shows that the undeformed strain vector, denoted $\buref$, is equal to the Darboux vector of the Frenet-Serret frame:
\beq
\buref = \kapref\bbref + \tauref\btref = \frac{2\pi h}{\lamref}\be_z \qquad \mathrm{where} \qquad \kapref = \frac{2\pi \sin\aref}{\lamref}, \quad \tauref = \frac{2\pi h\cos\aref}{\lamref}. \label{eqn:undeformedstraindim}
\eeq 
The quantities $\kapref$ and $\tauref$ are the Frenet curvature and torsion of $\brref$, respectively.

\subsection{Mechanics}

Let $\bfo(s,t)$ be the resultant force and $\bmo(s,t)$ be the resultant moment attached to the filament centreline, obtained by averaging the internal elastic stresses over the cross-section at position $s$. Balancing linear and angular momentum, we obtain the inextensible, unshearable Kirchhoff rod equations \citep{audoly2010}
\begin{align}
& \pd{\bfo}{s}+\bfe = \rho_s A \pdd{\br}{t}, \label{eqn:forcebalancedim} \\
& \pd{\bmo}{s}+\bd_3\times\bfo+\bme = \rho_s I\left(\bd_1\times\pdd{\bd_1}{t}+\bd_2\times\pdd{\bd_2}{t}\right), \label{eqn:momentbalancedim}
\end{align}
where $\bfe$ and $\bme$ are, respectively, the external force and moment exerted on the rod per unit arclength, $A = \pi a^2$ is the cross-sectional area, and $I = \pi a^4/4$ is the second moment of area of the cross-section. (We neglect fictitious forces that may arise when the helix frame is non-inertial and accelerates relative to the laboratory frame.) The above equations are supplemented with the isotropic, Hookean (linearly elastic) constitutive law
\beq
\bmo = B\left[u_1\bd_1 + \left(u_2-\kapref\right)\bd_2\right] + C\left(u_3-\tauref\right)\bd_3, \label{eqn:clawdim}
\eeq
where $B = EI$ is the bending modulus and $C = \mu_s J$ is the twist modulus ($\mu_s  = E/[2(1+\nu)]$ is the shear modulus and $J$ is the torsion constant) \citep{howell2009}. We neglect warping of the cross-section (justified by its circular shape), which gives $J = 2I =  \pi a^4/2$ and hence
\beq
C = \frac{B}{1+\nu}. \label{eqn:elimC}
\eeq
The appearance of $\kapref$ and $\tauref$ in Eq.~\eqref{eqn:clawdim} guarantees that the rod is unstressed in its undeformed shape when $\lbrace\bd_1,\bd_2,\bd_3\rbrace = \lbrace\bnref,\bbref,\btref\rbrace$ and $\bu = \buref = \kapref\bbref + \tauref\btref$.

\subsection{Boundary conditions}
\label{sec:formulationBCs}
Given the external forces and moments, the system is closed by appropriate boundary conditions and (if relevant) initial conditions. We shall restrict to situations where the filament tip $s = l$ remains free of forces and moments:
\beq
\bfo(l,t) = \bmo(l,t) = \mathbf{0}. \label{eqn:BCtipdim}
\eeq
In general, it is also necessary to provide boundary conditions at the filament base, for example specifying the position and orientation of the filament. However, the multiple-scales analysis presented in \S\ref{sec:multiscales} leads to equivalent-rod equations that are \emph{first-order} in space, so that the boundary conditions \eqref{eqn:BCtipdim} uniquely specify the solution (assuming the external forces and moments are known). To be compatible with the straight equivalent-rod approximation, we therefore suppose that the filament is supported at its base such that the centreline is located on the $x$-axis in the $z = 0$ plane; this is equivalent to the condition  
\beq
\br(0,t) \times \be_x = \mathbf{0}.  \label{eqn:BCbasedim}
\eeq
At leading order, we thus do not specify the orientation of the filament nor its $x$-coordinate at $z = 0$. Once the equivalent-rod equations are solved, the value of $\br(0,t)$ is determined and hence the centreline $\br(s,t)$ can be found by integrating the inextensibility constraint \eqref{eqn:inextensibilitydim}. 
 
\subsection{Non-dimensionalisation}
\label{sec:nondim}
To non-dimensionalise the problem, it is natural to scale all lengths by the undeformed contour wavelength, $\lamref$. From the kinematic equations \eqref{eqn:kinematicstrainspindim}, the strain vector scales as $|\bu| \sim 1/\lamref$. We scale the moment and force resultants by their typical magnitudes associated with bending over the lengthscale $\lamref$. Using the constitutive law \eqref{eqn:clawdim} and moment balance \eqref{eqn:momentbalancedim}, these are $|\bmo| \sim B/\lamref$ and $|\bfo| \sim B/({\lamref})^2$. We denote the typical magnitudes of the external force and moment by $[f]$ and $[m]$, respectively, which are determined by the physics governing the external loading. Keeping the timescale $[t]$ unspecified for now, we then introduce the dimensionless variables
\begin{gather}
\br = \lamref \bR,  \quad (x,y,z,s) = \lamref (X,Y,Z,S), \quad l = \lamref L, \quad (\kapref,\tauref) = \frac{1}{\lamref}(\hKref,\hTref), \quad \bu = \frac{1}{\lamref}\bU, \quad \buref = \frac{1}{\lamref}\bUref,  \quad u_i = \frac{1}{\lamref}U_i, \nonumber \\ 
t = [t] T, \quad \bomega = \frac{1}{[t]}\bOmega, \quad \omega_i = \frac{1}{[t]}\Omega_i, \quad \bmo = \frac{B}{\lamref}\bM, \quad \bfo = \frac{B}{({\lamref})^2}\bF, \quad \bfe = [f]\bFe, \quad \bme = [m]\bMe. \label{eqn:nondim}
\end{gather}

Under the above re-scalings, the inextensibility constraint \eqref{eqn:inextensibilitydim} becomes
\beq
\bd_3 = \pd{\bR}{S}. \label{eqn:inextensibility}
\eeq
In terms of the dimensionless strain vector $\bU = \sum_{i} U_i\bd_i$ and angular velocity vector $\bOmega = \sum_{i} \Omega_i\bd_i$, the kinematic equations \eqref{eqn:kinematicstrainspindim} are
\beq
\pd{\bd_i}{S} = \bU \times \bd_i, \quad \pd{\bd_i}{T} = \bOmega \times \bd_i, \qquad i = 1,2,3. \label{eqn:kinematicstrainspin}
\eeq

Using Eq.~\eqref{eqn:undeformedcentrelinedim}, the dimensionless natural shape, $\bRref = \brref/\lamref$, can be written as
\beq
\bRref(S)  = \hRref\be_r(S) + S\cos\aref\,\be_z \qquad  0 < S < L, \qquad \mathrm{where} \qquad \hRref \equiv \frac{\rref}{\lamref} = \frac{\sin\aref}{2\pi}. \label{eqn:refshape}
\eeq
In terms of dimensionless arclength, $S$, the undeformed winding angle is written $\Psiref(S) = 2\pi h S$. Using Eq.~\eqref{eqn:undeformedstraindim}, the dimensionless undeformed strain vector, Frenet curvature and torsion are
\beq
\bUref = \hKref\bbref+\hTref\btref = 2\pi h\be_z, \quad \hKref  =2\pi\sin\aref,\quad \hTref = 2\pi h\cos\aref. \label{eqn:undeformedstrain}
\eeq

The force and moment balances \eqref{eqn:forcebalancedim}--\eqref{eqn:momentbalancedim} now read
\begin{align}
& \pd{\bF}{S}+\epsilon\,\bFe = \left(\frac{{\tast}}{[t]}\right)^2\pdd{\bR}{T}, \label{eqn:forcebalance} \\
& \pd{\bM}{S}+\bd_3\times\bF+\epsilon\,\delta\,\bMe = \eta^2\left(\frac{{\tast}}{[t]}\right)^2\left(\bd_1\times\pdd{\bd_1}{T}+\bd_2\times\pdd{\bd_2}{T}\right), \label{eqn:momentbalancefull}
\end{align}
where we define 
\beq
\epsilon = \frac{({\lamref})^3 [f]}{B}, \quad  \delta = \frac{[m]}{[f]\lamref}, \quad \tast = \sqrt{\frac{\rho_s A (\lamref)^4}{B}}, \quad  \eta = \frac{\sqrt{I/A}}{\lamref}. \label{eqn:defnepsilon}
\eeq
Here $\epsilon$ and $\delta$ are dimensionless parameters measuring the relative importance of the external force and moment, respectively; $\tast$ is the timescale of inertial oscillations on the wavelength lengthscale; and $\eta$ is a slenderness parameter. Using $A = \pi a^2$ and $I = \pi a^4/4$, we have $\eta = a/(2\lamref)$; consistent with neglecting the effects of both axial extensibility and shear in the limit $a \ll \lamref$, we neglect the $\eta^2$ term on the right-hand side of Eq.~\eqref{eqn:momentbalancefull} so that the moment balance simplifies to 
\beq
\pd{\bM}{S}+\bd_3\times\bF+\epsilon\,\delta\,\bMe = \mathbf{0}. \label{eqn:momentbalance}
\eeq
We will assume that $\delta = O(1)$, i.e., the deformation is driven mainly by the external force in Eq.~\eqref{eqn:forcebalance}. However, this assumption is not essential and the averaging method we present in \S\ref{sec:multiscales} may be readily adapted to the case when the deformation is instead driven by the external moment.

We write the force and moment resultants in terms of components with respect to the director basis, $\bF =\sum_{i}F_i \bd_i$ and $\bM =\sum_{i}M_i \bd_i$. The constitutive law \eqref{eqn:clawdim} becomes (using Eq.~\eqref{eqn:elimC})
\beq
M_1 = U_1, \quad M_2 = U_2 - \hKref, \quad M_3 = \frac{U_3 - \hTref}{1+\nu}. \label{eqn:claw}
\eeq
Finally, the boundary conditions \eqref{eqn:BCtipdim}--\eqref{eqn:BCbasedim} become
\beq
\bR(0,T) \times \be_x = \mathbf{0}, \quad \bF(L,T) = \bM(L,T) = \mathbf{0}. \label{eqn:BCs}
\eeq
For unsteady problems, the relevant initial conditions are non-dimensionalised to complete the system. 

\subsection{Highly-coiled assumption}
\label{sec:highlycoiledassum}
The fundamental `highly-coiled' assumption that we make in this paper is that the impact of the external loading is small on the wavelength lengthscale, i.e., we assume that $\epsilon \ll 1$. We emphasise that our definition of the term `highly-coiled' should not be confused with the limit in which the helical pitch tends to zero, i.e., $\aref \to \pi/2$. Note from Eq.~\eqref{eqn:defnepsilon} that $\epsilon$ can be written as 
\beqn
\epsilon = \left(\frac{\lamref}{[s]}\right)^3,
\eeqn
where $[s] = (B/[f])^{1/3}$ is the typical lengthscale over which the filament bends significantly due to the external force $[f]$ (obtained by balancing terms in the force balance \eqref{eqn:forcebalancedim} with the scaling behaviour $|\bfo| \sim B/[s]^2$ and $\partial/\partial s \sim 1/[s]$). The highly-coiled assumption $\epsilon \ll 1$ then states that $\lamref \ll [s]$, i.e., there is negligible bending over each helical wavelength. In the next section, we show how this assumption can be exploited to construct an equivalent-rod theory.

\subsection{Numerical simulations of helical rods}
\label{sec:numerics}
In addition to the multiple-scales analysis presented in \S\ref{sec:multiscales}, we perform numerical simulations of the full Kirchhoff rod equations, i.e., Eqs.~\eqref{eqn:inextensibility}--\eqref{eqn:kinematicstrainspin}, \eqref{eqn:forcebalance} and \eqref{eqn:momentbalance}--\eqref{eqn:claw}. Later, in \S\ref{sec:scenario1} and \S\ref{sec:scenario2}, we will compare the results of these simulations with predictions of our equivalent-rod theory for specific loading scenarios. Because we consider either equilibrium solutions (\S\ref{sec:scenario1}) or dynamics dominated by viscous forces (\S\ref{sec:scenario2}), we neglect rod inertia in our numerical simulations, which is equivalent to considering a timescale $[t]\gg\tast$ so that the inertia term in Eq.~\eqref{eqn:forcebalance} is negligible. Moreover, we focus on external forces and moments in the form
\beq
\bFe = \bAe\pd{\bR}{T} + \bBe\bR + \bCe, \quad \bMe = \bDe\bOmega + \bEe, \label{eqn:loadingnumeric}
\eeq
for general second-order tensors $\bAe(S,T)$, $\bBe(S,T)$, $\bDe(S,T)$ and vectors $\bCe(S,T)$, $\bEe(S,T)$. The specific loading types studied in \S\ref{sec:scenario1} and \S\ref{sec:scenario2} can then be obtained by an appropriate choice of $\bAe$, $\bBe$, $\bCe$, $\bDe$ and $\bEe$.

Recall from the discussion surrounding Eq.~\eqref{eqn:BCbasedim} that we require additional boundary conditions at the filament base when solving the full Kirchhoff rod equations. In all numerical simulations reported in this paper, we suppose that the filament is rigidly clamped at its base such that the position and orientation remain unchanged from the undeformed configuration. (It should be noted that these boundary conditions are consistent with Eq.~\eqref{eqn:BCbasedim}). Together with the force-free and moment-free conditions at the filament tip, we therefore have the boundary conditions 
\beq
\bR(0,T) = \bRref(0) = \hRref\be_x, \quad \begin{cases} \bd_1(0,T) = \bnref(0) = -\be_x, \\ \bd_2(0,T) = \bbref(0) = -\cos\aref\,\be_y+ h\sin\aref\,\be_z, \\ \bd_3(0,T) = \btref(0) = h\sin\aref\,\be_y + \cos\aref\,\be_z, \end{cases} \quad \mathrm{and} \qquad \bF(L,T) = \bM(L,T) = \mathbf{0}. \label{eqn:BCsnumeric}
\eeq

With the external loading \eqref{eqn:loadingnumeric} and boundary conditions \eqref{eqn:BCsnumeric} (and appropriate initial conditions), we solve the dimensionless rod equations using the method of lines: we discretise the equations with respect to arclength $S\in[0,L]$, resulting in a set of ordinary differential equations (ODEs) in time that we integrate numerically. To this end, we first recast the Kirchhoff rod equations as a single integro-differential (vector) equation governing the orientation of the rod centreline. This is achieved by introducing Euler angles $\phi(S,T)$, $\theta(S,T)$, $\psi(S,T)$ that parameterise the directors $\bd_i$ with respect to the Cartesian basis in the helix frame; see, for example, \cite{coleman1993}. With this parameterisation, we then (i) determine the strain vector $\bU$ and angular velocity vector $\bOmega$, and hence $\bM$ and $\bMe$, in terms of $S$ and $T$ derivatives of $(\phi,\theta,\psi)$ (using Eqs.~\eqref{eqn:kinematicstrainspin}, \eqref{eqn:claw} and \eqref{eqn:loadingnumeric}); (ii) use Eqs.~\eqref{eqn:inextensibility} and \eqref{eqn:loadingnumeric} to express $\bR$ and $\bFe$ as single integrals involving $(\phi,\theta,\psi)$; (iii) integrate the force balance \eqref{eqn:forcebalance} to obtain $\bF$ as a double integral involving $(\phi,\theta,\psi)$; and (iv) insert the expressions from steps (i)--(iii) into the moment balance \eqref{eqn:momentbalance} to arrive at a single vector equation of integro-differential form, in which the Euler angles appear as the only unknowns. The integro-differential equation is then expressed in component form with respect to the Cartesian basis, and the boundary conditions \eqref{eqn:BCsnumeric} (and initial conditions if appropriate) are written in terms of $(\phi,\theta,\psi)$.

Our integro-differential formulation is an extension to three dimensions of previous work on dynamic \emph{elastica} simulations, which consider pure bending of an elastic rod in two dimensions \citep{gomez2018,liu2021,gutierrez2023}. The major benefit is that dynamic simulations using the method of lines may be performed efficiently, since there is no need to explicitly impose inextensibility of the rod centreline between neighbouring mesh points in the discretisation, thereby avoiding a large number of constraints that are typically encountered with exact inextensibility \citep{gomez2018}. 

We discretise the integro-differential equation using a similar procedure used for dynamic \emph{elastica} simulations \citep{gomez2018,liu2021,gutierrez2023}. We introduce a uniform mesh on $S\in[0,L]$ with spacing $\Delta S=L/N$, where $N$ is a fixed integer. We formulate a numerical scheme with second-order accuracy as the mesh size $\Delta S\to 0$: we use second-order finite differences to approximate spatial derivatives (centred differences at interior mesh points, with forward/backward differences at the boundaries), together with the trapezium rule for quadrature. The result is a system of $3(N-1)$ ODEs for the values of the Euler angles at the $(N-1)$ interior mesh points, which we express in matrix-vector form and implement in MATLAB. For steady solutions, the ODEs become algebraic equations which we solve using the MATLAB routine \texttt{fsolve} (error tolerances $10^{-10}$). We have verified that steady solutions converge to the solution obtained by directly solving the steady boundary-value problem (without first discretising) using a collocation method (\texttt{bvp4c} in MATLAB). For dynamic simulations, we integrate the ODEs (with appropriate initial conditions) using the routine \texttt{ode15s} in MATLAB (relative error tolerance $10^{-6}$, absolute error tolerance $10^{-2}$). We have verified second order convergence of successive solutions as $N$ is increased. In all dynamic simulations reported in this paper, we take $N = 200$ while for steady solutions we use $N = 1000$. We have verified that the results are insensitive to further increasing $N$ and decreasing error tolerances.

The simulations output the Euler angles parameterising the orientation of the rod centreline at each mesh point in the numerical discretisation. From these angles, the corresponding values of $\bR$, $\partial\bR/\partial T$, $\bU$, $\bM$ and $\bF$ are computed using the expressions obtained when formulating the integro-differential equation, i.e., steps (i)--(iii) above.

\section{Multiple-scales analysis of highly-coiled filaments}
\label{sec:multiscales}
In the limit of highly-coiled filaments ($\epsilon \ll 1$), we anticipate an approximate solution using the method of multiple scales, in which arclength is analogous to the time variable in a dynamical system. On the `fast' wavelength lengthscale, $S = O(1)$, the force and moment resultants are approximately constant. Provided that the helix axis remains straight, these resultants form a wrench aligned with the helix axis, for which a stationary solution to the Kirchhoff rod equations is another helix with (in general) modified pitch angle and wavelength~\citep{love1944}. The aim here is to obtain a system of equations governing the helical geometry that arise from `slow' changes in the force and moment resultants, as well as the constraints that must be satisfied by the external loading to ensure a straight helix axis.

\subsection{Method outline}
\label{sec:multiscalesmethodoutline}
\begin{figure}
\centering
\includegraphics[width=\textwidth]{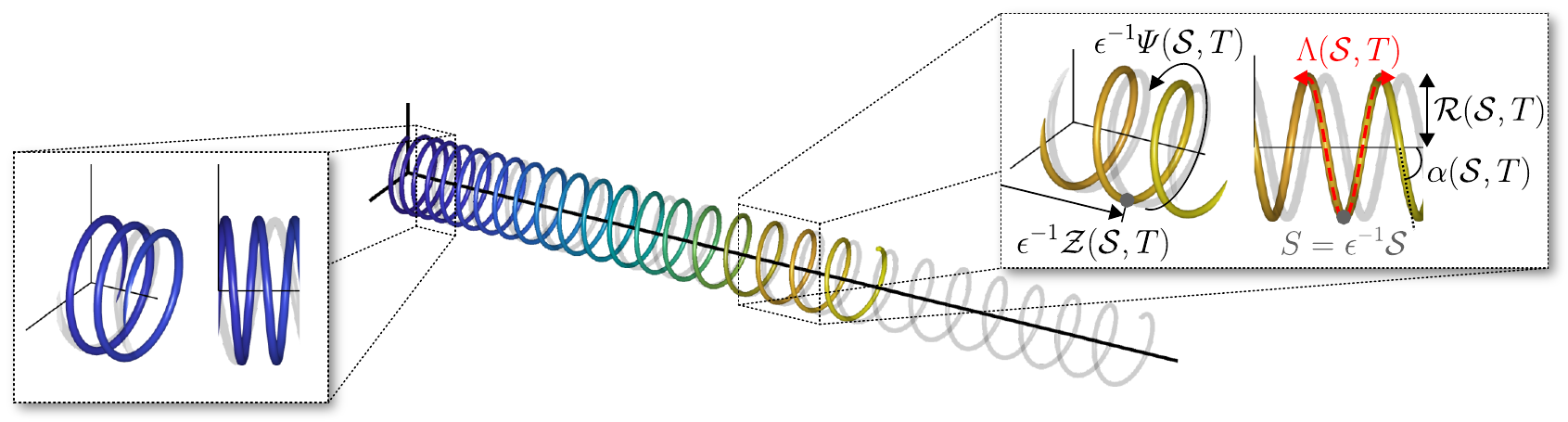} 
\caption{Multiple-scales analysis of highly-coiled helical filaments with a straight helix axis, presented using dimensionless quantities (lengths scaled by the undeformed contour wavelength, $\lamref$). Each close-up shows a helical wave centred around the point with `slow' variable $\hS$ (where $s = \epsilon^{-1}\lamref \hS$), together with the undeformed filament (grey curve). The deformed shape is locally another helix with pitch angle $\alpha$ and dimensionless wavelength $\Lambda = \lambda/\lamref$, where $\alpha$ and $\Lambda$ vary with $\hS$. Because the filament tip is free of forces and moments, $\alpha \approx \aref$ and $\Lambda \approx 1$ in its vicinity and the shape resembles a translation of the undeformed helix (right close-up). Elsewhere, the helical parameters may deviate significantly (left close-up).}
\label{fig:multiscalesschematic}
\end{figure}

As discussed in \S\ref{sec:papersummary}, in this paper we ignore bending of the helix axis. We seek a solution which, to leading order\footnote{Unless otherwise stated, the asymptotic limit we are considering is $\epsilon \to 0$. We use big-$O$ notation in the usual way: $f = O(g)$ means $|f/g| \leq C$ for some constant $C > 0$ as $\epsilon \to 0$. We also use the notation $f = \ord(g)$ to denote $|f/g| \to C$ for some constant $C > 0$ as $\epsilon \to 0$, and $f \sim g$ to denote asymptotic equivalence, i.e., $f/g \to 1$ as $\epsilon \to 0$.}, is a helix whose axis is parallel to the $Z$-axis with slowly-varying (unknown) pitch angle $\alpha$ and dimensionless contour wavelength $\Lambda$; see Fig.~\ref{fig:multiscalesschematic}. The undeformed filament corresponds to the values $\alpha=\aref$ and $\Lambda = 1$. We denote the slow lengthscale by $\hS$ (to be defined precisely below), so that $\alpha = \alpha(\hS,T)$ and $\Lambda = \Lambda(\hS,T)$.

Recall that we assume that the deformation is principally driven by the external force, and that the filament tip is free of forces and moments. Integrating Eq.~\eqref{eqn:forcebalance} backwards from the tip shows that the dimensionless force resultant at arclength $S$ is of magnitude $|\bF(S,T)| =\ord(\epsilon(L - S))$. We will show below that $\bF$ is of the same order as the change in helical parameters from their undeformed values, i.e. $(\alpha-\aref)$ and $(\Lambda -1)$. We therefore have
\beq
\left\lvert\alpha(\hS,T)-\aref\right\rvert = O(\epsilon(L - S)), \quad \left\lvert\Lambda(\hS,T) - 1\right\rvert = O(\epsilon(L - S)). \label{eqn:scalingestsalphalambda}
\eeq
The slow lengthscale $\hS$, over which the helical geometry varies significantly, is then defined by\footnote{We note that the lengthscale $\hS$ differs from the deformation lengthscale identified earlier ($s = [s]$ in dimensional terms), which, using $S = s/\lamref$ together with $\epsilon = (\lamref/[s])^3$, is given in terms of dimensionless arclength by $S = \ord(\epsilon^{-1/3})$. This is because $S = \ord(\epsilon^{-1/3})$ is the lengthscale over which the filament axis bends due to the external force, rather than the lengthscale over which the helical geometry (with straight axis) varies.}
\beq
S=\epsilon^{-1}\hS. \label{eqn:defnScal}
\eeq
In what follows we consider $\epsilon L= O(1)$, so that $\hS = O(1)$ throughout the filament and hence $O(1)$ changes to the helical geometry may occur. 

We also assume that any unsteadiness in the deformation is driven by the external force. Balancing the inertia term on the right-hand side of Eq.~\eqref{eqn:forcebalance} with the external force, and using the scaling behaviour $|\bR| = O(L) = O(\epsilon^{-1})$,  we obtain $\tast/[t] = O(\epsilon)$. Hence, we choose the timescale 
\beq
[t] = \epsilon^{-1}\tast. \label{eqn:defntscale}
\eeq


The basis of the multiple-scales method is to formally treat $S$ and $\hS$ as independent variables. The chain rule then implies that
\beq
\pd{}{S} = \pdS{} + \epsilon\pdhS{}. \label{eqn:chainrule}
\eeq
Throughout the following analysis, we consider $S$ varying over a general helical wave centred around the point with slow variable $\hS$, i.e., $S= \epsilon^{-1}\hS + \Delta S$ where $\Delta S \in (-\Lambda/2,\Lambda/2)$; see Fig.~\ref{fig:multiscalesschematic}. Later, we will integrate the force and moment balances with respect to $S$ over the helical wave, leading to evolution equations as $\hS$ and $T$ vary.

\subsection{Locally-helical kinematics}
\label{sec:multiscaleskinematics}
Before proceeding, we derive the kinematic quantities associated with our ansatz of a locally-helical shape. Let $\lbrace\bn,\bb,\bt\rbrace$ be the Frenet-Serret frame along the helical wave centred at $\hS$, associated with the local values of the pitch angle $\alpha$ and contour wavelength $\Lambda$. By analogy with the Frenet-Serret frame associated with the undeformed centreline, Eq.~\eqref{eqn:undeformedFSframe} (see also Fig.~\ref{fig:undeformedschematic}b), we postulate that
\begin{align}
\bn(S,\hS,T) & = -\be_r(S,\hS,T), \nonumber \\
\bb(S,\hS,T) & = -\cos\alpha(\hS,T)\,\be_\theta(S,\hS,T) + h\sin\alpha(\hS,T)\,\be_z, \nonumber \\
\bt(S,\hS,T) & = h\sin\alpha(\hS,T)\,\be_\theta(S,\hS,T) + \cos\alpha(\hS,T)\,\be_z, \label{eqn:FSframe}  
\end{align}
where $\be_r(S,\hS,T)$ and $\be_\theta(S,\hS,T)$ are the radial and azimuthal unit vectors, respectively, evaluated on the deformed centreline. (Here we are assuming that there is no change in chirality of the helix from its undeformed value, $h$.) These unit vectors are now parameterised by the \emph{unknown} winding angle $\Psi(S,\hS,T)$ of the deformed helix:
\beq
\be_r(S,\hS,T) = \cos\Psi(S,\hS,T)\,\be_x + \sin\Psi(S,\hS,T)\,\be_y, \quad \be_\theta(S,\hS,T) = -\sin\Psi(S,\hS,T)\,\be_x + \cos\Psi(S,\hS,T)\,\be_y. \label{eqn:defneRetheta} 
\eeq
Because the local contour wavelength is $\Lambda$, we have 
\beq
\pdS{\Psi} = \frac{2\pi h}{\Lambda}. \label{eqn:Psideriv}
\eeq
{After integrating, note that in general $\Psi \neq 2\pi h S/\Lambda$, i.e., the additive function of $\hS$ is generally non-zero. We will show below that this corresponds to a slowly-varying phase shift that accounts for variations in $\Lambda$.}

We write $\lbrace\bdonelead,\bdtwolead,\bdthreelead\rbrace$ for the leading-order directors associated with the locally-helical shape. Because we assume that the rod is unshearable, $\bdthreelead$ coincides with the unit tangent vector above: $\bdthreelead = \bt$. The other directors $\lbrace\bdonelead,\bdtwolead\rbrace$ are then a rotation of the Frenet-Serret basis vectors $\lbrace\bn,\bb\rbrace$ according to the excess axial twist in the rod. However, we note that the force balance \eqref{eqn:forcebalance} and moment balance \eqref{eqn:momentbalance} are both homogeneous at leading order in $\epsilon$ (using $[t] = \epsilon^{-1}\tast$ and $\delta = O(1)$). For a stationary helical rod in the absence of distributed loads, it can be shown that the excess twist must equal its undeformed value, in this case zero \citep{goriely1997c}. As a consequence, the leading-order directors are precisely the Frenet-Serret basis vectors reported above (Eq.~\eqref{eqn:FSframe}):
\beq
\bdonelead = \bn, \quad \bdtwolead = \bb, \quad \bdthreelead = \bt. \label{eqn:directorslead}
\eeq

We can determine the leading-order centreline, denoted $\bRlead$, and winding angle, $\Psi$, by solving the inextensibility constraint \eqref{eqn:inextensibility} with the tangent vector $\bdthreelead$ above. The calculation, detailed in \ref{sec:appendixcentreline}, yields
\begin{align}
\bRlead(S,\hS,T) & = \hR(\hS,T)\,\be_r(S,\hS,T) + \left[\epsilon^{-1}\hZ(\hS,T) + \Delta S\cos\alpha(\hS,T)\right]\be_z, \label{eqn:solnbRlead} \\
\Psi(S,\hS,T) & = \epsilon^{-1}\hPsi(\hS,T) + \frac{2\pi h \Delta S}{\Lambda(\hS,T)}, \label{eqn:solnPsilead}
\end{align}
where $\hR(\hS,T)$ is the slowly-varying helical radius, $\hZ(\hS,T)$ is the wavelength-averaged longitudinal coordinate, and $\hPsi(\hS,T)$ is the wavelength-averaged winding angle (see Fig.~\ref{fig:multiscalesschematic}). In terms of the slowly-varying pitch angle $\alpha$ and wavelength $\Lambda$, these are given by (\ref{sec:appendixcentreline})
\beq
\hR(\hS,T) = \frac{\Lambda(\hS,T)\sin\alpha(\hS,T)}{2\pi}, \quad \hZ(\hS,T) = \int_0^{\hS}\cos\alpha(\xi,T)\,\id\xi, \quad \hPsi(\hS,T) = 2\pi h \int_0^{\hS}\frac{\id\xi}{\Lambda(\xi,T)}, \label{eqn:solnshRavgZPsi}
\eeq
where we have made use of the boundary condition at the filament base, Eq.~\eqref{eqn:BCs}. These expressions show how changes to $\alpha$ and $\Lambda$ correspond to extensional (longitudinal) and torsional deformations about the helix axis. 

For later reference, we also calculate the centreline velocity and acceleration using the above expression for $\bRlead$:
\begin{align}
\pd{\bRlead}{T} & = \pd{\hR}{T}\be_r+\hR\left(\epsilon^{-1}\pd{\hPsi}{T}-\frac{2\pi h \Delta S}{\Lambda^2}\pd{\Lambda}{T}\right)\be_\theta+\left(\epsilon^{-1}\pd{\hZ}{T}-\Delta S \pd{\alpha}{T}\sin\alpha\right)\be_z \nonumber \\
& = \epsilon^{-1}\left(\hR\pd{\hPsi}{T}\be_{\theta}+\pd{\hZ}{T}\be_z\right) + O(1), \label{eqn:solnbRleaddot} \\
\pdd{\bRlead}{T} & = -\epsilon^{-2}\hR\left(\pd{\hPsi}{T}\right)^2\be_r + \epsilon^{-1}\left[\frac{4\pi h \Delta S}{\Lambda^2}\pd{\Lambda}{T}\pd{\hPsi}{T}\hR\be_r + \frac{1}{\hR}\pd{}{T}\left(\hR^2\pd{\hPsi}{T}\right)\be_\theta+\pdd{\hZ}{T}\be_z\right] + O(1). \label{eqn:solnbRleadddot}
\end{align}

In general, the external force $\bFe$ and moment $\bMe$ depend on the position and orientation of the rod. The above expressions, together with the expressions obtained in \S\ref{sec:multiscalesleadingorder} below, allow $\bFe$ and moment $\bMe$ to be expressed in terms of parameters governing the slowly-varying helical geometry. To be more concrete, suppose that $\bFe$ and $\bMe$ are known functions of the centreline position, centreline velocity, directors and angular velocity: $\bMe = \bMe(\bR,\partial\bR/\partial T,\bd_i,\bOmega)$ and $\bFe = \bFe(\bR,\partial\bR/\partial T,\bd_i,\bOmega)$. Formally, we can then expand
\beqn
\bFe \sim \bFelead, \quad \bMe \sim \bMelead \qquad \mathrm{where} \qquad \bFelead \equiv \bFe\left(\bRlead,\pd{\bRlead}{T},\bdilead,\bOmegalead\right), \quad \bMelead \equiv \bMe\left(\bRlead,\pd{\bRlead}{T},\bdilead,\bOmegalead\right).
\eeqn
Equations \eqref{eqn:directorslead}--\eqref{eqn:solnbRleaddot} (together with the solution for $\bOmegalead$ in Eq.~\eqref{eqn:strainspinlead} below) can then be used to express the external loads in terms $\alpha$ and $\Lambda$.

\subsection{Perturbation scheme}
\label{sec:perturbationscheme}
Following \cite{goriely1997a,goriely1997b}, we choose to not express the rod equations in component form with respect to a fixed external basis (using, for example, Euler angles to parameterise the directors) before perturbing quantities. Instead, we seek a regular perturbation expansion of the directors themselves in powers of $\epsilon$:
\beqn
 \bd_i = \bdilead + \epsilon\,\bdifirst + O(\epsilon^2), \qquad i = 1,2,3.
\eeqn
We need to ensure that the director basis remains orthonormal at each order of the expansion, i.e.~
\beqn
\delta_{ij} = \bd_i\cdot\bd_j =  \bdilead\cdot\bdjlead + \epsilon\left[\bdilead\cdot\bdjfirst  + \bdifirst\cdot\bdjlead \right] + O(\epsilon^2),
\eeqn
where $\delta_{ij}$ is the Kronecker delta. As explained by previous authors~\citep{goriely1997a,katsamba2019}, the $O(\epsilon)$ term in the above equation vanishes for all $i,j\in\lbrace 1,2,3\rbrace$ if and only if there exists a vector $\bPhi$ such that
\beq
\bdifirst = \bPhi \times \bdilead, \qquad i = 1,2,3. \label{eqn:directorsfirst}
\eeq

In addition, we expand the components (in the director basis) of the strain vector $\bU$, angular velocity vector $\bOmega$, resultant force $\bF$ and resultant moment $\bM$ in powers of $\epsilon$:
\beq
U_i = \Uilead + \epsilon\,\Uifirst + O(\epsilon^2), \quad \Omega_i  =\Omegailead + \epsilon\,\Omegaifirst + O(\epsilon^2), \quad M_i = \Milead + \epsilon\,\Mifirst + O(\epsilon^2), \quad F_i = \Filead + \epsilon\,\Fifirst + O(\epsilon^2). \label{eqn:expandcomponents}
\eeq
The advantage of this approach is that we can readily obtain the expansions of the \emph{vectors} $\bU$, $\bOmega$, $\bF$ and $\bM$ using Eqs.~\eqref{eqn:directorsfirst}--\eqref{eqn:expandcomponents}. Explicitly, for a general vector $\bW =\sum_{i}W_i \bd_i$ whose components have the regular expansion $W_i = \Wilead + \epsilon\,\Wifirst + O(\epsilon^2)$, we note that $\bW$ has the expansion \citep{goriely1997a}
\beq
\bW = \sum_{i = 1}^{3}W_i\bd_i = \bWlead + \epsilon\,\bWfirst + O(\epsilon^2) \qquad \mathrm{where} \qquad  \bWlead =  \sum_{i = 1}^{3}\Wilead\bdilead, \quad \bWfirst =  \sum_{i = 1}^{3}\Wifirst\bdilead + \bPhi \times \bWlead. \label{eqn:expandgeneralvector}
\eeq
In what follows, we will use this identity with $\bW = \bU$, $\bOmega$, $\bM$ and $\bF$. Our method is to substitute the above expansions into the dimensionless equations derived in \S\ref{sec:nondim}, expand spatial derivatives using the chain rule in Eq.~\eqref{eqn:chainrule}, and solve at each successive order in $\epsilon$.

\subsection{Leading-order problem}
\label{sec:multiscalesleadingorder}
The kinematic equations  \eqref{eqn:kinematicstrainspin} at leading order are
\beq
\pdS{\bdilead} = \bUlead \times \bdilead, \quad \pd{\bdilead}{T} = \bOmegalead \times \bdilead, \qquad i = 1,2,3. \label{eqn:kinematicstrainlead}
\eeq
Substituting the ansatz \eqref{eqn:directorslead} for the leading-order directors $\bdilead$, and evaluating derivatives of the Frenet-Serret basis vectors $\lbrace\bn,\bb,\bt\rbrace$ using Eqs.~\eqref{eqn:FSframe}--\eqref{eqn:Psideriv}, we obtain 
\beq
\bUlead =\hK\bb + \hT\bt = \frac{2\pi h}{\Lambda}\be_z, \quad \bOmegalead = h\pd{\alpha}{T}\bn + \pd{\Psi}{T}\be_z, \label{eqn:strainspinlead}
\eeq
i.e.,~$\bUlead$ is simply the Darboux vector associated with the Frenet-Serret frame $\lbrace\bn,\bb,\bt\rbrace$, where the slowly-varying Frenet curvature $\hK(\hS,T)$ and torsion $\hT(\hS,T)$ are
\beq
\hK(\hS,T) = \frac{2\pi\sin\alpha(\hS,T)}{\Lambda(\hS,T)}, \quad \hT(\hS,T) = \frac{2 \pi h \cos\alpha(\hS,T)}{\Lambda(\hS,T)}. \label{eqn:deformedDarboux}
\eeq
When the filament is undeformed, we have $\hK = \hKref$, $\hT = \hTref$, and $\lbrace\bn,\bb,\bt\rbrace = \lbrace\bnref,\bbref,\btref\rbrace$; the expression for $\bUlead$ then coincides with the undeformed strain vector, $\bUref = \hKref\bbref+\hTref\btref$, given earlier in Eq.~\eqref{eqn:undeformedstrain}.

The force balance \eqref{eqn:forcebalance} and moment balance \eqref{eqn:momentbalance} at leading-order are
\begin{align}
& \pdS{\bFlead} = \mathbf{0}, \label{eqn:forcebalancelead} \\
& \pdS{\bMlead} + \bdthreelead\times\bFlead = \mathbf{0}. \label{eqn:momentbalancelead}
\end{align}
The constitutive law \eqref{eqn:claw} with $\bUlead$ above (Eq.~\eqref{eqn:strainspinlead}) gives
\beq
\Monelead = 0, \quad \Mtwolead = \hK - \hKref, \quad \Mthreelead = \frac{\hT -\hTref}{1+\nu}. \label{eqn:momentlead}
\eeq
The force resultant satisfying Eqs.~\eqref{eqn:forcebalancelead}--\eqref{eqn:momentbalancelead} is then \citep{goriely1997c} 
\beq
\bFlead = \left(\frac{\hT-\hTref}{1+\nu}-\hT\frac{\hK-\hKref}{\hK}\right)\bUlead. \label{eqn:forcelead} 
\eeq
To see that $\bFlead$ is indeed independent of $S$, as required by Eq.~\eqref{eqn:forcebalancelead}, we note from Eq.~\eqref{eqn:strainspinlead} that $\bUlead = (2\pi h/\Lambda)\be_z$. We delay discussing the boundary conditions \eqref{eqn:BCs} until \S\ref{sec:multiscalessummary}.

For later reference, we also write the resultants in terms of the unit vectors in cylindrical polar coordinates. Making use of Eqs.~\eqref{eqn:FSframe} and \eqref{eqn:deformedDarboux}, we obtain
\beq
\bMlead = \hR F_Z\be_\theta + M_Z\be_z, \quad \bFlead = F_Z\be_z, \label{eqn:resultantsleadpolars}
\eeq
where the slowly-varying radius $\hR$ was defined in Eq.~\eqref{eqn:solnshRavgZPsi}, and $M_Z(\hS,T)$ and $F_Z(\hS,T)$ are given by 
\beq
M_Z =  \frac{\Lambda}{2\pi h}\left[\hK\left(\hK-\hKref\right) + \hT\frac{\hT-\hTref}{1+\nu}\right], \quad F_Z =  \frac{2\pi h}{\Lambda} \left(\frac{\hT-\hTref}{1+\nu}-\hT\frac{\hK-\hKref}{\hK}\right).
\label{eqn:resultantsleadZcomponents}
\eeq
We see that the moment resultant is composed of two terms: part of a wrench associated with torsion/winding about the helix axis ($Z$-axis), $M_Z\be_z$, and the torque produced by the force resultant, $-\bRlead\times\bFlead = \hR F_Z\be_{\theta}$ (from Eq.~\eqref{eqn:solnbRlead}).

We remark that an alternative method of solving the leading-order problem, which does not require posing the ansatz of a locally-helical shape \emph{a priori}, is to solve the homogeneous force and moment balances \eqref{eqn:forcebalancelead}--\eqref{eqn:momentbalancelead} directly. The general solution is a linear combination of six solutions whose coefficients depend on the slow variable, $\hS$. Two of these solutions correspond to a space-curve with slowly-varying Frenet curvature and torsion, and so yield a locally-helical solution; the other solutions correspond to bending about the $X$ and $Y$-axes, and so must vanish if the helix axis remains straight. Thus, the locally-helical form of the solution follows from the assumption of a straight helical axis. The disadvantage of this approach is that the interpretation of the slowly-varying coefficients in terms of physical parameters of the helix shape (e.g.~pitch angle and wavelength) is less clear than simply seeking a locally-helical solution to begin with.

\subsection{First-order problem}
\label{sec:multiscalesfirstorder}

\subsubsection{Derivation of the first-order equations}
At $O(\epsilon)$, the kinematic equations \eqref{eqn:kinematicstrainspin} for the strain vector become
\beqn
\pdS{\bdifirst} + \pdhS{\bdilead} = \bUlead \times \bdifirst + \bUfirst \times \bdilead, \qquad i = 1,2,3.
\eeqn
As well as $\hS$-derivatives now appearing, the vectors at this order contain both perturbations to the components (in the director basis) and perturbations to the directors; recall the identity \eqref{eqn:expandgeneralvector}. In particular, substituting $\bdifirst = \bPhi \times \bdilead$  and the expression for $\bUfirst$ (from setting $\bW = \bU$ in Eq.~\eqref{eqn:expandgeneralvector}), and making use of the leading-order kinematic equations \eqref{eqn:kinematicstrainlead}, the above simplifies to
\beq
\pdS{\bPhi}\times\bdilead + \pdhS{\bdilead} = \left(\sum_{j = 1}^{3}\Ujfirst\bdjlead\right) \times \bdilead, \qquad i = 1,2,3. \label{eqn:kinematicstrainfirst}
\eeq


We note from Eqs.~\eqref{eqn:FSframe}--\eqref{eqn:directorslead} that the leading-order directors $\bdilead$ depend on the slow variable $\hS$ and time $T$ only via the pitch angle $\alpha(\hS,T)$ and wavelength $\Lambda(\hS,T)$. Hence, $\hS$-derivatives of $\bdilead$ can be calculated analogously to time derivatives. By analogy with Eqs.~\eqref{eqn:kinematicstrainlead}--\eqref{eqn:strainspinlead}, we find that
\beq
\pdhS{\bdilead} = \bTheta\times\bdilead \qquad \mathrm{where} \qquad \bTheta \equiv h\pdhS{\alpha}\bn + \pdhS{\Psi}\be_z. \label{eqn:defnTheta}
\eeq
Substituting Eq.~\eqref{eqn:defnTheta} into Eq.~\eqref{eqn:kinematicstrainfirst}, and noting that $(\cdot)\times\bdilead = \mathbf{0}$ for $i=1,2,3$ implies that $(\cdot)=\mathbf{0}$, we obtain
\beq
\pdS{\bPhi} - \sum_{j = 1}^{3}\Ujfirst\bdjlead + \bTheta = \mathbf{0}. \label{eqn:Phideriv}
\eeq
Without the $\bTheta$-term, this is analogous to Eq.~(51) in \cite{goriely1997a} and Eq.~(27) in \cite{katsamba2019}. The $\bTheta$-term enters here because we are considering slow changes in the (currently unknown) leading-order shape, in addition to localised perturbations to the leading-order shape. From the scaling estimates \eqref{eqn:scalingestsalphalambda}, we have that $\left\lvert\alpha-\aref\right\rvert = O(\epsilon L)$ and $\left\lvert\Lambda - 1\right\rvert = O(\epsilon L)$ at the filament base, and hence $\bTheta = O(\epsilon L)$ from Eq.~\eqref{eqn:defnTheta}. It follows that the $\bTheta$-term can be neglected if $\epsilon L \ll 1$, corresponding to small deformations away from the undeformed shape, but must be considered in the case $\epsilon L = \ord(1)$.


With $\delta = O(1)$ and $[t]=\epsilon^{-1}\tast$ (recall Eq.~\eqref{eqn:defntscale}), the force balance \eqref{eqn:forcebalance} and moment balance \eqref{eqn:momentbalance} at $O(\epsilon)$ are
\begin{align}
& \pdS{\bFfirst} = -\pdhS{\bFlead} - \bFelead+\epsilon\pdd{\bRlead}{T}, \label{eqn:forcebalancefirst} \\
& \pdS{\bMfirst} + \bdthreelead\times\bFfirst + \bdthreefirst\times\bFlead = -\pdhS{\bMlead}-\delta\bMelead. \label{eqn:momentbalancefirst}
\end{align}
Equations \eqref{eqn:forcebalancefirst}--\eqref{eqn:momentbalancefirst} are linear in the first-order resultants $\bFfirst$ and $\bMfirst$; the inhomogeneous terms on the right-hand sides arise from inertia, external forces/moments and slow derivatives of the leading-order resultants. We insert the expressions for $\bFfirst$ and $\bMfirst$ (from setting $\bW = \bF$, $\bM$ in Eq.~\eqref{eqn:expandgeneralvector}) into Eqs.~\eqref{eqn:forcebalancefirst}--\eqref{eqn:momentbalancefirst} and evaluate $S$-derivatives of $\bdilead$, $\bFlead$ and $\bMlead$ using the leading-order equations \eqref{eqn:kinematicstrainlead} and \eqref{eqn:forcebalancelead}--\eqref{eqn:momentbalancelead}. After eliminating $\bPhi$ derivatives using Eq.~\eqref{eqn:Phideriv}, and expanding the $\hS$-derivatives of $\bFlead$ and $\bMlead$ using Eq.~\eqref{eqn:defnTheta}, the terms in $\bPhi$ and $\bTheta$ cancel yielding
\begin{align}
& \sum_{i = 1}^{3}\pdS{\Fifirst}\bdilead + \bUlead\times\left(\sum_{i = 1}^{3}\Fifirst\bdilead\right) + \left(\sum_{i = 1}^{3}\Uifirst\bdilead\right)\times\bFlead = -\sum_{i = 1}^{3}\pdhS{\Filead}\bdilead - \bFelead+\epsilon\pdd{\bRlead}{T}, \label{eqn:forcebalancefirstexpanded} \\
& \sum_{i = 1}^{3}\pdS{\Mifirst}\bdilead + \bUlead\times\left(\sum_{i = 1}^{3}\Mifirst\bdilead\right) + \left(\sum_{i = 1}^{3}\Uifirst\bdilead\right)\times\bMlead + \bdthreelead\times\left(\sum_{i = 1}^{3}\Fifirst\bdilead\right) = -\sum_{i = 1}^{3}\pdhS{\Milead}\bdilead -\delta\bMelead. \label{eqn:momentbalancefirstexpanded}
\end{align}
The $\left(\sum_{i = 1}^{3}\Uifirst\bdilead\right)\times\bFlead$ and $\left(\sum_{i = 1}^{3}\Uifirst\bdilead\right)\times\bMlead$ terms in Eqs.~\eqref{eqn:forcebalancefirstexpanded}--\eqref{eqn:momentbalancefirstexpanded} account for the corrections to the force and moment that arise from the perturbation $\bdifirst$ to the directors; the origin of these terms can be traced back to the $\bPhi \times \bFlead$ and $\bPhi \times \bMlead$ terms in the expressions for $\bFfirst$ and $\bMfirst$ (Eq.~\eqref{eqn:expandgeneralvector}). From Eqs.~\eqref{eqn:momentlead}--\eqref{eqn:forcelead} and the scaling estimates in Eq.~\eqref{eqn:scalingestsalphalambda}, we note that $\bMlead$ and $\bFlead$ are of size $O(\epsilon L)$. The $\left(\sum_{i = 1}^{3}\Uifirst\bdilead\right)\times\bFlead$ and $\left(\sum_{i = 1}^{3}\Uifirst\bdilead\right)\times\bMlead$ terms may therefore only be neglected if $\epsilon L \ll 1$, similar to the $\bTheta$-term above.

Finally, the constitutive law \eqref{eqn:claw} gives
\beq
\Monefirst = \Uonefirst, \quad \Mtwofirst = \Utwofirst, \quad \Mthreefirst = \frac{\Uthreefirst}{1+\nu}.  \label{eqn:clawfirst}
\eeq
Equations \eqref{eqn:Phideriv} and \eqref{eqn:forcebalancefirstexpanded}--\eqref{eqn:clawfirst} provide a closed system of equations for which solvability conditions can be formulated.


\subsubsection{Periodicity and solvability}
\label{sec:periodicitysolvability}
For general $\epsilon L = O(1)$, it can be shown that the homogeneous problem at first order, consisting of the homogeneous versions of Eqs.~\eqref{eqn:Phideriv} and \eqref{eqn:forcebalancefirstexpanded}--\eqref{eqn:clawfirst}, has non-trivial solutions. From the Fredholm Alternative Theorem \citep{keener1988}, solutions to the inhomogeneous problem then exist only if the inhomogeneous terms satisfy certain solvability conditions. These solvability conditions take the form of partial differential equations (PDEs) for the slowly-varying helix geometry.

To formulate the solvability conditions, we assume that the first-order components --- namely $\Uifirst$, $\Fifirst$, $\Mifirst$ and the components of $\bPhi$ --- are locally periodic over the helical wave centred at $S=\epsilon^{-1}\hS$ (i.e., they are $\Lambda$-periodic). With this assumption, two approaches are then possible:
\begin{itemize}
\item{In the first approach, we directly integrate the first-order equations over the helical wave $S \in (\epsilon^{-1}\hS-\Lambda/2,\epsilon^{-1}\hS+\Lambda/2)$, exploiting periodicity of the first-order components to eliminate all unknown variables so that only the leading-order components $\Filead$ , $\Milead$ and the terms $\partial^2\bRlead/\partial T^2$, $\bFelead$ and $\bMelead$ remain. The resulting equations can then be written as a closed system for the variables governing the slowly-varying helical geometry.}
\item{In the second approach, we write the first-order equations as a linear system of equations for the nine-dimensional vector $\bY^{(1)} = \lbrace\bPhi,\Fonefirst,\Ftwofirst,\Fthreefirst,\Monefirst,\Mtwofirst,\Mthreefirst\rbrace$ (using Eq.~\eqref{eqn:clawfirst} to eliminate $\Uifirst$ in terms of $\Mifirst$). The Fredholm Alternative Theorem states that the inhomogeneous part of this linear system must be orthogonal to $\Lambda$-periodic solutions of the homogeneous adjoint problem, when multiplied and integrated over the helical wave. This is a necessary condition for the first-order solution to be locally periodic and hence bounded across the filament, and is analogous to removing secular terms in the classical Poincar\'{e}-Lindstedt method. The leading-order solution should then provide a uniformly valid approximation of the filament shape~\citep{goriely2001}.}
\end{itemize} 
Below, we present the second approach since it does not rely on naive averaging and more cleanly reveals the difference between the cases $\epsilon L =\ord(1)$ and $\epsilon L\ll 1$. We will show that the terms of size $O(\epsilon L)$ in the first-order equations --- namely $\bTheta$ in Eq.~\eqref{eqn:Phideriv} and the terms $\left(\sum_{i = 1}^{3}\Uifirst\bdilead\right)\times\bFlead$ and $\left(\sum_{i = 1}^{3}\Uifirst\bdilead\right)\times\bMlead$ in Eqs.~\eqref{eqn:forcebalancefirstexpanded}--\eqref{eqn:momentbalancefirstexpanded} --- play a key role in determining the number of relevant solutions of the homogeneous adjoint problem, and hence the number of solvability conditions obtained. Nevertheless, in \ref{sec:appendixsolvability} we show how the same conditions can be derived directly using the first approach.

\begin{DIFnomarkup} 

\subsection{Solvability conditions for the first-order problem: $\epsilon L = \ord(1)$}
\label{sec:multiscalessolvability}
We first consider the case $\epsilon L = \ord(1)$, corresponding to $\ord(1)$ changes to the helical geometry. It is convenient to express vectors as a matrix of components (referred to as a \emph{triple}) with respect to the leading-order directors $\lbrace\bdonelead,\bdtwolead,\bdthreelead\rbrace$; a similar strategy was employed by \cite{goriely1997c} to obtain dispersion relations for small-amplitude oscillations of helical rods. To avoid ambiguity in what follows, we use the convention that \textsf{sans serif} fonts denote a matrix of components. Using the expressions in Eqs.~\eqref{eqn:FSframe}, \eqref{eqn:directorslead}--\eqref{eqn:solnbRlead}, \eqref{eqn:solnbRleadddot}, \eqref{eqn:strainspinlead}, \eqref{eqn:resultantsleadpolars} and \eqref{eqn:defnTheta}, the triples corresponding to the unit vectors $\be_r$, $\be_\theta$, $\be_z$,  tangent vector $\bdthreelead = \bt$, centreline position $\bRlead$, centreline acceleration $\partial^2\bRlead/\partial T^2$, strain vector $\bUlead$, resultant moment $\bMlead$, resultant force $\bFlead$ and vector $\bTheta$ are, respectively,
\begin{gather}
\sse_r = \begin{pmatrix} -1 \\ 0 \\ 0 \end{pmatrix}, \quad
\sse_{\theta} = \begin{pmatrix} 0 \\ -\cos\alpha \\ h\sin\alpha \end{pmatrix}, \quad
\sse_z = \begin{pmatrix} 0 \\ h\sin\alpha \\ \cos\alpha \end{pmatrix}, \quad
\ssdthreelead = \begin{pmatrix} 0 \\ 0 \\ 1 \end{pmatrix}, \nonumber \\
\ssRlead = \hR\sse_r + \left(\epsilon^{-1}\hZ+\Delta S\cos\alpha\right)\sse_z, \quad
\epsilon\pdd{\ssRlead}{T} = \hR\left[\frac{4\pi h\Delta S}{\Lambda^2}\pd{\Lambda}{T}\pd{\hPsi}{T}-\epsilon^{-1}\left(\pd{\hPsi}{T}\right)^2\right]\sse_r+\frac{1}{\hR}\pd{}{T}\left(\hR^2\pd{\hPsi}{T}\right)\sse_{\theta}+\pdd{\hZ}{T}\sse_z, \nonumber \\
\ssUlead  = \begin{pmatrix} 0 \\ \hK \\ \hT \end{pmatrix}, \quad
\ssMlead = \hR F_Z \sse_{\theta} + M_Z\sse_z,  \quad 
\ssFlead =  F_Z\sse_z, \quad 
\ssTheta = -h\pdhS{\alpha}\sse_r + \pdhS{\Psi}\sse_z.  \label{eqn:defntriples}
\end{gather}
We define the triples $\sstUfirst$, $\sstMfirst$, $\sstFfirst$ as the components of $\bUfirst$, $\bMfirst$, $\bFfirst$ that are independent of $\bPhi$ (recall the identity in Eq.~\eqref{eqn:expandgeneralvector}). Explicitly,
\beqn
\sstUfirst  = \begin{pmatrix} \Uonefirst \\ \Utwofirst \\ \Uthreefirst \end{pmatrix}, \quad \sstMfirst  = \begin{pmatrix} \Monefirst \\ \Mtwofirst \\ \Mthreefirst \end{pmatrix}, \quad \sstFfirst  = \begin{pmatrix} \Fonefirst \\ \Ftwofirst \\ \Fthreefirst \end{pmatrix}.
\eeqn
We also write $\ssPhi$, $\ssMelead$ and $\ssFelead$ for the triples corresponding to, respectively, $\bPhi$, $\bMelead$ and $\bFelead$. The first-order problem, consisting of Eqs.~\eqref{eqn:Phideriv} and \eqref{eqn:forcebalancefirstexpanded}--\eqref{eqn:clawfirst}, can then be expressed as
\begin{align}
    & \pdS{\ssPhi} + \ssUlead\times\ssPhi - \sstUfirst + \ssTheta = \sszero_{3\times 1}, \label{eqn:Phiderivcomp} \\
    & \pdS{\sstMfirst} + \ssUlead\times\sstMfirst  + \sstUfirst\times\ssMlead + \ssdthreelead\times\sstFfirst = -\pdhS{\ssMlead}-\delta\ssMelead, \label{eqn:momentbalancefirstexpandedcomp} \\
    & \pdS{\sstFfirst} + \ssUlead\times\sstFfirst  + \sstUfirst\times\ssFlead =  -\pdhS{\ssFlead} - \ssFelead + \epsilon\pdd{\ssRlead}{T}, \label{eqn:forcebalancefirstexpandedcomp} \\
    & \sstMfirst = \ssK_{\nu}\sstUfirst, \label{eqn:clawfirstcomp}
\end{align}
where, in the final equation, we introduce the diagonal stiffness matrix $\ssK_{\nu} = \diag{(1,1,1/[1+\nu])}$. 


We eliminate $\sstUfirst$ using Eq.~\eqref{eqn:clawfirstcomp}. Equations \eqref{eqn:Phiderivcomp}--\eqref{eqn:forcebalancefirstexpandedcomp} can then be written as the linear system
\beq
\pdS{\ssYfirst} + \ssA\,\ssYfirst = \begin{pmatrix} -\ssTheta \\ -\pdhSt{\ssMlead}-\delta\ssMelead \\ -\pdhSt{\ssFlead} - \ssFelead+\epsilon\,\pdd{\ssRlead}{T} \end{pmatrix}, \label{eqn:systemfirst}
\eeq
where we have introduced the  $9\times 1$ column vector $\ssYfirst$ and the $9\times 9$ block matrix $\ssA$:
\beqn
\ssYfirst = \begin{pmatrix} \ssPhi \\ \sstMfirst \\ \sstFfirst \end{pmatrix}, \qquad
\ssA = \begin{pmatrix} \skewmat{\ssUlead} & -\ssK_{\nu}^{-1} & \sszero_{3\times 3} \\ 
\sszero_{3\times 3} & \skewmat{\ssUlead} - \skewmat{\ssMlead}\ssK_{\nu}^{-1} & \skewmat{\ssdthreelead} \\ 
\sszero_{3\times 3} & -\skewmat{\ssFlead}\ssK_{\nu}^{-1} & \skewmat{\ssUlead}  \end{pmatrix}.
\eeqn
Here we use the notation $\skewmat{\ssV}$ for the $3\times 3$ skew-symmetric matrix whose off-diagonal elements are the components of the triple $\ssV$, so that for any triples $\ssV$, $\ssW$ we have the identity $\skewmat{\ssV}\ssW = \ssV\times\ssW$. 


\subsubsection{Homogeneous adjoint problem and solvability conditions}
To formulate solvability conditions on the first-order problem, we consider the homogeneous adjoint problem associated with Eq.~\eqref{eqn:systemfirst}. In component form, with solution vector $\ssYadj$, this reads
\beq
-\pdS{\ssYadj} + \ssA^{\mathrm{T}} \ssYadj = \sszero_{9\times 1}. \label{eqn:systemfirstadj}
\eeq
Using skew-symmetry of the matrix blocks, we calculate
\beqn
\ssA^{\mathrm{T}} = -\begin{pmatrix} \skewmat{\ssUlead} & \sszero_{3\times 3} & \sszero_{3\times 3} \\
\ssK_{\nu}^{-1} & \skewmat{\ssUlead} - \ssK_{\nu}^{-1}\skewmat{\ssMlead} & - \ssK_{\nu}^{-1}\skewmat{\ssFlead} \\ 
 \sszero_{3\times 3} & \skewmat{\ssdthreelead} & \skewmat{\ssUlead}  \end{pmatrix}.
\eeqn
Writing $\ssYadj = (\ssPhiadj,\ssMadj,\ssFadj)^{\mathrm{T}}$, the homogeneous adjoint problem can be written as
\begin{align}
& \pdS{\ssPhiadj} + \ssUlead\times\ssPhiadj = \sszero_{3\times 1}, \nonumber \\
& \pdS{\ssMadj} + \ssUlead\times\ssMadj + \ssK_{\nu}^{-1}\left(\ssPhiadj -\ssMlead\times\ssMadj - \ssFlead\times\ssFadj\right)  = \sszero_{3\times 1}, \nonumber \\
& \pdS{\ssFadj} + \ssUlead\times\ssFadj + \ssdthreelead\times\ssMadj  =  \sszero_{3\times 1}. \label{eqn:systemfirstadjexpand}
\end{align}

Without the $\ssK_{\nu}^{-1}$ term, the second and third equations in \eqref{eqn:systemfirstadjexpand} are equivalent to the leading-order force and moment balances (Eqs.~\eqref{eqn:forcebalancelead}--\eqref{eqn:momentbalancelead}) when the force and moment resultants are swapped, i.e., with $\ssMadj = \ssFlead$ and $\ssFadj = \ssMlead$. We also note that the first equation is trivially satisfied by $\ssPhiadj = \sszero_{3\times 1}$. Hence, seeking solutions for which $\ssPhiadj = \sszero_{3\times 1}$ and $\ssMlead\times\ssMadj + \ssFlead\times\ssFadj = \sszero_{3\times 1}$ (which ensures that the $\ssK_{\nu}^{-1}$ term is zero), we obtain two linearly independent solutions:
\beqn
\ssPhiadj = \sszero_{3\times 1}, \quad \ssMadj = \sszero_{3\times 1}, \quad \ssFadj = \ssFlead \qquad \mathrm{and} \qquad \ssPhiadj = \sszero_{3\times 1}, \quad \ssMadj = \ssFlead, \quad \ssFadj = \ssMlead.
\eeqn
Substituting for $\ssFlead$ and $\ssMlead$ using the expressions in Eq.~\eqref{eqn:defntriples} and re-scaling, we have the linearly independent solutions for $\ssYadj$: 
\beq
\ssYadj_1 =  \begin{pmatrix} \sszero_{3\times 1} \\ \sszero_{3\times 1} \\ \sse_z \end{pmatrix}, \quad \ssYadj_2 =  \begin{pmatrix} \sszero_{3\times 1} \\ \sse_z \\ \hR\sse_{\theta} \end{pmatrix} = \begin{pmatrix} \sszero_{3\times 1} \\ \sse_z \\ \sse_z\times\ssRlead \end{pmatrix}, \label{eqn:adjsoln1,2}
\eeq
where, in the second solution, we have taken a linear combination to remove the $\sse_z$-component of $\ssMlead$ (the second equality follows from the expression for $\ssRlead$ in Eq.~\eqref{eqn:defntriples}). Since these solutions are independent of $S$, they must correspond to eigenvectors of $\ssA^{\mathrm{T}}$ with eigenvalue zero, as can be readily verified directly.

To find other solutions to Eq.~\eqref{eqn:systemfirstadjexpand}, we note that if $\ssMadj = \sszero_{3\times 1}$ the second equation in \eqref{eqn:systemfirstadjexpand} simplifies to
\beqn
\ssPhiadj - \ssFlead\times\ssFadj =  \sszero_{3\times 1}.
\eeqn
With $\ssMadj = \sszero_{3\times 1}$, the other equations in \eqref{eqn:systemfirstadjexpand} are equivalent to the statement that $\ssPhiadj$ and $\ssFadj$ are triples corresponding to constant vectors, and so they must be a linear combination of the Cartesian unit vectors $\sse_x$, $\sse_y$ and $\sse_z$. Using Eqs.~\eqref{eqn:defneRetheta} and \eqref{eqn:defntriples}, we have
\beq
\sse_x = \cos\Psi\sse_r - \sin\Psi\sse_{\theta} = \begin{pmatrix} -\cos\Psi \\ \cos\alpha\sin\Psi \\ -h\sin\alpha\sin\Psi \end{pmatrix}, \quad 
\sse_y = \sin\Psi\sse_r + \cos\Psi\sse_{\theta} = \begin{pmatrix} -\sin\Psi \\ -\cos\alpha\cos\Psi \\ h\sin\alpha\cos\Psi \end{pmatrix}. \label{eqn:defnssexey}
\eeq
We can therefore generate linearly independent solutions of the form $\ssYadj = (\ssFlead\times\ssFadj,\sszero_{3\times 1},\ssFadj)^{\mathrm{T}}$ by choosing $\ssFadj = \sse_x$, $\ssFadj = \sse_y$ and $\ssFadj = \sse_z$. The choice $\ssFadj = \sse_z$ yields $\ssYadj_1$ above, while choosing $\ssFadj = \sse_x$ and $\ssFadj = \sse_y$ yields
\beq
\ssYadj_3 =  \begin{pmatrix} \ssFlead\times\sse_x \\ \sszero_{3\times 1} \\ \sse_x \end{pmatrix}, \quad \ssYadj_4 =  \begin{pmatrix} \ssFlead\times\sse_y \\\sszero_{3\times 1} \\ \sse_y \end{pmatrix}. \label{eqn:adjsoln3,4}
\eeq
These solutions can also be derived by considering the eigenvectors of $\ssA^{\mathrm{T}}$ with eigenvalue $\pm 2\pi h i/\Lambda$; these eigenvalues arise as the non-zero eigenvalues of the $\skewmat{\ssUlead}$ blocks on the diagonal of $\ssA^{\mathrm{T}}$.

Other linearly independent solutions of Eq.~\eqref{eqn:systemfirstadj} can be determined by considering the remaining eigenvalues and eigenvectors of $\ssA^{\mathrm{T}}$. For general $\alpha$ and $\Lambda$, there are four distinct remaining eigenvalues that have a non-zero real part and an imaginary part that is generally not equal to $\pm 2\pi h/\Lambda$. This is a consequence of the $\skewmat{\ssMlead}$ and $\skewmat{\ssFlead}$ terms in $\ssA$, which makes the first-order operator different to that obtained in the leading-order problem. Since these solutions are not $\Lambda$-periodic, they are not relevant in determining solvability conditions so we do not discuss them here\footnote{An additional solution can be determined from a generalised eigenvector corresponding to the eigenvalue $0$, which is generally defective (algebraic multiplicity $3$, geometric multiplicity $2$). However, this solution is also not $\Lambda$-periodic. It may be shown that the eigenvalues $\pm 2\pi h i/\Lambda$ are non-defective (each has algebraic multiplicity $1$ and geometric multiplicity $1$), and so only give rise to the solutions $\ssYadj_3$ and $\ssYadj_4$.}.


We are now in a position to formulate the solvability conditions for the first-order problem, Eq.~\eqref{eqn:systemfirst}. Dotting Eq.~\eqref{eqn:systemfirst} with $\ssYadj_i$ and integrating by parts over the wavelength centred at $S = \epsilon^{-1}\hS$, we obtain ($i = 1,2,3,4$) 
\begin{align}
\intwave{\ssYadj_i\cdot\begin{pmatrix} -\ssTheta \\ -\pdhSt{\ssMlead}-\delta\ssMelead \\ -\pdhSt{\ssFlead} - \ssFelead+\epsilon\,\pdd{\ssRlead}{T} \end{pmatrix}} & = \intwave{\ssYadj_i\cdot\left(\pdS{\ssYfirst} + \ssA\,\ssYfirst\right)} \nonumber \\
& = \left[\ssYadj_i\cdot\ssYfirst\right]_{\epsilon^{-1}\hS-\frac{\Lambda}{2}}^{\epsilon^{-1}\hS+\frac{\Lambda}{2}} + \intwave{\ssYfirst\cdot\left(-\pdS{\ssYadj_i} + \ssA^{\mathrm{T}} \ssYadj_i\right)} \nonumber \\
& = 0, \label{eqn:solvability}
\end{align}
where, in the final equality, the boundary terms vanish since $\ssYadj_i$ and $\ssYfirst$ are $\Lambda$-periodic\footnote{As discussed in \S\ref{sec:periodicitysolvability}, we are assuming that $\ssYfirst$ is $\Lambda$-periodic. We note from Eqs.~\eqref{eqn:defntriples}, \eqref{eqn:adjsoln1,2} and \eqref{eqn:adjsoln3,4} that the components of the adjoint solutions $\ssYadj_i$ are combinations on the triples $\sse_z$, $\sse_{\theta}$, $\sse_x$ and $\sse_y$, with coefficients that depend only on the slow variable, $\hS$; hence the $\ssYadj_i$ are also $\Lambda$-periodic.}, while the integrand is zero by definition of the adjoint solution, $\ssYadj_i$. Inserting the homogeneous adjoint solutions $\ssYadj_i$ from Eqs.~\eqref{eqn:adjsoln1,2} and \eqref{eqn:adjsoln3,4} into Eq.~\eqref{eqn:solvability}, and permuting the scalar triple products that appear, we obtain the solvability conditions ($i=1,2,3,4$)
\begin{align}
& \intwave{\sse_z\cdot\left(\pdhS{\ssFlead}+\ssFelead-\epsilon\pdd{\ssRlead}{T}\right)} = 0, \label{eqn:solvability1component} \\
& \intwave{\sse_z\cdot\left[\pdhS{\ssMlead}+\delta\ssMelead + \ssRlead\times\left(\pdhS{\ssFlead}+\ssFelead-\epsilon\pdd{\ssRlead}{T}\right)\right]} = 0, \label{eqn:solvability2component} \\
& \intwave{\sse_x\cdot\left(\pdhS{\ssFlead}+\ssTheta\times\ssFlead + \ssFelead-\epsilon\pdd{\ssRlead}{T}\right)} = \intwave{\sse_y\cdot\left(\pdhS{\ssFlead}+\ssTheta\times\ssFlead + \ssFelead-\epsilon\pdd{\ssRlead}{T}\right)} = 0. \label{eqn:solvability3,4component}
\end{align}

\subsubsection{Simplification to the (straight) equivalent-rod equations and discussion}
\label{sec:simplifytoeffectivecol}
To express Eqs.~\eqref{eqn:solvability1component}--\eqref{eqn:solvability3,4component} in the form of PDEs for the slowly-varying helical geometry, we substitute the expressions for $ \ssFlead$, $\ssMlead$ and $\partial^2\ssRlead/\partial T^2$ in Eq.~\eqref{eqn:defntriples} and simplify using the following identities (which follow from 
Eqs.~\eqref{eqn:solnPsilead}, \eqref{eqn:defnssexey} and $\Delta S =S - \epsilon^{-1}\hS$):
\begin{gather}
\intwave{\sse_x\cdot\sse_r} = 0, \quad \intwave{\sse_x\cdot\sse_\theta} = 0, \quad \intwave{\sse_y\cdot\sse_r} = 0, \quad\intwave{\sse_y\cdot\sse_\theta} = 0, \nonumber \\
\intwave{\sse_x\cdot\Delta S\sse_r} = -\frac{h \Lambda}{2\pi}\sin\left(\epsilon^{-1}\hPsi\right), \qquad \intwave{\sse_y\cdot\Delta S\sse_r} = \frac{h \Lambda}{2\pi}\cos\left(\epsilon^{-1}\hPsi\right). \label{eqn:integralidentities}
\end{gather}
In addition, we use bars to denote the average over the helical wave centred at $S=\epsilon^{-1}\hS$, i.e., for a function $f(S,\hS,T)$,
\beqn
\overline{f}(\hS,T) \equiv \frac{1}{\Lambda} \intwave{f(S,\hS,T)}.
\eeqn
After writing back in vector form, Eqs.~\eqref{eqn:solvability1component}--\eqref{eqn:solvability3,4component} become
\begin{align}
& \pd{F_Z}{\hS} + \overline{\be_z\cdot\bFelead} - \pdd{\hZ}{T}= 0,  \label{eqn:effectiveforcebalance} \\
& \pd{M_Z}{\hS} + \delta\overline{\be_z\cdot\bMelead} +\overline{\be_z\cdot\left(\bRlead\times\bFelead\right)} -\pd{}{T}\left(\hR^2\pd{\hPsi}{T}\right) = 0, \label{eqn:effectivemomentbalance} \\
& \overline{\be_x\cdot\bFelead} + \frac{2\hR}{\Lambda}\pd{\Lambda}{T}\pd{\hPsi}{T}	\sin\left(\epsilon^{-1}\hPsi\right) = 0,  \qquad \overline{\be_y\cdot\bFelead} - \frac{2\hR}{\Lambda}\pd{\Lambda}{T}\pd{\hPsi}{T}	\cos\left(\epsilon^{-1}\hPsi\right) = 0,  \label{eqn:effectiveforceX,Y}
\end{align}
where hereafter we drop the $|_{S}$ on $\hS$-derivatives whenever there is no ambiguity (i.e., for variables that have no explicit dependence on the fast variable, $S$). We refer to Eqs.~\eqref{eqn:effectiveforcebalance}--\eqref{eqn:effectiveforceX,Y} as the `(straight) equivalent-rod' equations. The first two equations correspond to wavelength-averaged force and moment balances, respectively, about the helix axis, $\be_z$. The final two equations correspond to wavelength-averaged force balances in the off-axis directions $\be_x$, $\be_y$; these equations state that a locally-helical solution with straight axis is only possible if the external force exactly balances the off-axis components of the filament acceleration (when averaged over the wavelength). In particular, for equilibrium solutions, a straight helix axis is only possible if the off-axis components of the external force average to zero. 

Remarkably, wavelength-averaged \emph{moment} balances in the off-axis directions have not arisen in our analysis: there are no additional conditions for the external moment needed for a straight helix axis (to leading order). The reason for this can be understood from the first-order force and moment balances, Eqs.~\eqref{eqn:forcebalancefirstexpanded}--\eqref{eqn:momentbalancefirstexpanded}. We note that any off-axis moments, which tend to cause axis bending, are of size $O(\delta) = O(1)$, while the $\left(\sum_{i = 1}^{3}\Uifirst\bdilead\right)\times\bFlead$ and $\left(\sum_{i = 1}^{3}\Uifirst\bdilead\right)\times\bMlead$ terms are of size $O(\epsilon L)$ (recall the discussion below Eq.~\eqref{eqn:momentbalancefirstexpanded}). Thus, for $\epsilon L =\ord(1)$, any off-axis moments can be balanced by choosing the unknown \emph{first-order} strain components $\Uifirst$ appropriately, to ensure that there is no net bending moment. Crucially, when $\epsilon L \ll 1$, this is no longer possible and so additional conditions on the \emph{leading-order} quantities are required to ensure a balance of moments in the off-axis directions. We consider this case in the next subsection.


\subsection{Solvability conditions for the first-order problem: $\epsilon L \ll 1$}
\label{sec:multiscalessolvabilitylinear}
We return to the first-order problem, consisting of Eqs.~\eqref{eqn:Phideriv} and \eqref{eqn:forcebalancefirstexpanded}--\eqref{eqn:clawfirst}. The key difference here is that, in the limit $\epsilon L\ll 1$, we may neglect
\begin{itemize}
    \item the $\bTheta$ term in Eq.~\eqref{eqn:Phideriv}; and
    \item the $\left(\sum_{i = 1}^{3}\Uifirst\bdilead\right)\times\bFlead$ and $\left(\sum_{i = 1}^{3}\Uifirst\bdilead\right)\times\bMlead$ terms in Eqs.~\eqref{eqn:forcebalancefirstexpanded}--\eqref{eqn:momentbalancefirstexpanded},
\end{itemize}
since these terms are a factor $O(\epsilon L)\ll 1$
smaller than the other terms; recall the discussions immediately below Eqs.~\eqref{eqn:Phideriv} and \eqref{eqn:momentbalancefirstexpanded}. Without these terms, Eqs.~\eqref{eqn:forcebalancefirstexpanded}--\eqref{eqn:momentbalancefirstexpanded} do not depend on the first-order strain components $\Uifirst$, so that Eq.~\eqref{eqn:Phideriv} and the constitutive law \eqref{eqn:clawfirst} decouple from the problem. We therefore consider only Eqs.~\eqref{eqn:forcebalancefirstexpanded}--\eqref{eqn:momentbalancefirstexpanded} in what follows.

We proceed similarly to \S\ref{sec:multiscalessolvability}, using \textsf{sans serif} fonts to denote a matrix of components with respect to the leading-order directors $\lbrace\bdonelead,\bdtwolead,\bdthreelead\rbrace$. When $\epsilon L\ll 1$, to leading order we have $\alpha \sim \aref$, $\Lambda \sim 1$, $\Psi \sim \Psiref$ and so we can identify the leading-order directors with their undeformed counterparts: $\lbrace\bdonelead,\bdtwolead,\bdthreelead\rbrace \sim \lbrace\bdoneref,\bdtworef,\bdthreeref\rbrace$. In addition, from Eq.~\eqref{eqn:solnshRavgZPsi}, we may expand the wavelength-averaged longitudinal coordinate and winding angle about the undeformed configuration:
\beq
\hZ(\hS,T) = \hZref(\hS) + \Delta\hZ(\hS,T), \quad \hPsi(\hS,T) = \hPsiref(\hS) + \Delta\hPsi(\hS,T) \qquad \mathrm{where} \qquad \hZref(\hS)\equiv\hS\cos\aref, \quad \hPsiref(\hS) \equiv 2\pi h\hS, \label{eqn:defnDeltahZhPsi}
\eeq
where $\Delta\hZ,\ \Delta\hPsi = O(\epsilon L) \ll 1$. From Eqs.~\eqref{eqn:defntriples} and \eqref{eqn:defnssexey}, we then have
\begin{gather}
\sse_{\theta} \sim \begin{pmatrix} 0 \\ -\cos\aref \\ h\sin\aref \end{pmatrix}, \quad
\sse_z = \begin{pmatrix} 0 \\ h\sin\aref \\ \cos\aref \end{pmatrix}, \quad
\ssUlead  \sim \begin{pmatrix} 0 \\ \hKref \\ \hTref \end{pmatrix}, \quad
\sse_x \sim \cos\Psiref\sse_r - \sin\Psiref\sse_{\theta}, \quad 
\sse_y \sim \sin\Psiref\sse_r + \cos\Psiref\sse_{\theta}, \nonumber \\
\ssRlead \sim \ssRref \equiv \hRref\sse_r + S\cos\aref\sse_z, \quad
\epsilon\pdd{\ssRlead}{T} \sim \frac{\sin\aref}{2\pi}\pdd{\Delta\hPsi}{T}\sse_{\theta}+\pdd{\Delta\hZ}{T}\sse_z. \label{eqn:defntripleslinear}
\end{gather}

 In component form, the first-order problem, consisting of Eqs.~\eqref{eqn:forcebalancefirstexpanded}--\eqref{eqn:momentbalancefirstexpanded} without the $\Uifirst$ terms, can be written as the $6$-dimensional linear system of equations
\beq
\pdS{\sstYfirst} + \sstA\,\sstYfirst = \begin{pmatrix} -\pdhSt{\ssMlead}-\delta\ssMelead \\ -\pdhSt{\ssFlead} - \ssFelead+\epsilon\,\pdd{\ssRlead}{T} \end{pmatrix} \qquad \mathrm{where} \qquad \sstYfirst = \begin{pmatrix} \sstMfirst \\ \sstFfirst \end{pmatrix}, \quad
\sstA = \begin{pmatrix} \skewmat{\ssUlead} & \skewmat{\ssdthreelead} \\ 
0_{3\times 3} & \skewmat{\ssUlead}  \end{pmatrix}. \label{eqn:systemfirstlinear}
\eeq


\subsubsection{Homogeneous adjoint problem and solvability conditions}
The homogeneous adjoint problem associated with Eq.~\eqref{eqn:systemfirstlinear} is 
\beqn
-\pdS{\sstYadj} + \sstA^{\mathrm{T}}\sstYadj = 0_{6\times1} \qquad \mathrm{where} \qquad \sstYadj = \begin{pmatrix} \ssMadj \\ \ssFadj \end{pmatrix}, \quad
\sstA^{\mathrm{T}} = -\begin{pmatrix} \skewmat{\ssUlead} & 0_{3\times 3} \\ 
    \skewmat{\ssdthreelead} & \skewmat{\ssUlead}  \end{pmatrix}.
\eeqn
By considering the eigenvalues and eigenvectors of $\sstA^{\mathrm{T}}$, we find the $6$ real linearly-independent solutions:
\beq
\sstYadj_1 = \begin{pmatrix} \sszero_{3\times 1} \\ \sse_z \end{pmatrix}, \quad
\sstYadj_2 =  \begin{pmatrix} \sse_z \\ \sse_z\times\ssRref \end{pmatrix}, \quad
\sstYadj_3 =  \begin{pmatrix} \sszero_{3\times 1} \\ \sse_x \end{pmatrix}, \quad \sstYadj_4 =  \begin{pmatrix} \sszero_{3\times 1} \\ \sse_y \end{pmatrix}, \quad
\sstYadj_5 = \begin{pmatrix} \sse_x \\ \sse_x\times\ssRref \end{pmatrix}, \quad
\sstYadj_6 = \begin{pmatrix} \sse_y \\ \sse_y\times\ssRref \end{pmatrix}. \label{eqn:adjsolnslinear}
\eeq
The solutions $\sstYadj_1$ and $\sstYadj_2$ arise as eigenvectors corresponding to the eigenvalue $0$, which is always non-defective (it may be shown that the algebraic and geometric multiplicity are both equal to $2$). The solutions $\sstYadj_3$ and $\sstYadj_4$ are obtained from eigenvectors of the eigenvalues $\pm 2\pi h i$, each of which is defective (algebraic multiplicity $2$, geometric multiplicity $1$); the final two solutions, $\sstYadj_5$ and $\sstYadj_6$, are derived from their generalised eigenvectors\footnote{We note that the solutions $\sstYadj_5$ and $\sstYadj_6$, while periodic along $\sse_z$, are not periodic in the off-axis directions $\sse_x$, $\sse_y$: we have $\sse_x\times\ssRref = \hRref\sin\Psiref\sse_z-S\cos\aref\sse_y$ and $\sse_y\times\ssRref = -\hRref\cos\Psiref\sse_z+S\cos\aref\sse_x$. However, we will show below that these solutions still give rise to physically-meaningful solvability conditions, and the unknown boundary terms that arise when formulating the solvability conditions (from integrating by parts) vanish under a reasonable assumption.}.

We formulate the solvability conditions for the first-order problem similarly to \S\ref{sec:simplifytoeffectivecol}: we dot Eq.~\eqref{eqn:systemfirstlinear} with $\sstYadj_i$ and integrate by parts over the helical wave centred at $S=\epsilon^{-1}\hS$, which, for $\Lambda\sim 1$, is $S\in(\epsilon^{-1}\hS-\Lambda/2,\epsilon^{-1}\hS+\Lambda/2)$. We obtain ($i=1,2,3,4,5,6$)
\begin{align}
\intwaveref{\sstYadj_i\cdot\begin{pmatrix} -\pdhSt{\ssMlead}-\delta\ssMelead \\ -\pdhSt{\ssFlead} - \ssFelead+\epsilon\,\pdd{\ssRlead}{T} \end{pmatrix}} & = \intwaveref{\sstYadj_i\cdot\left(\pdS{\sstYfirst} + \sstA\,\sstYfirst\right)} \nonumber \\
& = \left[\sstYadj_i\cdot\sstYfirst\right]_{\epsilon^{-1}\hS-\frac{1}{2}}^{\epsilon^{-1}\hS+\frac{1}{2}} + \intwaveref{\sstYfirst\cdot\left(-\pdS{\sstYadj_i} + \sstA^{\mathrm{T}} \sstYadj_i\right)} \nonumber \\
& = 0, \label{eqn:solvabilitylinear}
\end{align}
assuming that the boundary terms again vanish\footnote{For $i = 1,2,3,4$ the boundary terms are guaranteed to vanish by periodicity of $\sstYfirst$ and the triples $\sse_z$, $\sse_{\theta}$, $\sse_x$ and $\sse_y$ (periodicity $\Lambda\sim 1$), as in \S\ref{sec:multiscalessolvability}. For $i =5,6$ we require that $\left[(\sse_x\times\ssRref)\cdot\sstFfirst\right]_{\epsilon^{-1}\hS-\frac{1}{2}}^{\epsilon^{-1}\hS+\frac{1}{2}} = 0$ and $\left[(\sse_y\times\ssRref)\cdot\sstFfirst\right]_{\epsilon^{-1}\hS-\frac{1}{2}}^{\epsilon^{-1}\hS+\frac{1}{2}} = 0$. Permuting the scalar triple products shows that these are equivalent to requiring that the off-axis components of $\ssRref\times\sstFfirst$ are $1$-periodic, which is physically reasonable.}. Substituting the solutions from Eq.~\eqref{eqn:adjsolnslinear} and permuting scalar triple products yields
\begin{align}
    & \intwaveref{\ssV\cdot\left(\pdhS{\ssFlead}+\ssFelead-\epsilon\pdd{\ssRlead}{T}\right)} = 0, \label{eqn:solvability1componentlinear} \\
    & \intwaveref{\ssV\cdot\left[\pdhS{\ssMlead}+\delta\ssMelead + \ssRref\times\left(\pdhS{\ssFlead}+\ssFelead-\epsilon\pdd{\ssRlead}{T}\right)\right]} = 0, \label{eqn:solvability2componentlinear}  \qquad \mathrm{where} \qquad \ssV = \sse_x,~\sse_y,~\sse_z.
    \end{align}


\subsubsection{Simplification to the (straight) equivalent-rod equations and discussion}
\label{sec:simplifytoeffectivecollinear}
To simplify Eqs.~\eqref{eqn:solvability1componentlinear}--\eqref{eqn:solvability2componentlinear}, we insert the expressions in Eq.~\eqref{eqn:defntripleslinear} (together with $\ssMlead = \hR F_Z \sse_{\theta} + M_Z\sse_z$ and $\ssFlead =  F_Z\sse_z$) and simplify various integrals using the identities in Eq.~\eqref{eqn:integralidentities}. After writing back in vector form, again using bars to denote helical averages and dropping the $|_{S}$ on $\hS$-derivatives, we arrive at the (straight) equivalent-rod equations
\begin{align}
& \pd{F_Z}{\hS} + \overline{\be_z\cdot\bFelead} - \pdd{\Delta\hZ}{T}= 0,  \label{eqn:effectiveforcebalancelinear} \\
& \pd{M_Z}{\hS} + \delta\overline{\be_z\cdot\bMelead} + \overline{\be_z\cdot\left(\bRref\times\bFelead\right)}  -\left(\hRref\right)^2\pdd{\Delta\hPsi}{T} = 0, \label{eqn:effectivemomentbalancelinear} \\
& \overline{\be_x\cdot\bFelead} = 0,  \qquad \overline{\be_y\cdot\bFelead} = 0,  \label{eqn:effectiveforceX,Ylinear} \\
& \delta\overline{\be_x\cdot\bMelead} + \overline{\be_x\cdot\left(\bRref\times\bFelead\right)}-\frac{h\hRref\cos\aref}{2\pi}\pdd{\Delta\hPsi}{T}\sin\left(\epsilon^{-1}\hPsiref\right)  = 0, \label{eqn:effectivemomentXlinear} \\
& \delta\overline{\be_y\cdot\bMelead} + \overline{\be_y\cdot\left(\bRref\times\bFelead\right)}+\frac{h\hRref\cos\aref}{2\pi}\pdd{\Delta\hPsi}{T}\cos\left(\epsilon^{-1}\hPsiref\right)  = 0. \label{eqn:effectivemomentYlinear}
\end{align}
The first four equations are equivalent to those obtained in the case $\epsilon L=\ord(1)$, i.e., Eqs.~\eqref{eqn:effectiveforcebalance}--\eqref{eqn:effectiveforceX,Y}, when expanded in the small-deformation limit $\epsilon L \ll 1$ (the dynamic terms in Eq.~\eqref{eqn:effectiveforceX,Y} do not appear in Eq.~\eqref{eqn:effectiveforceX,Ylinear} since these terms are quadratic in the deformation). The final two equations, which only arise in the case $\epsilon L \ll 1$, correspond to wavelength-averaged moment balances in the off-axis directions $\be_x$, $\be_y$; these equations provide additional conditions for a straight helix axis.

\subsection{Summary}
\label{sec:multiscalessummary}
In this section, we have derived the dimensionless (straight) equivalent-rod equations using a multiple-scales analysis of the Kirchhoff rod equations. These equations --- which consist of Eqs.~\eqref{eqn:effectiveforcebalance}--\eqref{eqn:effectiveforceX,Y} in the case $\epsilon L=\ord(1)$, and Eqs.~\eqref{eqn:effectiveforcebalancelinear}--\eqref{eqn:effectivemomentYlinear} in the case $\epsilon L\ll1$ --- can be interpreted as force and moment balances averaged over the slowly-varying helical wavelength. In particular, we showed how the equations can be justified rigorously via solvability conditions on an appropriate first-order problem, when the solution is expanded in powers of the dimensionless parameter $\epsilon$. 

The equivalent-rod equations are supplemented with the expressions for the leading-order centreline $\bRlead$, helix radius $\hR$, wavelength-averaged longitudinal coordinate $\hZ$ and winding angle $\hPsi$ in Eqs.~\eqref{eqn:solnbRlead}--\eqref{eqn:solnshRavgZPsi}, together with the equations for the leading-order resultants $F_Z$ and $M_Z$ in Eq.~\eqref{eqn:resultantsleadZcomponents}. Furthermore, in terms of the slow variable $\hS$, the filament tip is located at $\hS = \epsilon L$ so that the boundary conditions \eqref{eqn:BCs} become
\beq
F_Z(\epsilon L,T) = M_Z(\epsilon L,T) = 0. \label{eqn:effectiveBCs}
\eeq
In \S\ref{sec:equivalent-rod}, we explicitly show how these additional equations allow us to write the equivalent-rod equations as a closed, quasi-linear system of equations for two independent variables that uniquely characterise the locally-helical shape.

\end{DIFnomarkup} 

\section{Analysis of the (straight) equivalent-rod equations}
\label{sec:equivalent-rod}
Up to this point, it has been useful to work with various parameters characterising the slowly-varying helical shape --- the pitch angle $\alpha$, contour wavelength $\Lambda$, Frenet curvature $\hK$, and Frenet torsion $\hT$ (defined in Eq.~\eqref{eqn:deformedDarboux}) --- despite the fact that only two parameters are needed to uniquely specify the local geometry. In this section, considering the case $\epsilon L=\ord(1)$, we show how the (straight) equivalent-rod equations derived in the previous section can be written as a closed system of PDEs for (i) the pair $(\alpha,\Lambda)$ and (ii) the wavelength-averaged longitudinal coordinate and winding angle, $(\hZ,\hPsi)$ (defined in Eq.~\eqref{eqn:solnshRavgZPsi}). For each pair of solution variables, we derive effective stiffness coefficients that depend nonlinearly on the variables. Depending on the nature of the external forces and moments under consideration, one of these formulations may be more convenient. Focussing on steady solutions in the $(\alpha,\Lambda)$-formulation, we analyse the Jacobian determinant associated with the system of differential equations; in particular, we determine where the equivalent-rod equations are singular and instabilities of the helical filament may occur. We then analyse the equivalent-rod equations in the small-deformation limit $\epsilon L \ll 1$, showing that we recover linearised stiffness coefficients reported previously \citep{phillips1972,costello1975,jiang1989,jiang1991}. We finish the section with a brief discussion of the limit of vanishing pitch angle $\aref \to 0$ in our equivalent-rod equations.

Throughout this section, we focus on the equivalent-rod equations that correspond to wavelength-averaged force and moment balances about the helix axis: Eqs.~\eqref{eqn:effectiveforcebalance}--\eqref{eqn:effectivemomentbalance}, and their counterparts \eqref{eqn:effectiveforcebalancelinear}--\eqref{eqn:effectivemomentbalancelinear} in the limit $\epsilon L \ll 1$. The remaining equations (Eq.~\eqref{eqn:effectiveforceX,Y} in the case $\epsilon L=\ord(1)$; Eqs.~\eqref{eqn:effectiveforceX,Ylinear}--\eqref{eqn:effectivemomentYlinear} in the case $\epsilon L \ll 1$) are constraints on the external loading needed for a straight helix axis, and will be assumed to be satisfied in what follows. We further discuss the significance of these remaining equations in \S\ref{sec:conclusiondiscussion}.
   
\subsection{Formulation in terms of the pitch angle and wavelength}
\label{sec:effectivehelixgeometry}
Expressions for the leading-order force and moment resultants, $F_Z$ and $M_Z$, were given earlier in Eq.~\eqref{eqn:resultantsleadZcomponents}. Using $\hK = 2\pi\sin\alpha/\Lambda$ and $\hT = 2\pi h\cos\alpha/\Lambda$ (recall Eq.~\eqref{eqn:deformedDarboux}), $F_Z$ and $M_Z$ can be written in terms of $\alpha$ and $\Lambda$ alone:
\begin{align}
F_Z & = \frac{4\pi^2\csc\alpha\left[\nu\cos\alpha\left(\Lambda\sin\aref - \sin\alpha\right) -\Lambda\sin\left(\alpha-\aref\right)\right]}{(1+\nu)\Lambda^2}, \nonumber \\
M_Z & = \frac{2\pi h\left[1 -\nu\sin\alpha\left(\Lambda\sin\aref-\sin\alpha\right) - \Lambda\cos\left(\alpha-\aref\right)\right]}{(1+\nu)\Lambda}. \label{eqn:resultantsleadZcomponentshelixgeometry}
\end{align}
We insert the above into Eqs.~\eqref{eqn:effectiveforcebalance}--\eqref{eqn:effectivemomentbalance} and expand the $\hS$-derivatives using the chain rule. Using $\overline{\be_z\cdot\left(\bRlead\times\bFelead\right)} = \overline{\hR\be_{\theta}\cdot\bFelead}$ (from Eq.~\eqref{eqn:solnbRlead}), and substituting the expressions in Eq.~\eqref{eqn:solnshRavgZPsi} for $\hR$, $\hZ$ and $\hPsi$, we obtain the following system of equations for $\alpha(\hS,T)$ and $\Lambda(\hS,T)$:
\begin{align}
& C_1 \pd{\alpha}{\hS} + C_2\pd{\Lambda}{\hS} + \overline{\be_z\cdot\bFelead} + \int_0^{\hS}\left[\pdd{\alpha}{T}\sin\alpha+\left(\pd{\alpha}{T}\right)^2\cos\alpha\right]\Bigg\lvert_{\hS=\xi}\id\xi = 0, \label{eqn:effectiveforcebalancehelixgeometry} \\
& C_3 \pd{\alpha}{\hS} + C_4\pd{\Lambda}{\hS} + \delta\overline{\be_z\cdot\bMelead} + \frac{\Lambda\sin\alpha}{2\pi}\overline{\be_\theta\cdot\bFelead}+\frac{h}{2\pi}\pd{}{T}\left[\Lambda^2\sin^2\alpha\int_0^{\hS}\left(\frac{1}{\Lambda^2}\pd{\Lambda}{T}\right)\Bigg\lvert_{\hS=\xi}\id\xi\right] = 0, \label{eqn:effectivemomentbalancehelixgeometry}
\end{align}
where the dimensionless stiffness coefficients $C_i = C_i\left(\alpha,\Lambda\right)$ ($i=1,2,3,4$) are the partial derivatives
\begin{align}
C_1 & = \pd{F_Z}{\alpha} = \frac{4\pi^2\left[\nu\sin\alpha - (1+\nu)\Lambda\sin\aref\csc^2\alpha\right]}{(1+\nu)\Lambda^2}, \nonumber \\
C_2 & = \pd{F_Z}{\Lambda} = \frac{4\pi^2\left\lbrace 2\nu\cos\alpha + \Lambda\csc\alpha\left[\sin\left(\alpha-\aref\right) - \nu\sin\aref\cos\alpha\right]\right\rbrace}{(1+\nu)\Lambda^3}, \nonumber \\
C_3 & = \pd{M_Z}{\alpha} = \frac{2\pi h\sin\alpha\left\lbrace 2\nu\cos\alpha+\Lambda\csc\alpha\left[\sin\left(\alpha-\aref\right) - \nu\sin\aref\cos\alpha\right]\right\rbrace}{(1+\nu)\Lambda} = \frac{h\Lambda^2\sin\alpha}{2\pi}C_2,\nonumber \\
C_4 & = \pd{M_Z}{\Lambda} = -\frac{2\pi h\left(1+ \nu\sin^2\alpha\right)}{(1+\nu)\Lambda^2}. \label{eqn:defnCi}
\end{align}
As discussed in \S\ref{sec:multiscaleskinematics}, provided that the external force and moment are known functions of the leading-order centreline and orientation of the rod, the external force and moment can, in principle, be expressed in terms of $\alpha$ and $\Lambda$ using Eqs.~\eqref{eqn:directorslead}--\eqref{eqn:solnbRleaddot}.

The boundary conditions \eqref{eqn:effectiveBCs} at the filament tip imply that
\beq
\alpha(\epsilon L,T) = \aref, \quad \Lambda(\epsilon L,T) = 1. \label{eqn:effectiveBCshelixgeometry}
\eeq
The system is closed by appropriate initial conditions (if considering unsteady deformations).

For the sake of completeness, in \ref{sec:appendixFrenetformulation} we also provide the formulation of Eqs.~\eqref{eqn:effectiveforcebalance}--\eqref{eqn:effectivemomentbalance} in terms of the Frenet curvature and torsion, $(\hK,\hT)$. Nevertheless, the $(\alpha,\Lambda)$-formulation above has the advantage that the dependence of the stiffness coefficients \eqref{eqn:defnCi} on $\Lambda$ is particularly simple, compared to the dependence of the corresponding coefficients on $\hK$ and $\hT$ (reported in \ref{sec:appendixFrenetformulation}). In \S\ref{sec:effectivejacobian}, we show how this allows us to analytically determine the region of the $(\alpha,\Lambda)$-plane where the Jacobian of the system vanishes, indicating that Eqs.~\eqref{eqn:effectiveforcebalance}--\eqref{eqn:effectivemomentbalance} are singular.

\subsection{Formulation in terms of the wavelength-averaged longitudinal coordinate and winding angle}
\label{sec:effectivehZhPsi}
The above formulation is convenient if considering steady solutions, in which the external force and moment are independent of the deformation (e.g.~for gravitational loading) or depend only on the local orientation of the filament. However, if considering unsteady deformations, or if  the external loads depend on the centreline position or its time derivatives (as is generally the case with hydrodynamic loading, for example), then Eqs.~\eqref{eqn:effectiveforcebalancehelixgeometry}--\eqref{eqn:effectivemomentbalancehelixgeometry} take the form of integro-differential equations for $(\alpha,\Lambda)$. This is because the leading-order centreline $\bRlead$ is expressed in terms of integrals of $\alpha$ and $\Lambda$ via Eqs.~\eqref{eqn:solnbRlead}--\eqref{eqn:solnshRavgZPsi}. In such scenarios, it may be more appropriate to write the evolution equations in terms of the wavelength-averaged longitudinal coordinate, $\hZ$, and winding angle, $\hPsi$. 

Using the expressions in Eq.~\eqref{eqn:solnshRavgZPsi}, the helical parameters $\alpha$, $\Lambda$ and $\hR$ can be expressed in terms of $\hZ$ and $\hPsi$:
\beq
\alpha = \arccos\left(\pd{\hZ}{\hS}\right), \quad \Lambda = \frac{2\pi h}{\partial\hPsi/\partial\hS}, \quad \hR = \frac{h\sqrt{1-\left(\partial\hZ/\partial\hS\right)^2}}{\partial\hPsi/\partial\hS}. \label{eqn:alphalambdaavgZPsi}
\eeq
Inserting these expressions into $F_Z$ and $M_Z$ (Eq.~\eqref{eqn:resultantsleadZcomponentshelixgeometry}), the equivalent-rod equations \eqref{eqn:effectiveforcebalance}--\eqref{eqn:effectivemomentbalance} can then be written in terms of $\hZ(\hS,T)$ and $\hPsi(\hS,T)$:
\begin{align}
& K_1 \pdd{\hZ}{\hS} + K_2\pdd{\hPsi}{\hS} + \overline{\be_z\cdot\bFelead} - \pdd{\hZ}{T}= 0, \label{eqn:forceavgZPsi} \\
& K_3 \pdd{\hZ}{\hS} + K_4\pdd{\hPsi}{\hS} + \delta\overline{\be_z\cdot\bMelead} + \overline{\be_z\cdot\left(\bRlead\times\bFelead\right)} - \pd{}{T}\left[\frac{1-\left(\partial\hZ/\partial\hS\right)^2}{\left(\partial\hPsi/\partial\hS\right)^2}\pd{\hPsi}{T}\right] = 0, \label{eqn:momentavgZPsi}
\end{align}
where the dimensionless stiffness coefficients $K_i = K_i\left(\hZ,\hPsi\right)$ ($i=1,2,3,4$) are
\begin{align}
& K_1 = \pd{\hPsi}{\hS}\left\lbrace\frac{2\pi h\sin\aref}{\left[1-\left(\partial\hZ/\partial\hS\right)^2\right]^{3/2}} - \frac{\nu}{1+\nu}\pd{\hPsi}{\hS}\right\rbrace, \nonumber \\
& K_2 = K_3 = 2\pi h\sin\aref\frac{\partial\hZ/\partial\hS}{\sqrt{1-\left(\partial\hZ/\partial\hS\right)^2}} - \frac{2}{1+\nu}\left(\pi h\cos\aref + \nu\pd{\hZ}{\hS}\pd{\hPsi}{\hS}\right), \qquad K_4 = 1 - \frac{\nu}{1+\nu}\left(\pd{\hZ}{\hS}\right)^2. \label{eqn:defnKi}
\end{align}
We emphasise that the coupling coefficients $K_2$ and $K_3$ are equal in this formulation. The force and moment-free conditions \eqref{eqn:effectiveBCs} now correspond to Neumann conditions
\beq
\pd{\hZ}{\hS}(\epsilon L,T) = \cos\aref, \quad \pd{\hPsi}{\hS}(\epsilon L,T) = 2\pi h.  \label{BCstipavgZPsi}
\eeq
Because Eqs.~\eqref{eqn:forceavgZPsi}--\eqref{eqn:momentavgZPsi} are second order in $\hS$, we also impose the boundary conditions at the filament base (Eq.~\eqref{eqn:BCs}):
\beq
\hZ(0,T) = \hPsi(0,T) = 0, \label{BCsbaseavgZPsi}
\eeq
together with initial conditions (if relevant).

\subsection{Steady solutions: Jacobian determinant and singular behaviour}
\label{sec:effectivejacobian}
In each of the formulations above, the stiffness coefficients are given by partial derivatives of the resultants $F_Z$ and $M_Z$ with respect to the solution variables (or their first-order derivatives in the case of $\hZ$ and $\hPsi$). The matrix of stiffness coefficients therefore corresponds to the Jacobian matrix $J$ of the vector-valued function $(F_Z,M_Z)$. In particular, for the $(\alpha,\Lambda)$-formulation,
\beqn
J = \frac{\partial(F_Z,M_Z)}{\partial(\alpha,\Lambda)} = \begin{pmatrix} C_1 & C_2 \\ C_3 & C_4 \end{pmatrix},
\eeqn
where the coefficients $C_i\left(\alpha,\Lambda\right)$ are given in Eq.~\eqref{eqn:defnCi}. As discussed at the end of \S\ref{sec:effectivehelixgeometry}, these stiffness coefficients have a particularly simple form compared to the corresponding coefficients in other formulations. The Jacobian determinant simplifies to
\beq
\det J = C_1 C_4 - C_2 C_3 = \frac{4\pi^3 h\left(C_a \Lambda^2 + C_b \Lambda + C_c\right)}{(1+\nu)^2\Lambda^4}, \label{eqn:Jacobdet}
\eeq
where we have introduced the coefficients $C_{a,b,c}(\alpha;\aref,\nu)$:
\begin{align*}
C_a & = -2\csc\alpha\left[\sin\left(\alpha-\aref\right) - \nu\sin\aref\cos\alpha\right]^2, \\
C_b & = 2(1+\nu)\sin\aref\csc^2\alpha\left(1+\nu\sin^2\alpha\right) - 8\nu\cos\alpha\left[\sin\left(\alpha-\aref\right) - \nu\sin\aref\cos\alpha\right], \\
C_c & = -2\nu\sin\alpha\left(1 + 4\nu - 3\nu\sin^2\alpha\right).
\end{align*}

Crucially, the Jacobian determinant can be zero for certain values of $\alpha$ and $\Lambda$, which correspond to critical points on the $(\alpha,\Lambda)$ phase-plane where the steady equivalent-rod equations \eqref{eqn:effectiveforcebalancehelixgeometry}--\eqref{eqn:effectivemomentbalancehelixgeometry} are singular. Because the coefficients $C_{a,b,c}$ are independent of $\Lambda$, the determinant vanishes if and only if $\Lambda$ is a root of the quadratic polynomial in the numerator of Eq.~\eqref{eqn:Jacobdet}:
\beq
\Lambda = \Lambda_{\pm} \equiv \frac{-C_b \pm \sqrt{C_b^2 - 4C_a C_c}}{2C_a}. \label{eqn:defnLambdapm}
\eeq
As $\alpha$ varies, the real roots $\Lambda_{\pm}$ trace out branches of critical points on the $(\alpha,\Lambda)$ phase-plane.

\begin{figure}
\centering
\includegraphics[width=0.7\textwidth]{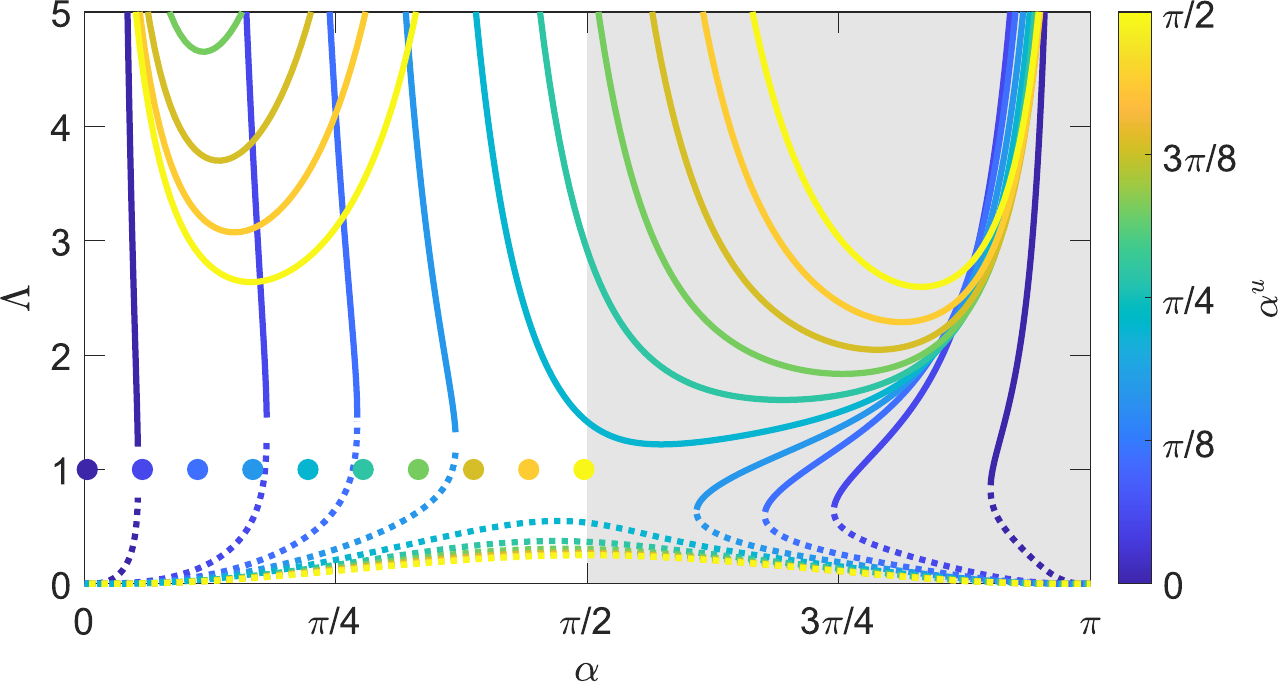} 
\caption{Branches of critical points on the $(\alpha,\Lambda)$ phase-plane where the matrix of stiffness coefficients is singular: for each value of $\aref$ (shown by the colourbar), we plot the real roots $\Lambda_{-}$ (solid curves) and $\Lambda_{+}$ (dotted curves), defined in Eq.~\eqref{eqn:defnLambdapm}, as $\alpha$ varies (here $\nu = 1/3$). The point $(\alpha,\Lambda) = (\aref,1)$, corresponding to the undeformed configuration, is also plotted (circles). The unphysical region $\alpha \geq \pi/2$ is shaded.}
\label{fig:determinantroots}
\end{figure}

In Fig.~\ref{fig:determinantroots} we plot the branches of critical points for various values of $\aref$ (with fixed $\nu = 1/3$). Generally speaking, we find that, for sufficiently small $\aref$, there is an interval around $\alpha = \pi/2$ where the discriminant $(C_b^2 - 4C_a C_c)$ is negative and hence the roots $\Lambda_{\pm}$ in Eq.~\eqref{eqn:defnLambdapm} are complex. In particular, part of this interval lies in the physical range $0 < \alpha < \pi/2$ in which the filament does not intersect itself (neglecting a small correction due to its finite cross-section). The size of the interval decreases as $\aref$ increases (or $\nu$ decreases), eventually shrinking to zero as both fold points --- where the discriminant is zero and $\Lambda_{+} = \Lambda_{-}$ --- collide and disappear. For larger values of $\aref$, the discriminant $(C_b^2 - 4C_a C_c)$ is positive over the interval $0 <\alpha <\pi/2$: the branches $\Lambda_{\pm}$ are disconnected and the roots are real across the interval. (The vertical asymptotes to the $\Lambda_{-}$ branches in Fig.~\ref{fig:determinantroots} correspond to where the leading coefficient $C_a = 0$.)

More precisely, it may be shown that the discriminant attains its minimum value at $\alpha = \pi/2$, for any values of $\aref$ and $\nu$. By considering the minimum value as a function of $\aref$, we find that the discriminant is positive at $\alpha = \pi/2$ (and hence the roots are real across the entire interval $0 <\alpha <\pi/2$) if and only if
\beqn
\aref > \arctan\left[\frac{2\nu^{1/2}}{(1+\nu)^{3/2}}\right].
\eeqn
In this regime, as $\aref$ increases further for fixed $\nu$, the local minimum that can be observed in the $\Lambda_{-}$ branches (solid curves in Fig.~\ref{fig:determinantroots}) for $\alpha < \pi/2$ decreases significantly, eventually reaching values around $2.5$ when $\aref \approx \pi/2$. In contrast, the $\Lambda_{+}$ branches (dotted curves) generally remain smaller than unity and decrease slightly as $\aref$ increases. This behaviour as $\aref$ varies has implications for the physical relevance of the critical points, as discussed later in \S\ref{sec:conclusiondiscussion}. 
 
\subsection{Linearised equations: $\epsilon L \ll 1$}
\label{sec:linearisedeqns}
The above formulations apply to the case $\epsilon L = \ord(1)$, in which the changes to the helical parameters are comparable to the undeformed values. This means that while the evolution equations are quasi-linear (i.e., linear in the first-order derivatives of $\alpha$ and $\Lambda$, or the second-order derivatives of $\hZ$ and $\hPsi$), the coefficients are nonlinear functions of the solution variables, so that analytical progress is generally not possible. We consider here the limit $\epsilon L \ll 1$, corresponding to relatively short or stiff filaments, for which the deformation is small and the equivalent-rod equations are linear.

Considering the formulation in terms of helix angle and contour wavelength, we write
\beqn
\alpha(\hS,T) = \aref + \Delta\alpha(\hS,T), \quad \Lambda(\hS,T) = 1+\Delta\Lambda(\hS,T),
\eeqn
where $\Delta\alpha,\ \Delta\Lambda = O(\epsilon L) \ll 1$. Using Eq.~\eqref{eqn:solnshRavgZPsi}, these are related to the perturbations $(\Delta\hZ,\Delta\hPsi)$, introduced in \S\ref{sec:multiscalessolvabilitylinear}, by
\beq
\Delta\hZ= -\sin\aref\int_0^{\hS}\Delta\alpha(\xi,T)\,\id\xi, \quad \Delta\hPsi = -2\pi h\int_0^{\hS}\Delta\Lambda(\xi,T)\,\id\xi. \label{eqn:relateexpandhelixgeometrytoavgZPsi}
\eeq
Neglecting terms of $O(\epsilon L)^2$ in Eq.~\eqref{eqn:resultantsleadZcomponentshelixgeometry}, the linearised force and moment resultants are given by
\beq
F_Z \sim \frac{4\pi^2\left[\nu\Delta\Lambda\cos\aref - \Delta\alpha\csc\aref\left(1+\nu\cos^2\aref\right)\right]}{1+\nu}, \quad 
M_Z \sim \frac{2\pi h\left[\nu\Delta\alpha\cos\aref\sin\aref - \Delta\Lambda\left(1+\nu\sin^2\aref\right)\right]}{1+\nu}. \label{eqn:resultantsleadZcomponentshelixgeometrylinearised}
\eeq
Substituting the above expressions into the equivalent-rod equations \eqref{eqn:effectiveforcebalancelinear}--\eqref{eqn:effectivemomentbalancelinear} and boundary conditions \eqref{eqn:effectiveBCs}, and using $\overline{\be_z\cdot\left(\bRref\times\bFelead\right)} = \overline{\hRref\be_{\theta}\cdot\bFelead}$ where $\hRref = \sin\aref/(2\pi)$, we obtain the linear system of equations:
\begin{align}
& C_1^u \pd{\Delta\alpha}{\hS} + C_2^u\pd{\Delta\Lambda}{\hS}+ \overline{\be_z\cdot\bFelead} + \sin\aref\int_0^{\hS}\pdd{\Delta\alpha}{T}\Bigg\lvert_{\hS=\xi}\id\xi = 0, \label{eqn:linearisedforcehelixgeometry} \\
& C_3^u \pd{\Delta\alpha}{\hS} + C_4^u\pd{\Delta\Lambda}{\hS} + \delta\overline{\be_z\cdot\bMelead} + \frac{\sin\aref}{2\pi}\overline{\be_\theta\cdot\bFelead} +\frac{h\sin^2\aref}{2\pi}\int_0^{\hS}\pdd{\Delta\Lambda}{T}\Bigg\lvert_{\hS=\xi}\id\xi  = 0, \label{eqn:linearisedmomenthelixgeometry} \\
& \Delta\alpha(\epsilon L,T) = \Delta\Lambda(\epsilon L,T) = 0, \label{eqn:linearisedBCshelixgeometry}
\end{align}
where the linearised stiffness coefficients are
\beq
C_1^u = -\frac{4\pi^2\csc\aref \left(1+\nu\cos^2\aref\right)}{1+\nu}, \quad C_2^u = \frac{4\pi^2\nu\cos\aref}{1+\nu}, \quad C_3^u = \frac{2 \pi h\nu\cos\aref\sin\aref}{1+\nu}, \quad C_4^u = -\frac{2\pi h\left(1+\nu\sin^2\aref\right)}{1+\nu}. \label{eqn:defnCiref}
\eeq
Equations \eqref{eqn:linearisedforcehelixgeometry}--\eqref{eqn:defnCiref} can also be obtained directly from \eqref{eqn:effectiveforcebalancehelixgeometry}--\eqref{eqn:effectiveBCshelixgeometry} by linearising in the perturbations $\Delta\alpha$, $\Delta\Lambda$.

We note that the Jacobian determinant of the linearised system, $\Jref$, is
\beq
\Jref =C_1^u C_4^u - C_2^u C_3^u = \frac{8\pi^3 h \csc\aref}{1+\nu} \neq 0. \label{eq:Jlin}
\eeq
The (steady) linearised system therefore does not exhibit singular behaviour, as is also evidenced by the fact that, for each $\aref$, the undeformed point $(\aref,1)$ lies away from the branches of critical points in Fig.~\ref{fig:determinantroots}. Moreover, when the inertia terms are negligible, we can solve the system \eqref{eqn:linearisedforcehelixgeometry}--\eqref{eqn:linearisedmomenthelixgeometry} uniquely for $\partial\Delta\alpha/\partial\hS$ and $\partial\Delta\Lambda/\partial\hS$, and formally integrate with the boundary conditions \eqref{eqn:linearisedBCshelixgeometry}. After substituting the expressions for the coefficients $C_i^u$, we obtain
 \begin{align}
\Delta\alpha & = -\int_{\hS}^{\epsilon L}\left[\frac{\sin\aref\left(1+\nu\sin^2\aref\right)}{4\pi^2}\overline{\be_z\cdot\bFelead}\Big\lvert_{\hS = \xi} + \frac{h\nu\cos\aref\sin\aref}{2\pi} \left(\delta\overline{\be_z\cdot\bMelead} + \frac{\sin\aref}{2\pi}\overline{\be_{\theta}\cdot\bFelead}\right)\Bigg\lvert_{\hS = \xi}\right]\id\xi, \nonumber \\
\Delta\Lambda & = -\int_{\hS}^{\epsilon L}\left[\frac{\nu\cos\aref\sin^2\aref}{4\pi^2}\overline{\be_z\cdot\bFelead}\Big\lvert_{\hS = \xi} + \frac{h\left(1+\nu\cos^2\aref\right)}{2\pi}\left(\delta\overline{\be_z\cdot\bMelead} + \frac{\sin\aref}{2\pi}\overline{\be_{\theta}\cdot\bFelead}\right)\Bigg\lvert_{\hS = \xi}\right]\id\xi. \label{eqn:linearisedsolnhelixgeometry}
\end{align}
Using Eq.~\eqref{eqn:relateexpandhelixgeometrytoavgZPsi}, the corresponding (linearised) wavelength-averaged longitudinal and rotational displacement are
 \begin{align}
\Delta\hZ & = \int_0^{\hS}\int_{\eta}^{\epsilon L}\left[\frac{\sin^2\aref\left(1+\nu\sin^2\aref\right)}{4\pi^2}\overline{\be_z\cdot\bFelead}\Big\lvert_{\hS = \xi} +  \frac{h\nu\cos\aref\sin^2\aref}{2\pi}\left(\delta\overline{\be_z\cdot\bMelead} + \frac{\sin\aref}{2\pi}\overline{\be_{\theta}\cdot\bFelead}\right)\Bigg\lvert_{\hS = \xi}\right]\id\xi\id\eta, \nonumber  \\
\Delta\hPsi & = \int_0^{\hS}\int_{\eta}^{\epsilon L}\left[\frac{h\nu\cos\aref\sin^2\aref}{2\pi}\overline{\be_z\cdot\bFelead}\Big\lvert_{\hS = \xi} + \left(1+\nu\cos^2\aref\right)\left(\delta\overline{\be_z\cdot\bMelead} + \frac{\sin\aref}{2\pi}\overline{\be_{\theta}\cdot\bFelead}\right)\Bigg\lvert_{\hS = \xi}\right]\id\xi\id\eta.
\label{eqn:linearisedsolndisplacement}
\end{align}

\subsection{Comparison with past work}
We show in this section that the linearised equivalent-rod equations, when written in terms of $\Delta\hZ$ and $\Delta\hPsi$, recover the equations previously proposed for helical coil springs \citep{phillips1972,jiang1989,jiang1991}.  

Substituting expressions for the linearised resultants $F_Z$ and $M_Z$ in terms of $\Delta\hZ$ and $\Delta\hPsi$ (using Eqs.~\eqref{eqn:relateexpandhelixgeometrytoavgZPsi}--\eqref{eqn:resultantsleadZcomponentshelixgeometrylinearised}) into Eqs.~\eqref{eqn:effectiveforcebalancelinear}--\eqref{eqn:effectivemomentbalancelinear} and \eqref{eqn:effectiveBCs}, or alternatively directly linearising Eqs.~\eqref{eqn:forceavgZPsi}--\eqref{BCsbaseavgZPsi}, we obtain
\begin{align}
& K_1^u\pdd{\Delta\hZ}{\hS} + K_2^u\pdd{\Delta\hPsi}{\hS}+ \overline{\be_z\cdot\bFelead} -\pdd{\Delta\hZ}{T}= 0, \label{eqn:linearisedforceavgZPsi} \\
& K_3^u\pdd{\Delta\hZ}{\hS} + K_4^u\pdd{\Delta\hPsi}{\hS} + \delta\overline{\be_z\cdot\bMelead} + \overline{\be_z\cdot\left(\bRref\times\bFelead\right)} -\left(\hRref\right)^2\pdd{\Delta\hPsi}{T}= 0,  \label{eqn:linearisedmomentavgZPsi} \\
& \Delta\hZ(0,T) = \Delta\hPsi(0,T) = 0, \quad \pd{\Delta\hZ}{\hS}(\epsilon L,T) = \pd{\Delta\hPsi}{\hS}(\epsilon L,T) = 0, \label{eqn:linearisedBCsavgZPsi}
\end{align}
where
\beq
K_1^u = \frac{4\pi^2\csc^2\aref \left(1+\nu\cos^2\aref\right)}{1+\nu}, \quad K_2^u = K_3^u = -\frac{2\pi h\nu\cos\aref}{1+\nu}, \quad K_4^u = \frac{1+\nu\sin^2\aref}{1+\nu}. \label{eqn:defnKiref}
\eeq
Recall from Eqs.~\eqref{eqn:solnbRlead}--\eqref{eqn:solnPsilead} that $\hZ$ and $\hPsi$ represent the $O(\epsilon^{-1})$ contribution to the dimensionless longitudinal coordinate and winding angle, respectively. Hence, in dimensional variables, the wavelength-averaged extensional and rotational displacements are
\beqn
\Delta z = \frac{\lamref}{\epsilon}\Delta \hZ, \quad \Delta\Psi = \frac{1}{\epsilon}\Delta\hPsi.
\eeqn
From the re-scalings introduced in \S\ref{sec:nondim}, and the definitions of $\hS$ and $[t]$ in \S\ref{sec:multiscalesmethodoutline}, we have $s = \lamref S=\epsilon^{-1}\lamref\hS$ and $t = [t]T = \epsilon^{-1}\tast T$. Setting $\bfelead = [f]\bFelead$ and $\bmelead = [m]\bMelead$, and making use of the expressions \eqref{eqn:defnepsilon} for $\epsilon$, $\delta$ and $\tast$, Eqs.~\eqref{eqn:linearisedforceavgZPsi}--\eqref{eqn:linearisedBCsavgZPsi} in terms of dimensional variables are
\begin{align}
& k_1^u \pdd{\Delta z}{s} + k_2^u\pdd{\Delta\Psi}{s}+ \overline{\be_z\cdot\bfelead} - \rho_s A\pdd{\Delta z}{t} = 0, \label{eqn:linearisedforceavgdimzPsi} \\
& k_3^u \pdd{\Delta z}{s} + k_4^u\pdd{\Delta\Psi}{s} + \overline{\be_z\cdot\bmelead} + \overline{\be_z\cdot\left(\brref\times\bfelead\right)} - \rho_s A\left(\rref\right)^2\pdd{\Delta\Psi}{t} = 0, \label{eqn:linearisedmomentavgdimzPsi} \\
& \Delta z(0,t) = \Delta\Psi(0,t) = 0, \quad \pd{\Delta z}{s}(l,t) = \pd{\Delta\Psi}{s}(l,t) = 0, \label{eqn:linearisedBCsavgdimzPsi}
\end{align}
where, using $\rref = \lamref\sin\aref/(2\pi)$, 
\beq
k_1^u = \frac{B K_1^u}{\left(\lamref\right)^2} = \frac{B\left(1+\nu\cos^2\aref\right)}{(1+\nu)\left(\rref\right)^2}, \quad k_2^u = k_3^u = \frac{B K_2^u}{\lamref} = -\frac{h\nu B\cos\aref\sin\aref}{(1+\nu)\rref}, \quad k_4^u = B K_4^u = \frac{B\left(1+\nu\sin^2\aref\right)}{1+\nu}. \label{eqn:defndimKiref}
\eeq
In the absence of external loads, Eqs.~\eqref{eqn:linearisedforceavgdimzPsi}--\eqref{eqn:linearisedmomentavgdimzPsi} are equivalent to the linearised equations proposed by \cite{phillips1972} to model the free vibrations of helical coil springs; we also recover the linearised equations later reported by \cite{jiang1989,jiang1991}, once their stiffness coefficients are expanded in the inextensible limit $a\ll \rref$ considered here\footnote{To map Eqs.~\eqref{eqn:linearisedforceavgdimzPsi}--\eqref{eqn:linearisedmomentavgdimzPsi} onto the equations proposed by \cite{phillips1972} and \cite{jiang1989,jiang1991}, the axial coordinate $z$ ($\sim s\cos\aref$) rather than arclength $s$ is used as the independent variable. We also note that the pitch angle is defined by \cite{phillips1972} and \cite{jiang1989,jiang1991} as the angle between the centreline tangent and the plane perpendicular to the helix axis, i.e., $\pi/2-\alpha$ in our notation.}. In these studies, the dynamic equations were obtained by considering the equilibrium solution of a helical filament under a constant wrench aligned with the helix axis, determining effective stiffness coefficients from this solution, then making the ad hoc assumption that the stiffness coefficients can be applied \emph{locally} when balancing linear and angular momentum for each infinitesimal spring element. Our multiple-scales analysis therefore rigorously justifies this assumption for problems involving unsteady deformations and distributed loads.

\subsection{Equivalent-rod equations in the straight-rod limit, $\aref\to 0$}
\label{sec:straightrodlimit}
In this final subsection, we briefly discuss the limit of vanishing pitch angle. For simplicity, we focus on the linearised equivalent-rod equations in terms of the (dimensional) wavelength-averaged displacements, i.e., Eqs.~\eqref{eqn:linearisedforceavgdimzPsi}--\eqref{eqn:linearisedmomentavgdimzPsi}. As $\aref\to 0$ for fixed $\lamref$, the coefficients $k_i^u$ in Eq.~\eqref{eqn:defndimKiref} take the limiting form
\beqn
k_1^u \sim \frac{4\pi^2 B}{\left(\lamref\aref\right)^2}, \quad k_2^u =  k_3^u \sim -\frac{2\pi h \nu B}{(1+\nu)\lamref} , \quad  k_4^u \sim \frac{B}{1+\nu}.
\eeqn
Because $k_1^u \to \infty$, if we assume that the external force $\bfelead$ and accelerations $\partial^2\Delta z/\partial t^2$ remain bounded as $\aref\to 0$, Eq.~\eqref{eqn:linearisedforceavgdimzPsi} requires that $\partial^2\Delta z/\partial s^2 \approx 0$  and hence, from the boundary conditions \eqref{eqn:linearisedBCsavgdimzPsi}, $\Delta z\approx 0$ throughout the filament. Equation \eqref{eqn:linearisedmomentavgdimzPsi} then reduces to
\beqn
C \pdd{\Delta\Psi}{s}+\overline{\be_z\cdot\bmelead} = 0,
\eeqn
recalling from Eq.~\eqref{eqn:elimC} that $C = B/(1+\nu)$ is the twist modulus. We see that the effective moment resultant (about the helix axis) is $m_z = C\left(\partial\Delta\Psi/\partial s\right)$; hence, it is precisely the gradient of the winding angle, $\partial\Delta\Psi/\partial s$, which maps onto axial (excess) twist of the limiting straight rod. This is a consequence of the fact that the straight-rod limit $\aref\to 0$ here is taken with $\lamref$ and $a$ fixed, i.e., the helix radius $\rref = \lamref\sin\aref/(2\pi) \to 0$ while the cross-section radius is constant, so that rotation of the centreline about the helix axis becomes equivalent to axial twist. We note that a term $\partial^2\Delta\Psi/\partial t^2$, which is present when modelling torsional vibrations in a straight rod, does not appear here because we neglected rotary inertia in the moment balance \eqref{eqn:momentbalance}.

We also deduce that, under the assumption that the filament tip is free of forces and moments, non-trivial equilibrium solutions are only possible as $\aref\to 0$ (with $\lamref$ and $a$ fixed) if the external moment $\bmelead$ is non-zero. This can be traced back to the inextensibility assumption: there cannot be any longitudinal displacement of the straight rod and hence the force balance is not relevant in the straight-rod limit.

\section{Physical scenario I: The heavy helical column}
\label{sec:scenario1}

The first specific physical scenario we analyse is a helical filament deforming under its own weight. As in previous sections, the filament is supported at its base such that the helix axis is directed along $\be_z$, with the other end free. We assume that the gravitational field is parallel to $\be_z$ and we focus on equilibrium solutions; the conditions for a straight helix axis (Eq.~\eqref{eqn:effectiveforceX,Y} in the case $\epsilon L=\ord(1)$; Eqs.~\eqref{eqn:effectiveforceX,Ylinear}--\eqref{eqn:effectivemomentYlinear} in the case $\epsilon L \ll 1$) are then satisfied. This scenario is analogous to the classic problem studied by \cite{greenhill1881} for a straight column, though we do not address the stability of solutions here. 

\subsection{Governing equations}
The external force and moment (per unit arclength) are $\bfe = \rho_s A g\,\be_z$ and $\bme = \mathbf{0}$, respectively, where $g$ is the gravitational acceleration in the direction of increasing $z$: if $g > 0$ the filament is under tension, and if $g < 0$ the filament is under compression. To non-dimensionalise, we use the force scale $[f] = \rho_s A|g|$ in the re-scalings introduced in \S\ref{sec:nondim}, which gives
\beq
\epsilon = \frac{\rho_s A|g|\left(\lamref\right)^3}{B}, \quad \delta = 0, \quad \bFe = \bFelead = \sgn g\,\be_z. \label{eqn:scenarioIdefnepsilon}
\eeq
We note that the highly-coiled assumption ($\epsilon\ll 1$) in this context reads $\lamref \ll [s] = \left[B/\left(\rho_s A|g|\right)\right]^{1/3}$, where $[s]$ is the elasto-gravity length that frequently arises in problems involving rods bending under self-weight \citep{wang1986}.

We use the dimensionless formulation of the equivalent-rod equations in terms of the slowly-varying pitch angle, $\alpha$, and wavelength, $\Lambda$, i.e., Eqs.~\eqref{eqn:effectiveforcebalancehelixgeometry}--\eqref{eqn:effectivemomentbalancehelixgeometry}. Neglecting time derivatives, these equations become
\begin{align}
& C_1 \pd{\alpha}{\hS} + C_2\pd{\Lambda}{\hS} + \sgn g = 0, \label{eqn:scenarioIeffectiveforcebalance} \\
& C_3 \pd{\alpha}{\hS} + C_4\pd{\Lambda}{\hS} = 0,  \label{eqn:scenarioIeffectivemomentbalance}
\end{align}
to be solved with the boundary conditions \eqref{eqn:effectiveBCshelixgeometry}.

\subsection{Linearised solution: $\epsilon L \ll 1$}
While the deformation remains small, the general solution of the linearised equations, Eq.~\eqref{eqn:linearisedsolnhelixgeometry} (determined earlier in \S\ref{sec:linearisedeqns}), yields
\beq
\Delta\alpha = -\frac{\sin\aref\left(1+\nu\sin^2\aref\right)\sgn g}{4\pi^2}\left(\epsilon L-\hS\right), \quad
 \Delta\Lambda = -\frac{\nu\cos\aref\sin^2\aref\sgn g}{4\pi^2}\left(\epsilon L-\hS\right). \label{eqn:linearisedsolngravity}
\eeq
A few features of this solution are noteworthy. Firstly, due to linearisation and the fact that the external force is independent of the solution, the perturbations have the symmetry $(\Delta\alpha,\Delta\Lambda)\to-(\Delta\alpha,\Delta\Lambda)$ as $g\to -g$. Because the filament tip at $\hS=\epsilon L$ is free and there are no other boundary conditions for $\alpha$ and $\Lambda$ (in particular, there are no unknown parameters that depend on the value of the solution away from the tip), the linearised solution is always valid near the tip, even if $\epsilon L=O(1)$. In particular, $\Delta\alpha$ and $\Delta\Lambda$ remain much smaller than unity provided $(\epsilon L-\hS) \ll 1$. If $\epsilon L \ll 1$, we then expect that the linearised solution is valid throughout the entire filament. Furthermore, we note that, in the straight-rod limit $\aref \to 0$ (with $\lamref$ fixed), the perturbation $\Delta\alpha \to 0$ because of rod inextensibility, and $\Delta\Lambda \to 0$ because no external moment is applied (recall the discussion in \S\ref{sec:straightrodlimit}). 

From Eq.~\eqref{eqn:linearisedsolndisplacement}, the corresponding wavelength-averaged longitudinal and rotational displacement are given by
\beqn
\Delta\hZ = \frac{\sin^2\aref\left(1+\nu\sin^2\aref\right)\sgn g}{8\pi^2}\:\hS\left(2\epsilon L - \hS\right), \quad \Delta\hPsi = \frac{h \nu\cos\aref\sin^2\aref\sgn g}{4\pi}\:\hS\left(2\epsilon L - \hS\right).
\eeqn
We see that the displacements vary quadratically with the slow variable, $\hS$, and, as expected, their magnitude is largest at the filament tip, $\hS=\epsilon L$. 


\subsection{General solution: $\epsilon L = \ord(1)$}
When the deformation is no longer small, it is not possible to solve the equivalent-rod equations \eqref{eqn:scenarioIeffectiveforcebalance}--\eqref{eqn:scenarioIeffectivemomentbalance} analytically so we appeal to numerical integration\footnote{Note that, alternatively, Eqs.~\eqref{eqn:scenarioIeffectiveforcebalance}--\eqref{eqn:scenarioIeffectivemomentbalance} have the first integrals $F_Z = \sgn g(\epsilon L-\hS)$ and $M_Z = 0$, which provide two algebraic equations for $\alpha$ and $\Lambda$ via Eq.~\eqref{eqn:resultantsleadZcomponentshelixgeometry}. However, these equations must also be solved numerically (e.g., using a root-finding algorithm) in general.}. We recast the equations as an initial-value problem by introducing
\beqn
 \Xi = \epsilon L - \hS,  \qquad \Xi\in[0,\epsilon L],
\eeqn
so that the boundary conditions \eqref{eqn:effectiveBCshelixgeometry} at the filament tip become initial conditions at $\Xi = 0$. For specified values of the dimensionless parameters (namely $\sgn g$, $\aref$, $\nu$ and $h$), we numerically integrate Eqs.~\eqref{eqn:scenarioIeffectiveforcebalance}--\eqref{eqn:scenarioIeffectivemomentbalance} up to some maximum value $\Xi=\Xi_{\mathrm{max}}$; the solution for any $\epsilon L \leq \Xi_{\mathrm{max}}$ can then be found by truncating the trajectories at $\Xi = \epsilon L$.

\begin{figure}[t]
\centering
\includegraphics[width=0.8\textwidth]{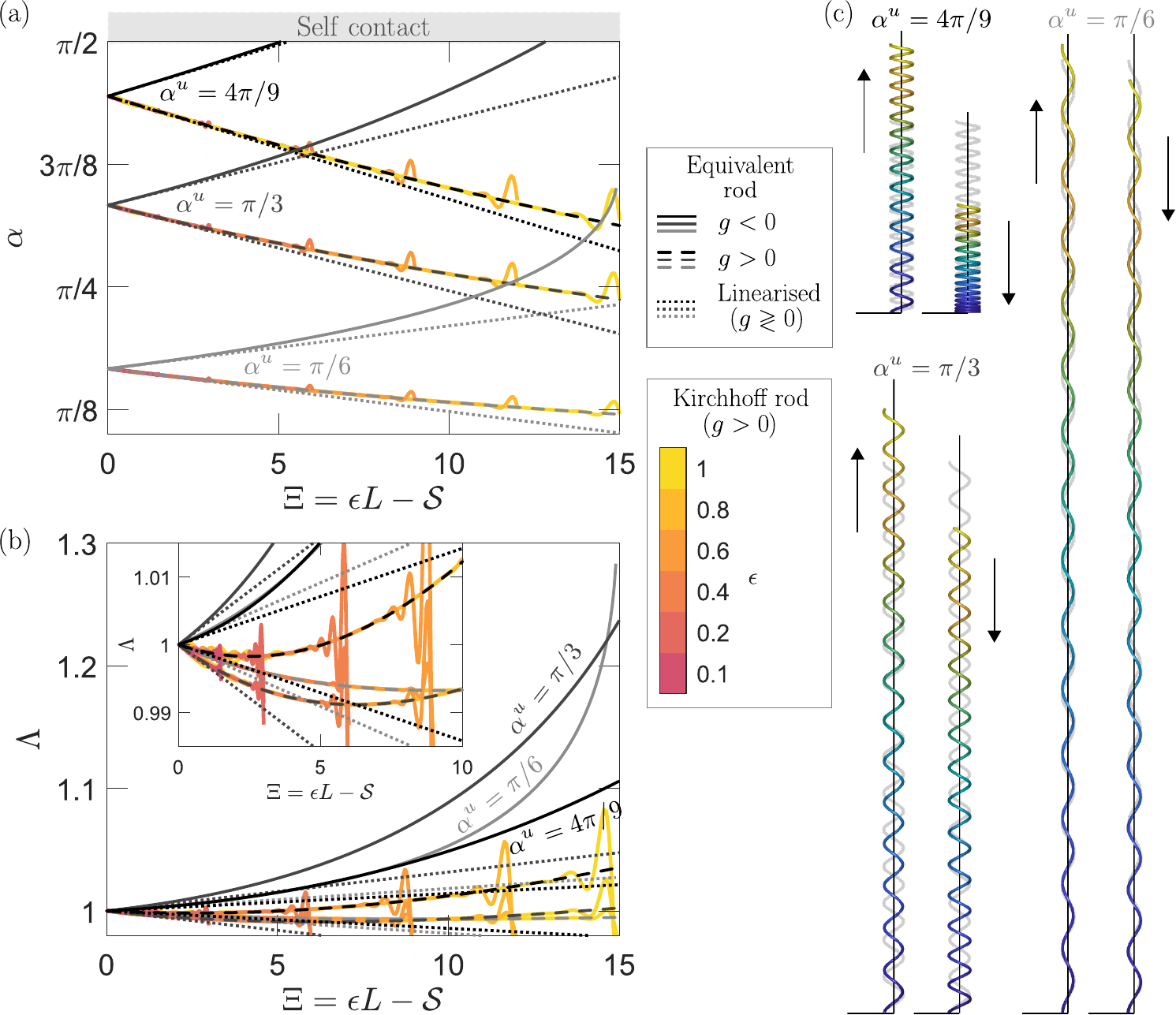}
\caption{The heavy helical column: the deformation of a highly-coiled helical rod due to a gravitational field parallel to the helix axis (here $\nu = 1/3$ and $h = -1$). We plot trajectories of (a) the slowly-varying pitch angle, $\alpha$; and (b) the dimensionless helical wavelength, $\Lambda = \lambda/\lamref$, obtained by numerically integrating the (straight) equivalent-rod equations \eqref{eqn:scenarioIeffectiveforcebalance}--\eqref{eqn:scenarioIeffectivemomentbalance} with boundary conditions \eqref{eqn:effectiveBCshelixgeometry}. In panels (a)--(b), the solutions are plotted as a function of dimensionless arclength from the filament tip, $\Xi = \epsilon L -\hS$, for pitch angles $\aref\in\lbrace\pi/6,\pi/3,4\pi/9\rbrace=\lbrace 30^{\circ},60^{\circ},80^{\circ}\rbrace$ and either $g > 0$ (dashed curves) or $g < 0$ (solid curves); see legend. The corresponding linearised solutions \eqref{eqn:linearisedsolngravity}, formally valid for $\Xi\ll 1$, are plotted as dotted lines. For later reference, we have also superimposed solutions of the Kirchhoff rod equations for $g > 0$, $L = 15$ and $\epsilon\in\lbrace 0.1,0.2,0.4,0.6,0.8,1\rbrace$ (solid coloured curves; see legend). (c)~Corresponding filament shapes (side view), determined from the equivalent-rod solutions in (a)--(b) for $L = 15$ and $\epsilon = 0.3$. Arrows indicate the direction of gravity. The undeformed filament is shown as a grey curve.}
\label{fig:ScenarioISol}
\end{figure}

In Figs.~\ref{fig:ScenarioISol}a--b, we plot typical trajectories of $\alpha$ and $\Lambda$ as a function of $\Xi$, both for $g > 0$ (filament under tension; grey dashed curves) and $g < 0$ (filament under compression; grey solid curves). (We have also superimposed solutions of the full Kirchhoff rod equations as solid coloured curves; these are discussed in \S\ref{sec:scenario1comparekirchhoff} below.) Even though the linearised solution in Eq.~\eqref{eqn:linearisedsolngravity} (dotted lines in Figs.~\ref{fig:ScenarioISol}a--b) is formally valid only for $\Xi\ll 1$, we see that it performs excellently up to $\Xi \approx 1$. In the regime $\Xi\lesssim 1$, the symmetry in the perturbations $\Delta\alpha = \alpha-\aref$ and $\Delta\Lambda = \Lambda-1$ as $g\to -g$ is also evident. However, at larger values of $\Xi$, the perturbations are no longer small and the trajectories deviate significantly from the linearised solution. In particular, because the coefficients $C_i$ in Eqs.~\eqref{eqn:scenarioIeffectiveforcebalance}--\eqref{eqn:scenarioIeffectivemomentbalance} depend nonlinearly on the values of $\alpha$ and $\Lambda$, the symmetry in the solutions as $g\to -g$ is no longer present. The nonlinear dependence of the coefficients $C_i$ is also responsible for the non-monotonic variation of the wavelength $\Lambda$ observed when $g > 0$ (Fig.~\ref{fig:ScenarioISol}b inset): $\Lambda$ decreases slightly before increasing again, corresponding to winding then unwinding, which is a generic phenomenon for helical filaments under longitudinal forces~\citep{goriely2017}. The shapes of the filament predicted by the equivalent-rod model for $L = 15$ and $\epsilon = 0.3$ are shown in Fig.~\ref{fig:ScenarioISol}c; these shapes are determined from the numerical solutions for $\alpha$ and $\Lambda$ by integrating the tangent vector $\bdthreelead=\bt$ in Eq.~\eqref{eqn:FSframe}.

\subsubsection{Singular behaviour}
\begin{figure}
\centering
\includegraphics[width=0.9\textwidth]{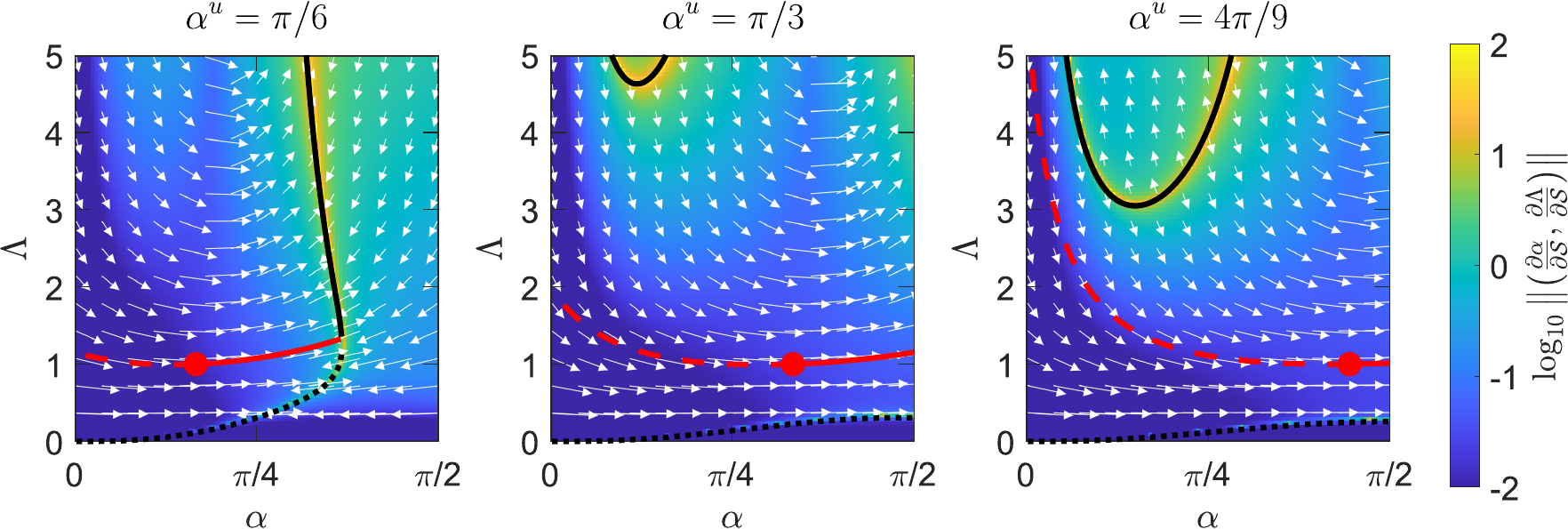} 
\caption{The heavy helical column: the phase plane of Eqs.~\eqref{eqn:scenarioIeffectiveforcebalance}--\eqref{eqn:scenarioIeffectivemomentbalance} ($\nu = 1/3$ and $h = -1$). In each panel, corresponding to a different undeformed pitch angle $\aref$, phase trajectories (white arrows) are shown; arrows point in the direction of increasing (decreasing) arclength from the filament tip when $g < 0$ ($g > 0$), corresponding to the filament under compression (tension). The trajectory emerging from the undeformed point $(\aref,1)$ (red circle) is highlighted red. These trajectories are superimposed on a density plot of the local gradient vector $(\partial\alpha/\partial\hS,\partial\Lambda/\partial\hS)$, coloured according to the logarithm of its magnitude (see colourbar). Also shown are the branches of critical points $\Lambda_{-}$ (black solid curves) and $\Lambda_{+}$ (black dotted curves) from Eq.~\eqref{eqn:defnLambdapm}, at which the matrix of stiffness coefficients is singular and the magnitude of the local gradient vector is infinite.}
\label{fig:ScenarioIPhasePlanes}
\end{figure}

Another key feature observed in Figs.~\ref{fig:ScenarioISol}a--b is that the solution of the equivalent-rod equations becomes singular if $g < 0$ and $\aref$ is sufficiently small: the trajectory for $\aref=\pi/6$ reaches a point of vertical tangency at $\Xi \approx 15$ and it is not possible to integrate further. (For larger values of $\aref$, self-contact at $\alpha=\pi/2$ occurs before any singularity.) To gain further understanding, Fig.~\ref{fig:ScenarioIPhasePlanes} shows a density plot (`heat map') of the magnitude of the gradient vector associated with Eqs.~\eqref{eqn:scenarioIeffectiveforcebalance}--\eqref{eqn:scenarioIeffectivemomentbalance}, i.e., $\|(\partial\alpha/\partial\hS,\partial\Lambda/\partial\hS)\|$, which is plotted on the $(\alpha,\Lambda)$-plane together with phase trajectories (white arrows). When $g < 0$, these arrows point in the direction of increasing $\Xi$, i.e., moving away from the filament tip; while when $g > 0$, the arrows point in the direction of decreasing $\Xi$. Figure \ref{fig:ScenarioIPhasePlanes} confirms that the singularity observed in Figs.~\ref{fig:ScenarioISol}a--b corresponds to a critical point where the Jacobian determinant $\det J = C_1 C_4 - C_2 C_3 $ vanishes (black curves) and the magnitude of the gradient vector is infinite; we discussed the general existence of such critical points in \S\ref{sec:effectivejacobian}. Also plotted on Fig.~\ref{fig:ScenarioIPhasePlanes} is the trajectory emerging from $(\alpha,\Lambda)=(\aref,1)$ (red curves), corresponding to the solution satisfying the boundary conditions \eqref{eqn:effectiveBCshelixgeometry} at the filament tip, $\Xi = 0$; this solution evidently only becomes singular in the physical region $\Xi > 0$ when $g < 0$, i.e., under compression. Nevertheless, we emphasise that near critical points, the solution varies rapidly with $\Xi$ and our multiple-scales analysis is not asymptotically valid (we discuss this point further in \S\ref{sec:conclusiondiscussion}).

While the singularity in the system only occurs under compression for large deformations $\epsilon L\gtrsim 15$, in practice the helix axis will buckle first, meaning our assumption of a straight axis is no longer valid. More precisely, in \ref{appendix:gravitybuckling} we estimate the buckling threshold of a helical filament using an ad hoc effective-beam approximation, which yields
\beq
\epsilon L \approx 7.84 \frac{L^{-2}\sec\aref}{1+\left(\nu/2\right)\sin^2\aref} \qquad \mathrm{(buckling\ threshold)}. \label{eqn:approxbucklethresh}
\eeq
Because our multiple-scales analysis requires $L = l/\lamref \gg 1$, we expect that significant axis bending occurs well before $\epsilon L\approx 15$ for $\aref \lesssim \pi/6$. This prediction is confirmed by the solutions of the full rod equations discussed below. However, it may be possible to reach values $\epsilon L\gtrsim 15$ with a straight helix axis if axial bending is prevented, for example by confining the filament radially. As discussed in \S\ref{sec:conclusiondiscussion}, a radial contact force could be incorporated into our equivalent-rod model without much difficulty.

\subsection{Comparison with full Kirchhoff rod simulations}
\label{sec:scenario1comparekirchhoff}
We also compare the (straight) equivalent-rod theory with equilibrium solutions of the full Kirchhoff rod equations (discussed earlier in \S\ref{sec:numerics}). In particular, for the gravitational loading considered here, comparing Eq.~\eqref{eqn:scenarioIdefnepsilon} with the general form in Eq.~\eqref{eqn:loadingnumeric} shows that 
\beqn
\bAe = \mathbf{0}, \quad \bBe = \mathbf{0}, \quad \bCe = \sgn g\,\be_z, \quad \bDe = \mathbf{0}, \quad \bEe = \mathbf{0}. 
\eeqn
Using the numerical implementation described in \S\ref{sec:numerics}, we perform simple continuation in the loading parameter $\epsilon$: we increase $\epsilon$ in small steps $\Delta\epsilon$, using the solution at each step as the initial guess for the next value of $\epsilon$. The undeformed solution is used as the first guess to begin the continuation. (For $g > 0$, the solver converges without issue for $\Delta\epsilon \lesssim 10^{-1}$. For $g < 0$, due to convergence issues at the buckling onset, we use a small step size $\Delta\epsilon = 0.002\epsilon_{\mathrm{buckle}}$, where $\epsilon_{\mathrm{buckle}}(L,\aref)$ is the value of $\epsilon$ at buckling predicted by Eq.~\eqref{eqn:approxbucklethresh}.) To compare simulation results with the equivalent-rod theory, we note from Eqs.~\eqref{eqn:strainspinlead}--\eqref{eqn:deformedDarboux} that the leading-order strain components $U_1$, $U_2$, $U_3$ in the equivalent-rod theory are given in terms of the pitch angle $\alpha$ and wavelength $\Lambda$ by
\beqn
U_1\approx 0, \quad U_2 \sim \hK = \frac{2\pi\sin\alpha}{\Lambda}, \quad U_3 \sim \hT = \frac{2\pi h\cos\alpha}{\Lambda}.
\eeqn
These expressions allow us to calculate the strain components predicted by the equivalent-rod model from solutions for $(\alpha,\Lambda)$; while inverting the expressions allows us to determine effective values of $(\alpha,\Lambda)$ from numerical simulations.

Simulation results for $\alpha$ and $\Lambda$, plotted as a function of $\Xi$, are superimposed (as solid coloured curves) on Figs.~\ref{fig:ScenarioISol}a--b; here we take $L = 15$ and various $\epsilon$ in the range $[0.1,1]$, and we show results only for $g > 0$ (as discussed further below, for $g < 0$ the filament buckles for $\epsilon L \ll 1$, i.e., $\Xi \ll 1$). We observe that the numerical solutions are generally in excellent agreement with the predictions from the equivalent-rod theory, even up to $\epsilon = 1$, despite the fact that the theory is formally valid only for $\epsilon \ll 1$. However, in the vicinity of the filament base (i.e., for $\Xi \approx \epsilon L$, which varies with $\epsilon$ for fixed $L$), oscillations in the numerical curves occur. These oscillations are boundary effects resulting from the rigid clamping condition applied to the filament base: as discussed in \S\ref{sec:numerics}, these additional boundary conditions are necessary when solving the full Kirchhoff rod equations, but are not included in the equivalent-rod theory to leading order. The oscillations are confined to a few wavelengths of the filament base, and their amplitude generally decreases as $\epsilon$ decreases, indicating that they are indeed a higher-order effect as $\epsilon \to 0$.

\begin{figure}[t]
    \centering
    \includegraphics[width=0.9\textwidth]{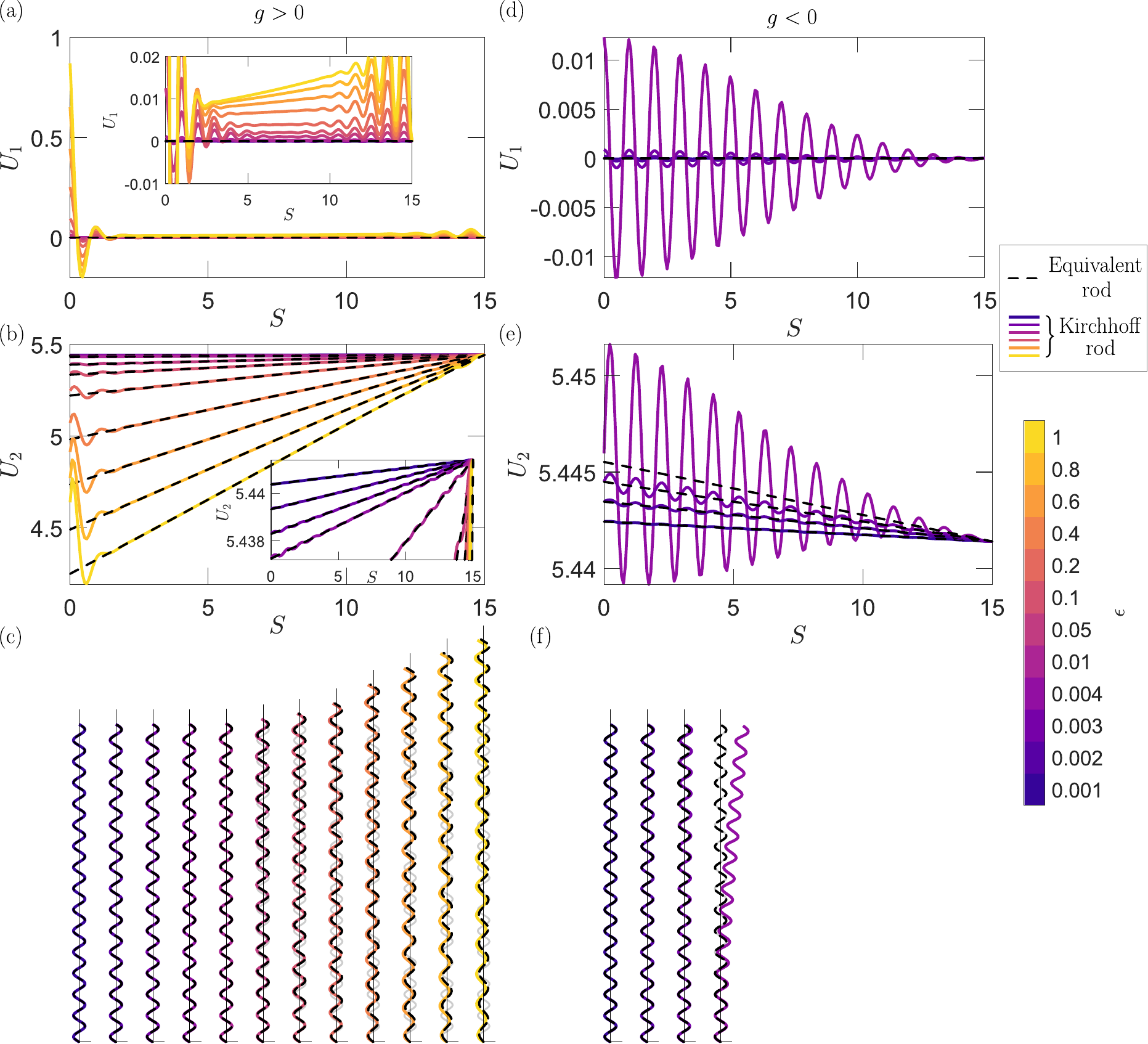}
    \caption{The heavy helical column: comparison between (straight) equivalent-rod theory (dashed black curves) and Kirchhoff rod simulations (solid coloured curves); here we use $\aref = \pi/3$, $L = 15$, $\nu = 1/3$, $h = -1$ and various $\epsilon \in [0.001,1]$ (see colourbar). (a)--(b): Strain components $U_1$ and $U_2$ as a function of dimensionless arclength $S$ for $g > 0$. (c) Corresponding filament shapes (projecting the centreline in the $y$-$z$ plane). The undeformed filament is shown as a grey curve. (d)--(f): As in (a)--(c), though with $g < 0$, showing results only until the onset of axis buckling. }
    \label{fig:ScenarioICompareKirchhoff}
\end{figure}

To further understand these boundary effects and the buckling behaviour for $g < 0$, in Fig.~\ref{fig:ScenarioICompareKirchhoff} we compare simulation results (solid coloured curves) with equivalent-rod solutions (black dashed curves) for $\aref=\pi/3$ and both $g > 0$ (left column) and $g < 0$ (right column); similar to Fig.~\ref{fig:ScenarioISol}, we take $L = 15$ and various $\epsilon$ in the range $[0.001,1]$ (see legend). In both columns, we plot the strain components $U_1$ and $U_2$ as a function of dimensionless arclength $S$ (top panels), together with the filament shapes (bottom panels). (The plots of the strain component $U_3$ are similar to $U_2$, so we omit them here.) From Fig.~\ref{fig:ScenarioICompareKirchhoff}, we observe the following:
\begin{itemize}
    \item{In the case $g > 0$ (filament under tension), we generally have $U_1 \approx 0$ (Fig.~\ref{fig:ScenarioICompareKirchhoff}a) as predicted by the (straight) equivalent-rod model. However, we observe significant oscillations in $U_1$ near the boundaries, which decay significantly at a distance of around three wavelengths. We see that the variation in $U_2$ along the filament length is in excellent agreement with the equivalent-rod model (Fig.~\ref{fig:ScenarioICompareKirchhoff}b); the boundary oscillations are more noticeable at the filament base since the change in $U_2$ is largest here. We also observe small-amplitude oscillations throughout the filament length (see inset of Fig.~\ref{fig:ScenarioICompareKirchhoff}b), which presumably arise due to the finite helical wavelength: further simulations (not shown) indicate that these oscillations disappear as $L=l/\lamref\to \infty$ and $\epsilon\to 0$. Despite these oscillations, we observe excellent agreement in the filament shape up to $\epsilon = 1$ (Fig.~\ref{fig:ScenarioICompareKirchhoff}c).}
    \item{In the case $g < 0$ (filament under compression), we observe similar behaviour to the case $g > 0$ provided that $\epsilon$ is sufficiently small ($\epsilon \lesssim 0.003$); see Figs.~\ref{fig:ScenarioICompareKirchhoff}d--f. However, for larger values $\epsilon \gtrsim 0.0035$, the filament buckles and undergoes significant axis bending, as illustrated by the filament shape for $\epsilon=0.004$ in Fig.~\ref{fig:ScenarioICompareKirchhoff}f. As a result, large-amplitude oscillations in the strain components occur throughout the filament, though we note that the equivalent-rod model still captures the average value of these oscillations. (For clarity, on Figs.~\ref{fig:ScenarioICompareKirchhoff}d--f we do not plot the curves for values $\epsilon > 0.004$ when the amplitude of the oscillations becomes very large.)}
\end{itemize}

\begin{figure}[t]
  \centering
  \includegraphics[width=\textwidth]{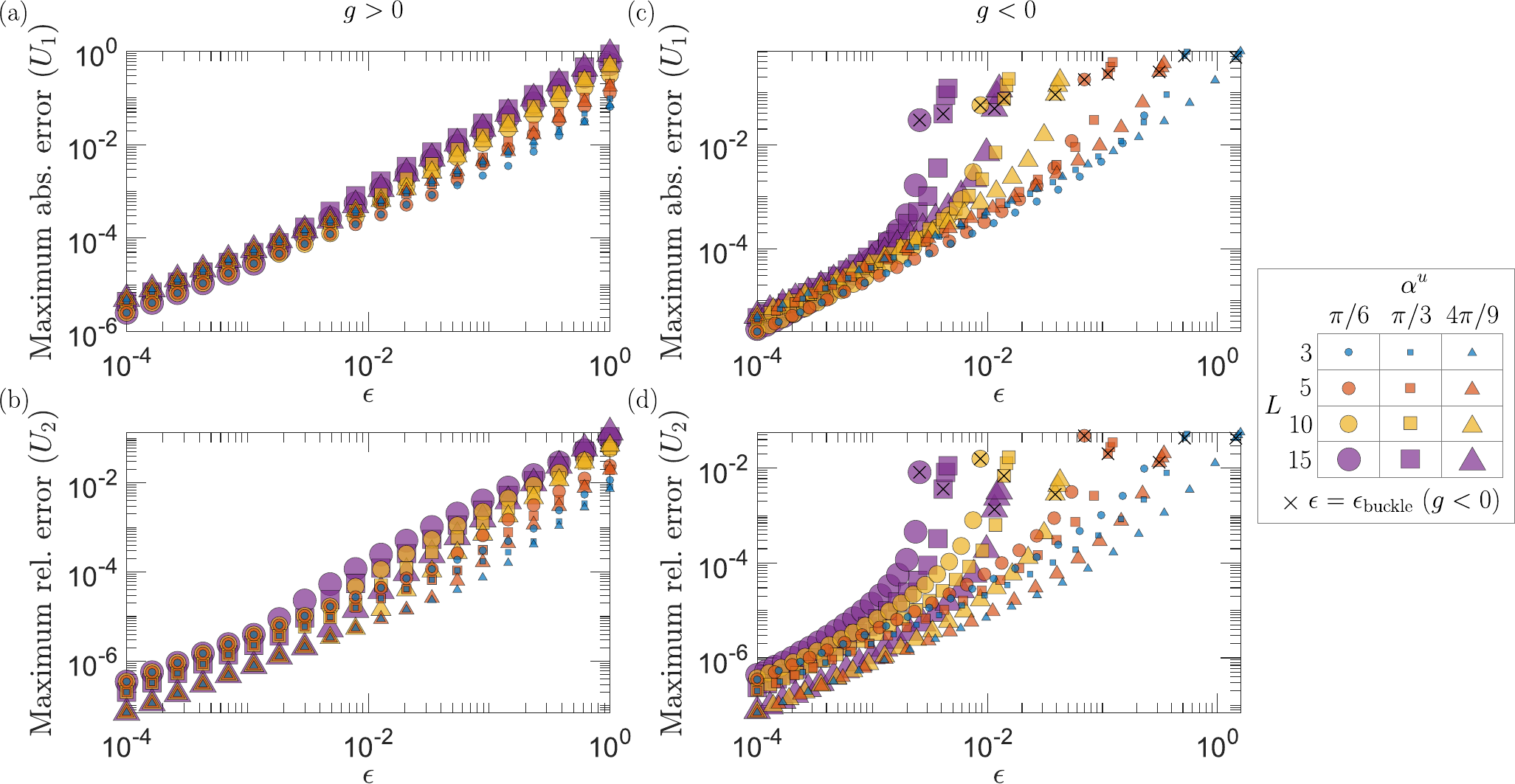}
  \caption{The heavy helical column: error between (straight) equivalent-rod theory and Kirchhoff rod simulations, plotted as a function of $\epsilon$ for various pitch angles $\aref$ and filament lengths $L$ (indicated by symbol type and size/colour, respectively; see legend). (a)--(b): Maximum absolute error in the strain component $U_1$, and the maximum relative error in the strain component $U_2$, for $g > 0$ (here $\nu = 1/3$, $h = -1$). Results are shown at $20$ values of $\epsilon$ equally spaced on a logarithmic scale between $10^{-4}$ and $1$. (c)--(d): As in panels (a)--(b), though with $g < 0$. Results are shown at $20$ values of $\epsilon$ equally spaced on a logarithmic scale between $10^{-4}$ and $1.1\epsilon_{\mathrm{buckle}}$, as well as at $\epsilon = \epsilon_{\mathrm{buckle}}$ (indicated by black crosses) and $\epsilon = 1.1\epsilon_{\mathrm{buckle}}$, where $\epsilon_{\mathrm{buckle}}(L,\aref)$ is the buckling value predicted by Eq.~\eqref{eqn:approxbucklethresh}. (For some values of $\aref$ and $L$, the numerical solver aborted due to convergence issues beyond the buckling onset, so that not all values of $\epsilon$ are plotted.)}
  \label{fig:ScenarioIErrorWithKirchhoff}
\end{figure}

We observe similar trends for other values of the pitch angle $\aref$ and filament length $L$. To be more quantitative, in Fig.~\ref{fig:ScenarioIErrorWithKirchhoff} we plot the error between solutions of the Kirchhoff rod equations and the equivalent-rod model for both $g > 0$ (left column) and $g < 0$ (right column). In each column, we plot the maximum absolute error in the strain component $U_1$ (top panels), and the maximum relative error in the strain component $U_2$ (bottom panels), where the maximum is taken over the entire filament; these errors are plotted as a function of $\epsilon$ for different pitch angles $\aref\in\lbrace\pi/6,\pi/3,4\pi/9\rbrace$ (symbol type; see legend) and filament lengths $L\in\lbrace 3,5,10,15\rbrace$ (symbol size/colour). (As in Fig.~\ref{fig:ScenarioICompareKirchhoff}, we omit the corresponding plots for $U_3$ as they are similar to $U_2$.) Figure \ref{fig:ScenarioIErrorWithKirchhoff} confirms that the equivalent-rod model performs excellently for $g > 0$, even when $\epsilon$ is not strictly small, as well as for $g < 0$ provided that the system is not near the onset of buckling: both absolute and relative errors decrease systematically as $\epsilon$ decreases, reaching values of around $10^{-6}$ when $\epsilon = 10^{-4}$. For sufficiently small $\epsilon$, the error scales linearly with $\epsilon$; we note that a linear scaling is expected based on the regular asymptotic expansion in $\epsilon$ that we used in \S\ref{sec:perturbationscheme}. In addition, provided $\epsilon L \ll 1$, the error is independent of $L$, as evidenced by the collapse of the symbols for each value of $\aref$. This collapse breaks down when $\epsilon L =\ord(1)$, though in this regime we are able to collapse the data by instead plotting the error as a function of $\epsilon L$. While a detailed analysis of axis buckling and the stability of solutions is beyond the scope of this paper, in Figs.~\ref{fig:ScenarioIErrorWithKirchhoff}c--d we observe that the point where the error starts to increase rapidly for $g > 0$ (due to significant axis bending) matches well the buckling threshold predicted by Eq.~\eqref{eqn:approxbucklethresh} (black crosses).

\section{Physical scenario II: Axial rotation (twirling) in viscous fluid}
\label{sec:scenario2}
\begin{figure}
\centering
\includegraphics[width=0.47\textwidth]{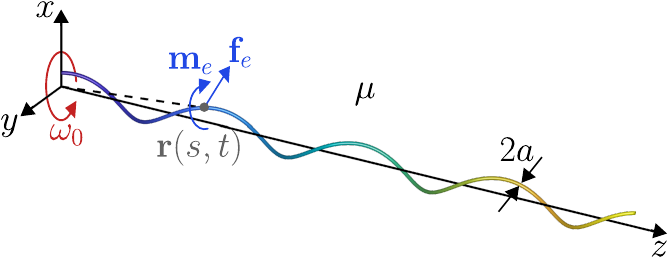}
\caption{Schematic diagram of axial rotation (twirling) in viscous fluid: the helical filament is driven at its base at a prescribed angular frequency $\omega_0$ relative to fluid (viscosity $\mu$) in the laboratory frame. The hydrodynamic loading comprises a distributed force $\bfe$ and moment $\bme$.}
\label{fig:scenarioIIschematic}
\end{figure}

In the second physical scenario, we assume that the filament is immersed in viscous fluid and is rotated at its base about the helix axis at some prescribed angular frequency, $\omega_0$, while its other end is free; see Fig.~\ref{fig:scenarioIIschematic}. This may be considered as a simple model for a bacteria flagellar filament in which (i) the filament base is tethered in the laboratory frame, i.e., not freely swimming; (ii) we neglect the flexible `hook' joint and tapered part connecting the rotary motor to the filament base~\citep{higdon1979,park2017}, but incorporate their effect as transmitting a wrench to the filament needed to achieve the prescribed frequency, $\omega_0$; and (iii) polymorphic transformations away from the so-called `normal' helical form do not occur~\citep{Hasegawa1998}. Here we describe how the (straight) equivalent-rod equations can be applied to this problem. 

\subsection{Governing equations}
 Since inertial forces are dominated by viscous forces at the small lengthscales characteristic of bacterial flagella \citep{powers2010}, we neglect inertia of both the fluid and rod. We may then use resistive-force theory to compute the hydrodynamic forces and moments exerted on the filament \citep{lauga2020}. While this theory neglects long-range hydrodynamic interactions arising from the curved geometry of the helix, it can be obtained as the asymptotic limit of slender-body theory when the cross-section radius $a$ is much smaller than the typical radius of curvature of the rod~\citep{cox1970,leal2007}. For the helical filament considered here, this radius of curvature is of size $\lamref$ so the asymptotic limit is equivalent to $a \ll\lamref$, which is precisely the assumption we made in \S\ref{sec:formulation} to neglect axial extensibility.
 
Recall from \S\ref{sec:formulationBCs} that the helix frame $Oxyz$ is defined such that the filament base (at $z = 0$) is always located on the $x$-axis. The velocity of a point $\br$ on the filament centreline, relative to the fluid in the laboratory frame, is then
\beq
\bv = \omega_0\be_z\times\br + \pd{\br}{t}. \label{eqn:velocitydim}
\eeq
The final term here, which accounts for motions relative to the helix frame, incorporates both spatial variation in the rotation rate and motions parallel to the helix axis. From resistive-force theory, the viscous drag force (per unit length) is given by \citep{lauga2020}
\beq
\bfe = -\zpa\left(\bd_3\cdot\bv\right)\bd_3-\zpe\left[\bv-\left(\bd_3\cdot\bv\right)\bd_3\right],
\label{eqn:dragforcedim}
\eeq
where, for a helical rod, the local drag coefficients for motion parallel and perpendicular to the rod centreline are given by Lighthill's corrected coefficients~\citep{lighthill1976}:
\beqn
\zpa = \frac{2\pi\mu}{\log\left(0.18\lamref/a\right)}, \quad \zpe = \frac{4\pi\mu}{\log\left(0.18\lamref/a\right)+1/2}.
\eeqn
Here $\mu$ is the dynamic viscosity of the fluid  and we neglect small changes to the drag coefficients arising from variations in the contour wavelength. The fluid also exerts a moment (per unit length) opposing rotation of the filament cross-section:
\beq
\bme = -4\pi\mu a^2\left(\bd_3\cdot\left[\omega_0\be_z+\bomega\right]\right)\bd_3,
\label{eqn:dragmomentdim}
\eeq
where $4\pi\mu a^2$ is the rotational drag coefficient (per unit length) of a rod with radius $a$~\citep{landau1987}, and $\bomega$ is the angular velocity vector (in the helix frame) introduced in Eq.~\eqref{eqn:kinematicstrainspindim}. We note that the distributed moment $\bme$ also has terms along $\bd_1$ and $\bd_2$, though these are a factor $O(a/l)^2 \ll 1$ smaller than the distributed force $\bfe$ and so can be safely neglected for the slender filament considered here \citep{garg2023}.


To non-dimensionalise, we expect that the relevant velocity scale is the tangential velocity associated with the imposed rotation, i.e., $|\bv| \sim \lamref|\omega_0|$. From Eqs.~\eqref{eqn:dragforcedim}--\eqref{eqn:dragmomentdim}, it is then natural to choose the force and moment scales
\beqn
[f] = \zpe\lamref|\omega_0|, \quad [m] = 4\pi\mu a^2|\omega_0|.
\eeqn
The dimensionless parameters $\epsilon$ and $\delta$, defined in Eq.~\eqref{eqn:defnepsilon}, are then given by
\beqn
\epsilon = \frac{\zpe\left(\lamref\right)^4|\omega_0|}{B}, \quad \delta = \frac{4\pi\mu a^2}{\zpe\left(\lamref\right)^2}.
\eeqn
With the above force scale $[f]$ and moment scale $[m]$, we introduce dimensionless variables as in \S\ref{sec:formulation} (Eq.~\eqref{eqn:nondim}). Because we neglect rod inertia here, we do not rescale time by the timescale $[t] = \epsilon^{-1}\tast$, as was done in \S\ref{sec:multiscales}--\ref{sec:equivalent-rod}. Instead, we consider the timescale over which the filament deforms due to the hydrodynamic load: balancing the $\partial\br/\partial t$ term in the velocity \eqref{eqn:velocitydim} with $|\bv| \sim \lamref|\omega_0|$ (noting that $|\br| = O(l) = O(\lamref L) = O(\epsilon^{-1}\lamref)$ for $\epsilon L = O(1)$), this timescale is $t \sim \epsilon^{-1}|\omega_0|^{-1}$. Re-scaling also the angular velocity by $1/t \sim \epsilon|\omega_0|$ here, we therefore set
\beq
t = \epsilon^{-1}|\omega_0|^{-1}\hat{T}, \quad \bv = \lamref|\omega_0|\bV, \quad \bomega = \epsilon|\omega_0|\,\hat{\bOmega}. \label{eqn:scenarioIInondim}
\eeq
Equations \eqref{eqn:velocitydim}--\eqref{eqn:dragmomentdim} then become
\beq
\bV = \sgn\omega_0\be_z\times\bR + \epsilon\pd{\bR}{\hat{T}}, \quad 
\bFe = -(1-\chi)\left(\bd_3\cdot\bV\right)\bd_3-\left[\bV-\left(\bd_3\cdot\bV\right)\bd_3\right], \quad 
\bMe = -\left[\bd_3\cdot\left(\sgn \omega_0\be_z+\epsilon\,\hat{\bOmega}\right)\right]\bd_3, \label{eqn:dragforcemoment}
\eeq
where the local drag anisotropy is characterised by the dimensionless parameter
\beqn
\chi = \frac{\zpe-\zpa}{\zpe} \approx \frac{1}{2}.
\eeqn

\subsubsection{Formulation of the (straight) equivalent-rod equations: $\epsilon L=\ord(1)$}
\label{sec:scenario2nonlinearform}
To formulate the (straight) equivalent-rod equations for $\epsilon L=\ord(1)$, we use the expressions in Eqs.~\eqref{eqn:solnbRleaddot} and \eqref{eqn:strainspinlead} for, respectively, the leading-order centreline velocity, $\partial\bRlead/\partial T$, and angular velocity vector, $\bOmegalead$, in terms of the slowly-varying helical radius, $\hR$, wavelength-averaged longitudinal coordinate, $\hZ$, and wavelength-averaged winding angle, $\hPsi$. Noting that $\partial/\partial T = (|\omega_0|\tast)\partial/\partial\hat{T}$ and $\bOmega = \epsilon^{-1}\tast\bomega = |\omega_0|\tast\hat{\bOmega}$, we obtain
\beq
\pd{\bR}{\hat{T}} \sim \pd{\bRlead}{\hat{T}} = \epsilon^{-1}\left(\hR\pd{\hPsi}{\hat{T}}\be_{\theta} + \pd{\hZ}{\hat{T}}\be_z\right) + O(1), \quad \hat{\bOmega} \sim \hat{\bOmega}^{(0)} = \epsilon^{-1} \pd{\hPsi}{\hat{T}}\be_z + O(1). \label{eqn:velocityangularvelocityleadingorder}
\eeq
Using $\bd_3\sim\bdthreelead = \bt = h\sin\alpha\,\be_{\theta}+\cos\alpha\,\be_z$ (recall Eqs.~\eqref{eqn:FSframe} and \eqref{eqn:directorslead}), the external loads \eqref{eqn:dragforcemoment} to leading order (i.e., neglecting terms of $O(\epsilon)$) can be written in cylindrical polar coordinates as
\begin{align*}
\bFelead = & -\left[\hR\left(1-\chi\sin^2\alpha\right)\left(\sgn \omega_0 +\pd{\hPsi}{\hat{T}}\right) - h\chi\cos\alpha\sin\alpha\pd{\hZ}{\hat{T}}\right]\be_{\theta}  \\
& \: - \left[\left(1-\chi\cos^2\alpha\right)\pd{\hZ}{\hat{T}} - h\chi\hR\cos\alpha\sin\alpha\left(\sgn \omega_0 + \pd{\hPsi}{\hat{T}}\right)\right]\be_z,\qquad \\
\bMelead = & -\cos\alpha\left(\sgn \omega_0 + \pd{\hPsi}{\hat{T}}\right)\left(h\sin\alpha\,\be_{\theta}+\cos\alpha\,\be_z\right).
\end{align*}
We then calculate the wavelength-averaged external force and moment along $\be_z$:
\beqn
\overline{\be_z\cdot\bFelead} = -\Apa\pd{\hZ}{\hat{T}} - \Bpa\left(\sgn \omega_0 + \pd{\hPsi}{\hat{T}}\right), \quad
\delta\overline{\be_z\cdot\bMelead} + \overline{\be_z\cdot\left(\bRlead\times\bFelead\right)} = -\Bpa \pd{\hZ}{\hat{T}} - \Cpa\left(\sgn \omega_0 + \pd{\hPsi}{\hat{T}}\right),
\eeqn
where we define the drag coefficients
\beqn
\Apa = 1-\chi\cos^2\alpha, \quad
\Bpa = -h\chi\hR\cos\alpha\sin\alpha, \quad
\Cpa = \delta\cos^2\alpha + \hR^2\left(1-\chi\sin^2\alpha\right).
\eeqn

Substituting the above expressions into the equivalent-rod equations \eqref{eqn:forceavgZPsi}--\eqref{eqn:momentavgZPsi} from \S\ref{sec:effectivehZhPsi} then yields a closed system of PDEs for $\hZ(\hS,T)$ and $\hPsi(\hS,T)$, which, in the absence of the inertia terms, take the form of coupled diffusion equations. Because the drag coefficients generally depend on the (unknown) helical geometry, these equations are nonlinear; the drag coefficients and helical radius $\hR$ can be expressed in terms of $\hZ$ and $\hPsi$ using Eq.~\eqref{eqn:alphalambdaavgZPsi}. We therefore focus on the case $\epsilon L \ll 1$, considered below, for which an analytical solution of the equivalent-rod equations is possible.

We note that the external loads $\bFelead$ and $\bMelead$ are consistent with our assumption of a straight helix axis. More precisely, since the coefficients of $\be_{\theta}$ and $\be_z$ in the above expressions for $\bFelead$ and $\bMelead$ are independent of the fast variable, $S$, the off-axis components of $\bFelead$ average to zero over each wavelength: $\overline{\be_x\cdot\bFelead} = 0$ and $\overline{\be_y\cdot\bFelead} = 0$. Because we also neglect inertial terms, the off-axis solvability conditions \eqref{eqn:effectiveforceX,Y}, needed for a straight helix axis in the case $\epsilon L=\ord(1)$, are then satisfied.

\subsubsection{Formulation of the (straight) equivalent-rod equations: $\epsilon L \ll 1$}
For the remainder of this section we restrict to small deformations, $\epsilon L \ll 1$. In terms of the perturbations $\Delta\hZ \equiv \hZ - \hS\cos\aref$ and  $\Delta\hPsi \equiv \hPsi - 2\pi h\hS$ (introduced in \S\ref{sec:multiscalessolvabilitylinear}), and using $\alpha \sim \aref$ and $\hR \sim \hRref$, the external loads are
\begin{align*}
    \bFelead \sim & -\left[\hRref\left(1-\chi\sin^2\aref\right)\left(\sgn \omega_0 +\pd{\Delta\hPsi}{\hat{T}}\right) - h\chi\cos\aref\sin\aref\pd{\Delta\hZ}{\hat{T}}\right]\be_{\theta}  \\
    & \: - \left[\left(1-\chi\cos^2\aref\right)\pd{\Delta\hZ}{\hat{T}} - h\chi\hRref\cos\aref\sin\aref\left(\sgn \omega_0 + \pd{\Delta\hPsi}{\hat{T}}\right)\right]\be_z,\qquad \\
    \bMelead \sim & -\cos\aref\left(\sgn \omega_0 + \pd{\Delta\hPsi}{\hat{T}}\right)\left(h\sin\aref\,\be_{\theta}+\cos\aref\,\be_z\right).
\end{align*}
Similar to the case $\epsilon L=\ord(1)$ above, we calculate the wavelength-averaged external force and moment along $\be_z$:
\beqn
\overline{\be_z\cdot\bFelead} \sim -\Aparef\pd{\Delta\hZ}{\hat{T}} - \Bparef\left(\sgn \omega_0 + \pd{\Delta\hPsi}{\hat{T}}\right), \quad
\delta\overline{\be_z\cdot\bMelead} + \overline{\be_z\cdot\left(\bRref\times\bFelead\right)} \sim -\Bparef \pd{\Delta\hZ}{\hat{T}} - \Cparef\left(\sgn \omega_0 + \pd{\Delta\hPsi}{\hat{T}}\right),
\eeqn
where the drag coefficients are now evaluated using the undeformed geometry:
\beq
\Aparef = 1-\chi\cos^2\aref, \quad
\Bparef = -h\chi\hRref\cos\aref\sin\aref, \quad
\Cparef = \delta\cos^2\aref + \left(\hRref\right)^2\left(1-\chi\sin^2\aref\right). \label{eqn:defnABCparref}
\eeq

Neglecting inertia terms, the linearised equivalent-rod equations  \eqref{eqn:linearisedforceavgZPsi}--\eqref{eqn:linearisedmomentavgZPsi} and boundary conditions \eqref{eqn:linearisedBCsavgZPsi} become
\begin{gather}
 K_1^u\pdd{\Delta\hZ}{\hS} + K_2^u\pdd{\Delta\hPsi}{\hS} = \Aparef\pd{\Delta\hZ}{\hat{T}} + \Bparef\left(\sgn \omega_0+\pd{\Delta\hPsi}{\hat{T}}\right), \label{eqn:linearisedforceavgZPsiScenarioII} \\
 K_2^u\pdd{\Delta\hZ}{\hS} + K_4^u\pdd{\Delta\hPsi}{\hS} = \Bparef\pd{\Delta\hZ}{\hat{T}} + \Cparef\left(\sgn \omega_0+\pd{\Delta\hPsi}{\hat{T}}\right),  \label{eqn:linearisedmomentavgZPsiScenarioII} \\
 \Delta\hZ(0,\hat{T}) = \Delta\hPsi(0,\hat{T}) = 0, \quad \pd{\Delta\hZ}{\hS}(\epsilon L,\hat{T}) = \pd{\Delta\hPsi}{\hS}(\epsilon L,\hat{T}) = 0. \label{eqn:linearisedBCsavgZPsiScenarioII}
\end{gather}
For initial conditions, we suppose that the filament is undeformed at $\hat{T}=0$ before the rotation is instantaneously applied for $\hat{T}>0$:
\beq
\Delta\hZ(\hS,0) = \Delta\hPsi(\hS,0) = 0. \label{eqn:linearisedICsavgZPsiScenarioII}
\eeq
It is worth noting that, in the straight-rod limit $\aref\to 0$ with $\lamref$ fixed, Eqs.~\eqref{eqn:linearisedforceavgZPsiScenarioII}--\eqref{eqn:linearisedmomentavgZPsiScenarioII} reduce to the classical equation governing twist diffusion in a straight rod~\citep{wolgemuth2000} when we identify $\partial\Delta\hPsi/\partial\hS$ with the axial twist (recall the discussion in \S\ref{sec:straightrodlimit}).

An important point is that, in contrast to the case $\epsilon L=\ord(1)$, the external loading does not satisfy all off-axis solvability conditions. In particular, the additional constraints \eqref{eqn:effectivemomentXlinear}--\eqref{eqn:effectivemomentYlinear}, which only arise in the case $\epsilon L\ll 1$, become (using $\overline{\be_x\cdot\bMelead} = \overline{\be_y\cdot\bMelead} = 0$)
\beq
\overline{\be_x\cdot\left(\bRref\times\bFelead\right)} = 0, \qquad \overline{\be_y\cdot\left(\bRref\times\bFelead\right)} = 0 \qquad \mathrm{(off{-}axis\ solvability\ conditions)}. \label{eqn:offaxissolvscenario2}
\eeq
Using $\bRref(S)  = \hRref\be_r + S\cos\aref\be_z$ and Eq.~\eqref{eqn:defneRetheta} (with $\Psi \sim \Psiref$), we calculate
\beqn
\overline{\be_x\cdot\left(\bRref\times\bFelead\right)} \sim -\cos\aref\overline{S\be_y\cdot\bFelead}, \qquad \overline{\be_y\cdot\left(\bRref\times\bFelead\right)} \sim \cos\aref\overline{S\be_x\cdot\bFelead}.
\eeqn
Because the coefficient of $\be_{\theta}$ in the expression for $\bFelead$ is independent of $S$, we have $\overline{S\be_y\cdot\bFelead}\propto\overline{S\be_y\cdot\be_{\theta}} = \overline{S\cos\Psiref} \neq 0$ and $\overline{S\be_x\cdot\bFelead}\propto\overline{S\be_x\cdot\be_{\theta}} = -\overline{S\sin\Psiref} \neq 0$, so that in general $\overline{\be_x\cdot\left(\bRref\times\bFelead\right)} \neq 0$ and $\overline{\be_y\cdot\left(\bRref\times\bFelead\right)} \neq 0$.

Because is not possible to satisfy Eq.~\eqref{eqn:offaxissolvscenario2}, we expect that the helix axis will not remain straight to leading order: bending strains will arise that are of comparable size to the extensional and torsional strains predicted by the equivalent-rod model. This axis bending is verified by solutions of the full Kirchhoff rod equations, discussed later in \S\ref{sec:scenario2comparekirchhoff}. Nevertheless, we will also show that the solution of the equivalent-rod equations, obtained in \S\ref{sec:scenario2analyticalsol} below, approximates the \emph{average} strains from simulations very well, i.e., with small relative error, provided that $\epsilon L$ is not too small.

\subsection{Analytical solution for $\epsilon L \ll 1$}
\label{sec:scenario2analyticalsol}
To solve the linearised equivalent-rod equations \eqref{eqn:linearisedforceavgZPsiScenarioII}--\eqref{eqn:linearisedICsavgZPsiScenarioII}, we write them in matrix-vector form as
\begin{gather}
J^u\pdd{\mathcal{Y}}{\hS} = \Upsilon^u\left[\pd{\mathcal{Y}}{\hat{T}} + \begin{pmatrix} 0 \\ \sgn\omega_0 \end{pmatrix}\right] \qquad \mathrm{where} \qquad
\mathcal{Y}\left(\hS,\hat{T}\right) = \begin{pmatrix} \Delta\hZ \\ \Delta\hPsi \end{pmatrix}, \quad 
J^u = \begin{pmatrix} K_1^u & K_2^u \\ K_2^u & K_4^u \end{pmatrix}, \quad 
\Upsilon^u = \begin{pmatrix}\Aparef & \Bparef \\ \Bparef & \Cparef \end{pmatrix}, \label{eqn:linearisedforcemomentmatrixformScenarioIIfull} \\
\mathcal{Y}\left(0,\hat{T}\right) = \sszero_{2\times 1}, \quad \pd{\mathcal{Y}}{\hS}\left(\epsilon L,\hat{T}\right) = \sszero_{2\times 1}, \quad \mathcal{Y}\left(\hS,0\right) = \sszero_{2\times 1}. \label{eqn:linearisedBCsICsmatrixformScenarioII}
\end{gather}
Note that the matrices of undeformed stiffness and drag coefficients have positive determinant: from Eqs.~\eqref{eqn:defnKiref} and \eqref{eqn:defnABCparref}, we calculate
\beqn
\det{J^u} = K_1^u K_4^u - \left(K_2^u\right)^2 = \frac{4\pi^2\csc^2\aref}{1+\nu}, \quad \det{\Upsilon^u} = \Aparef\Cparef-\left(\Bparef\right)^2 = \delta\cos^2\aref\left(1-\chi\cos^2\aref\right) + \left(1-\chi\right)\frac{\sin^2\aref}{4\pi^2}.
\eeqn
Hence, after pre-multiplying by $(\Upsilon^u)^{-1}$, Eq.~\eqref{eqn:linearisedforcemomentmatrixformScenarioIIfull} becomes
\beq
\mathcal{M}\pdd{\mathcal{Y}}{\hS} = \pd{\mathcal{Y}}{\hat{T}} + \begin{pmatrix} 0 \\ \sgn\omega_0 \end{pmatrix} \qquad \mathrm{where} \qquad \mathcal{M} = \left(\Upsilon^u\right)^{-1}J^u = \frac{1}{\Aparef\Cparef-(\Bparef)^2}\begin{pmatrix} K_1^u\Cparef-K_2^u\Bparef & K_2^u\Cparef-K_4^u\Bparef \\ K_2^u\Aparef-K_1^u\Bparef & K_4^u\Aparef-K_2^u\Bparef \end{pmatrix}. \label{eqn:linearisedforcemomentmatrixformScenarioII}
\eeq

Equation \eqref{eqn:linearisedforcemomentmatrixformScenarioII}, together with the boundary conditions and initial conditions in Eq.~\eqref{eqn:linearisedBCsICsmatrixformScenarioII}, can be solved using a variety of methods. We choose to decompose the solution into a steady part, $\mathcal{Y}^{\mathrm{S}}$ (with $\mathcal{Y}\to \mathcal{Y}^{\mathrm{S}}$ as $\hat{T}\to\infty$), and a transient part, $\mathcal{Y}^{\mathrm{D}}$; we then seek a separable solution for the transient part, noting that the boundary conditions in Eq.~\eqref{eqn:linearisedBCsICsmatrixformScenarioII} imply the spatial dependence of $\mathcal{Y}^{\mathrm{D}}$ is of the form $\sin\left[(2n+1)\pi\hS/(2\epsilon L)\right]$ ($n = 0,1,2,\ldots$). Writing $M_{ij}$ for the entries of $\mathcal{M}$, the final result can be written as\footnote{It may be verified that Eq.~\eqref{eqn:linearisedsolnmatrixformScenarioII} satisfies the initial condition in Eq.~\eqref{eqn:linearisedBCsICsmatrixformScenarioII} using the identity $\frac{32}{\pi^3}\sum_{n=0}^{\infty}\frac{1}{\left(2n+1\right)^3}\sin\left[\frac{(2n+1)\pi}{2}\frac{\hS}{\epsilon L}\right]=\frac{\hS}{\epsilon L}\left(2-\frac{\hS}{\epsilon L}\right)$ and $\det\mathcal{M}=\mu_{+}\mu_{-}$.}
\begin{gather}
\mathcal{Y}\left(\hS,\hat{T}\right) = \mathcal{Y}^{\mathrm{S}}\left(\hS\right) + \mathcal{Y}^{\mathrm{D}}\left(\hS,\hat{T}\right) \qquad \mathrm{where} \qquad
\mathcal{Y}^{\mathrm{S}}\left(\hS\right) = -\frac{\left(\epsilon L\right)^2\sgn\omega_0}{2\det\mathcal{M}}\frac{\hS}{\epsilon L}\left(2-\frac{\hS}{\epsilon L}\right)\begin{pmatrix} -M_{12} \\ M_{11}\end{pmatrix}, \label{eqn:linearisedsolnmatrixformScenarioII} \\
\mathcal{Y}^{\mathrm{D}}\left(\hS,\hat{T}\right) = \frac{16\left(\epsilon L\right)^2\sgn\omega_0}{\pi^3\left(\mu_{+}-\mu_{-}\right)}\sum_{n=0}^{\infty}\frac{\sin\left[\frac{(2n+1)\pi}{2}\frac{\hS}{\epsilon L}\right]}{\left(2n+1\right)^3}\left\lbrace \begin{pmatrix} M_{12}/\mu_{+} \\ 1-M_{11}/\mu_{+}\end{pmatrix}e^{-\tfrac{(2n+1)^2\pi^2}{4}\tfrac{\mu_{+}\hat{T}}{(\epsilon L)^2}} - \begin{pmatrix} M_{12}/\mu_{-} \\ 1-M_{11}/\mu_{-}\end{pmatrix}e^{-\tfrac{(2n+1)^2\pi^2}{4}\tfrac{\mu_{-}\hat{T}}{(\epsilon L)^2}}\right\rbrace, \nonumber
\end{gather}
where $\mu_{\pm}$ are the eigenvalues of $\mathcal{M}$:
\beqn
\mu_{\pm} \equiv \frac{K_1^u\Cparef-2 K_2^u\Bparef+K_4^u\Aparef}{2\left[\Aparef\Cparef-(\Bparef)^2\right]} \pm \sqrt{\left\lbrace\frac{K_1^u\Cparef-2 K_2^u\Bparef+K_4^u\Aparef}{2\left[\Aparef\Cparef-(\Bparef)^2\right]}\right\rbrace^2-\frac{K_1^u K_4^u - (K_2^u)^2}{\Aparef\Cparef-(\Bparef)^2}}.
\eeqn

\begin{figure}[t]
    \centering
    \includegraphics[width=0.9\textwidth]{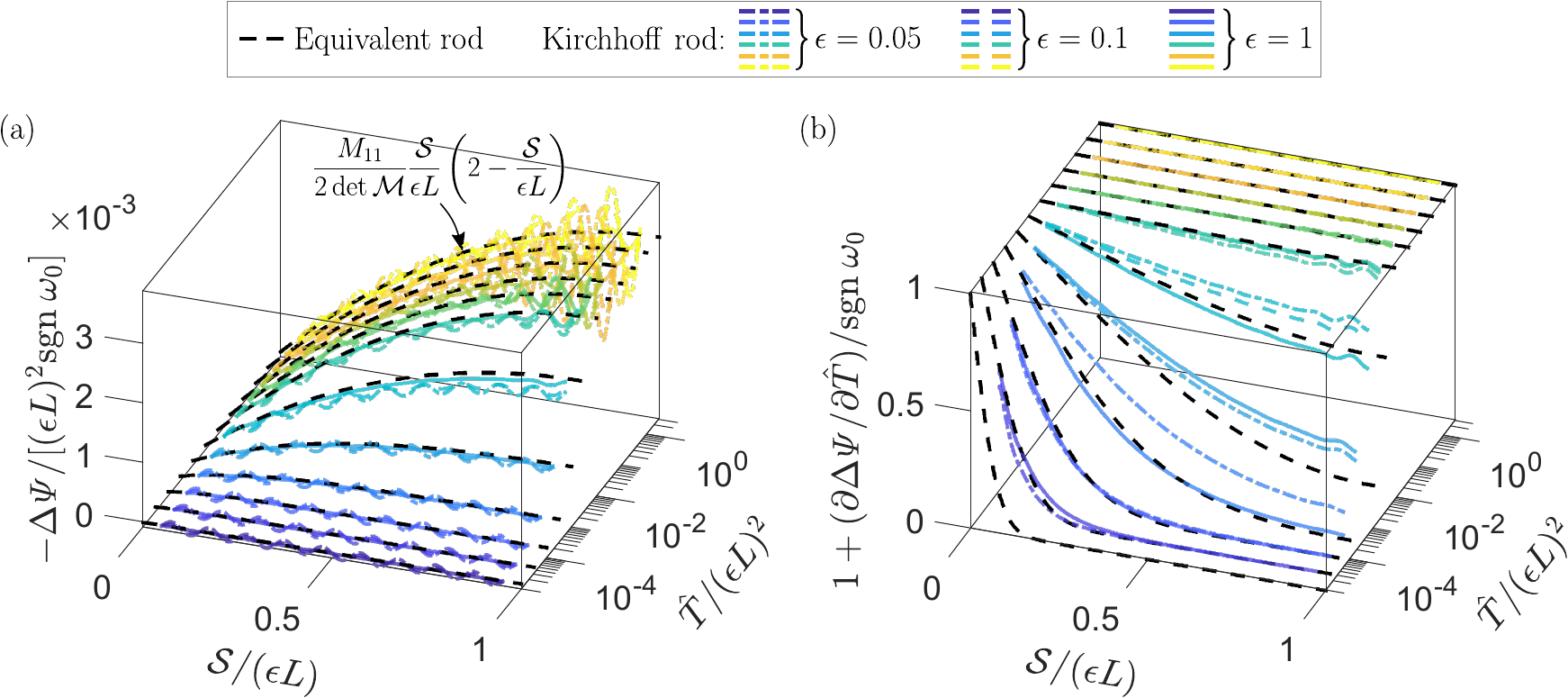}
    \caption{Axial rotation (twirling) in viscous fluid: viscous dynamics of a helical rod that is suddenly rotated at its base at a prescribed angular frequency $\omega_0$ (parameter values in Eq.~\eqref{eqn:scenario2paramvals}). We plot spatio-temporal diagrams of (a) the (negative) wavelength-averaged rotational displacement, $-\Delta\hPsi/[(\epsilon L)^2\sgn\omega_0]$; and (b) $1+(\partial\Delta\hPsi/\partial\hat{T})/\sgn\omega_0$, which corresponds to the leading-order angular velocity in the laboratory frame (normalised by $\omega_0$). In both panels, predictions of the linearised (straight) equivalent-rod model, evaluated using the analytical solution Eq.~\eqref{eqn:linearisedsolnmatrixformScenarioII}, are shown (black dashed curves), together with results of dynamic Kirchhoff rod simulations with $L = 10$ and $\epsilon \in \lbrace 0.05,0.1,1\rbrace$ (dash-dotted, dashed and solid coloured curves, respectively; see legend).}
    \label{fig:ScenarioIISol}
\end{figure}

In what follows, we consider typical parameter values for a bacteria flagellar filament in its normal left-handed helical form \citep{namba1997,vogel2012,son2013}:
\beq
\aref = 30^{\circ}, \quad \nu = 0, \quad h = -1, \quad \frac{a}{\lamref} = 10^{-2}. \label{eqn:scenario2paramvals}
\eeq
In Fig.~\ref{fig:ScenarioIISol}a we use the solution in Eq.~\eqref{eqn:linearisedsolnmatrixformScenarioII} to construct a spatio-temporal plot of the wavelength-averaged rotational displacement, $\Delta\hPsi$. (The corresponding plot for $\Delta\hZ$ is similar and will not be discussed here.) In particular, we plot curves of $-\Delta\hPsi/[(\epsilon L)^2\sgn{\omega_0}]$ as a function of $\hS/(\epsilon L)$, for several times $\hat{T}/(\epsilon L)^2$ (black dashed curves); these re-scalings are chosen according to Eq.~\eqref{eqn:linearisedsolnmatrixformScenarioII}, so that the plotted curves are independent of the values of $\epsilon L$ and $\sgn{\omega_0}$. We observe an initial transient in which the displacement varies from zero, after which the solution approaches the quadratic profile predicted by the steady part of the solution, $\mathcal{Y}^{\mathrm{S}}$, in Eq.~\eqref{eqn:linearisedsolnmatrixformScenarioII}. 

Figure \ref{fig:ScenarioIISol}b displays the corresponding spatio-temporal plot for $1+(\partial\Delta\hPsi/\partial\hat{T})/\sgn\omega_0$ (black dashed curves), which corresponds to the leading-order angular velocity (about the helix axis) in the \emph{laboratory frame}, normalised by $\omega_0$: from Eqs.~\eqref{eqn:scenarioIInondim} and \eqref{eqn:velocityangularvelocityleadingorder} (and using $\Delta\hPsi = \hPsi - 2\pi h\hS$), we have $1+(\partial\Delta\hPsi/\partial\hat{T})/\sgn\omega_0 \sim \be_z\cdot(\omega_0\be_z+\bomega)/\omega_0$. At early times, the equivalent-rod model predicts that the bulk of the filament is at rest relative to the fluid, with only a neighbourhood of the filament base (where the specified frequency is instantaneously applied for $\hat{T}>0$) rotating appreciably. As $\hat{T}$ increases, the rotation rate propagates along the filament until the filament reaches a state of uniform rotation, which is approximately attained for times $\hat{T}/(\epsilon L)^2 \gtrsim 10^{-2}$.

\subsubsection{Resultant force and moment at the filament base}
For locomotion driven by axial rotation of a helical filament in viscous fluid, a key quantity is the resultant force at the filament base, $F_Z(0,\hat{T})$: if the filament is no longer tethered but attached to a freely-swimming body, the swimming speed is determined by balancing this propulsive force with the total viscous drag on the body and filament (neglecting inertia). Thus, in absence of other forces, $F_Z(0,\hat{T})$ is proportional to the swimming velocity. The resultant moment $M_Z(0,\hat{T})$ corresponds to the torque required by the rotary motor to achieve the imposed frequency $\omega_0$.

\begin{figure}[t]
    \centering
    \includegraphics[width=0.9\textwidth]{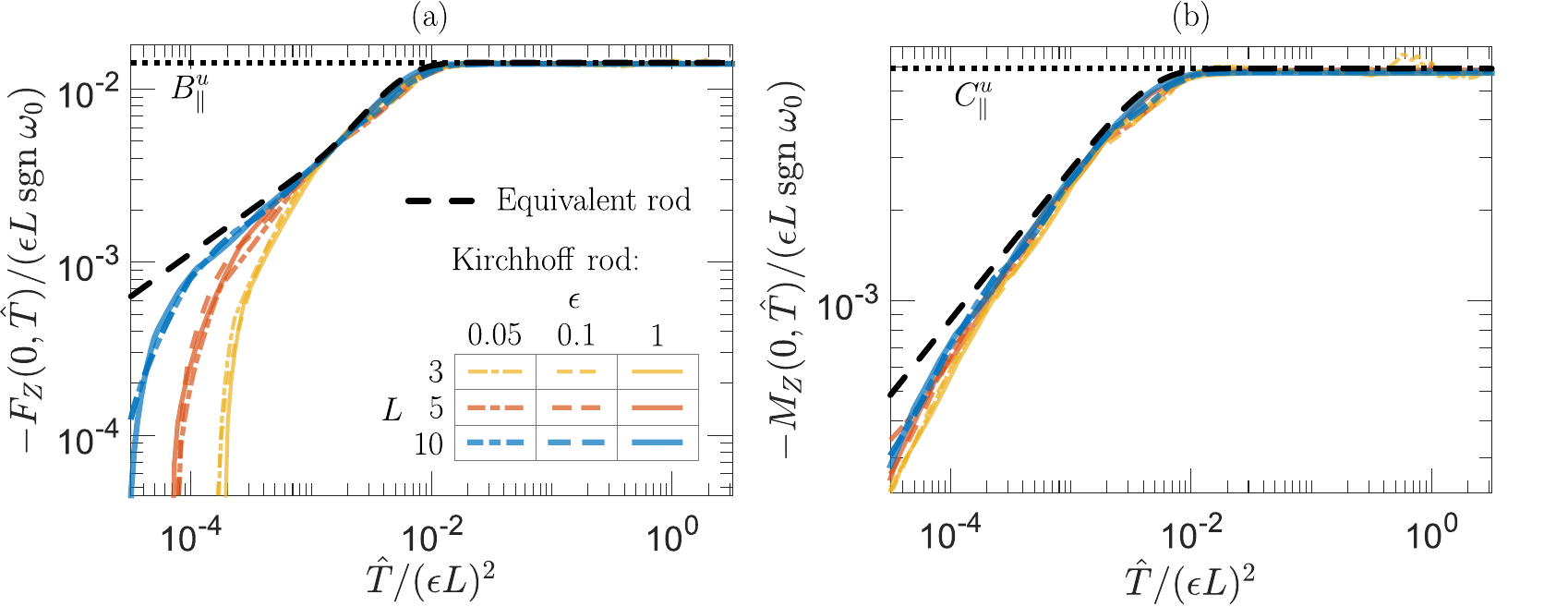}
    \caption{Axial rotation (twirling) in viscous fluid: (a) resultant force and (b) resultant moment at the filament base, re-scaled according to Eq.~\eqref{eqn:linearisedresultantsScenarioII} (parameter values in Eq.~\eqref{eqn:scenario2paramvals}). In both panels, we plot the analytical predictions \eqref{eqn:linearisedresultantsScenarioII} of the linearised (straight) equivalent-rod model (black dashed curves), corresponding to the solution shown in Fig.~\ref{fig:ScenarioIISol}. Also shown are results of dynamic Kirchhoff rod simulations using $L \in \lbrace 3,5,10\rbrace$ and $\epsilon \in \lbrace 0.05,0.1,1\rbrace$, as indicated by line thickness/colour and linestyle (see legend).}
    \label{fig:ScenarioIIForceMoment}
\end{figure}

Using the expressions for the linearised resultants $F_Z$ and $M_Z$ in terms of $\Delta\hZ$ and $\Delta\hPsi$ (from combining Eqs.~\eqref{eqn:relateexpandhelixgeometrytoavgZPsi}--\eqref{eqn:resultantsleadZcomponentshelixgeometrylinearised}), we obtain, after simplifying,
\begin{align}
F_Z(0,\hat{T}) & = -\epsilon L\sgn\omega_0\left\lbrace\Bparef -\frac{8}{\pi^2}\sum_{n=0}^{\infty}\frac{1}{\left(2n+1\right)^2}\left[\frac{K_2^u-\mu_{-}\Bparef}{\mu_{+}-\mu_{-}}e^{-\tfrac{(2n+1)^2\pi^2}{4}\tfrac{\mu_{+}\hat{T}}{(\epsilon L)^2}} - \frac{K_2^u-\mu_{+}\Bparef}{\mu_{+}-\mu_{-}}e^{-\tfrac{(2n+1)^2\pi^2}{4}\tfrac{\mu_{-}\hat{T}}{(\epsilon L)^2}} \right]\right\rbrace, \nonumber \\
M_Z(0,\hat{T}) & = -\epsilon L\sgn\omega_0\left\lbrace\Cparef -\frac{8}{\pi^2}\sum_{n=0}^{\infty}\frac{1}{\left(2n+1\right)^2}\left[\frac{K_4^u-\mu_{-}\Cparef}{\mu_{+}-\mu_{-}}e^{-\tfrac{(2n+1)^2\pi^2}{4}\tfrac{\mu_{+}\hat{T}}{(\epsilon L)^2}} -  \frac{K_4^u-\mu_{+}\Cparef}{\mu_{+}-\mu_{-}}e^{-\tfrac{(2n+1)^2\pi^2}{4}\tfrac{\mu_{-}\hat{T}}{(\epsilon L)^2}}\right]\right\rbrace. \label{eqn:linearisedresultantsScenarioII}
\end{align}
In Fig.~\ref{fig:ScenarioIIForceMoment}, we use these expressions to plot (as black dashed curves) the resultant force (Fig.~\ref{fig:ScenarioIIForceMoment}a) and resultant moment (Fig.~\ref{fig:ScenarioIIForceMoment}b) at the filament base, corresponding to the solution shown in Fig.~\ref{fig:ScenarioIISol}. Figure \ref{fig:ScenarioIIForceMoment} shows how both resultants grow like $\hat{T}^{1/2}$ (i.e., diffusively) at early times, $\hat{T}/(\epsilon L)^2 \lesssim 10^{-3}$. The resultants then approach the steady values (equal to the drag coefficients $\Bparef$ and $\Cparef$; black dotted lines) for times $\hat{T}/(\epsilon L)^2 \gtrsim 10^{-2}$, corresponding to when the rotation rate is close to being spatially uniform in Fig.~\ref{fig:ScenarioIISol}b.

\subsection{Comparison with dynamic Kirchhoff rod simulations}
\label{sec:scenario2comparekirchhoff}
To test our (straight) equivalent-rod theory, we perform dynamic simulations of the full Kirchhoff rod equations, based on the integro-differential formulation discussed in \S\ref{sec:numerics}. For the hydrodynamic loading considered here, i.e., Eq.~\eqref{eqn:dragforcemoment}, we set
\begin{gather*}
\bAe = -\epsilon\left(\mathbf{I}-\chi\bd_3\otimes\bd_3\right), \quad \bBe = -\sgn\omega_0\left(\mathbf{I}-\chi\bd_3\otimes\bd_3\right)\skewmat{\be_z}, \quad \bCe = \mathbf{0}, \\
\bDe = -\epsilon\,\bd_3\otimes\bd_3, \quad \bEe = -\sgn\omega_0\left(\bd_3\otimes\bd_3\right)\skewmat{\be_z}, 
\end{gather*}
in Eq.~\eqref{eqn:loadingnumeric} where $\mathbf{I}$ is the second-order identity tensor. To directly compare simulation results with the equivalent-rod theory, we compute the effective winding angle, $\Psi$, and longitudinal coordinate, $Z$, in our simulations; these are determined from values of the centreline $\bR$ on the numerical mesh using $\tan\Psi = \bR\cdot\be_y/(\bR\cdot\be_x)$ and $Z=\be_z\cdot\bR$. The wavelength-averaged rotational and longitudinal displacements are then found using a centred moving average, with window size equal to the undeformed wavelength, $1$ (we consider only mesh points where a full window size is available, i.e., for $S\in[0.5,L-0.5]$).

Simulations results for the wavelength-averaged rotational displacement are plotted in Fig.~\ref{fig:ScenarioIISol}a (coloured curves). Here three sets of simulations are shown, which use the parameter values reported in Eq.~\eqref{eqn:scenario2paramvals} together with a dimensionless length $L = 10$ and $\epsilon \in \lbrace 0.05,0.1,1\rbrace$ (dash-dotted, dashed and solid curves, respectively; see legend). Significant oscillations are present in the numerical curves for $\epsilon = 0.05$, whose amplitude generally increases in time until a steady configuration is reached. These oscillations are due to axis bending in the regime $\epsilon L\ll 1$, for which the off-axis solvability conditions are not satisfied; simulations for smaller $\epsilon$ (not shown) display larger relative oscillations when plotted on Fig.~\ref{fig:ScenarioIISol}a. Nevertheless, we see that the (straight) equivalent-rod solution captures well the average profile of the numerical curves. As $\epsilon$ is increased up to the regime where $\epsilon L=\ord(1)$, the oscillations in the curves decrease in amplitude; here the bending strains become asymptotically small, as predicted in \S\ref{sec:scenario2nonlinearform}. Remarkably, we obtain excellent agreement with the linearised equivalent-rod model in this regime up to $\epsilon = 1$, well beyond the limit of validity of the key assumptions underlying the theory: namely, a highly-coiled filament ($\epsilon \ll 1$) and small deformations ($\epsilon L\ll 1$).

In Fig.~\ref{fig:ScenarioIISol}b, we plot corresponding simulation results for the average angular velocity in the laboratory frame. While we do not observe significant oscillations in the numerical curves, the agreement with the equivalent-rod model is generally worse as $\epsilon$ decreases. Moreover, for all values of $\epsilon$, the agreement breaks down at early times when the bulk of the filament is at rest. In this regime, variations in the angular velocity occur on a lengthscale comparable to the helical wavelength, which we do not expect the equivalent-rod theory to capture adequately.

Corresponding simulation results for the resultant force and moment at the filament base are plotted in Fig.~\ref{fig:ScenarioIIForceMoment} (coloured dash-dotted, dashed and solid curves). In addition, we show results for simulations in which the filament length is reduced to $L = 5$ and $L = 3$, again with $\epsilon \in \lbrace 0.05,0.1,1\rbrace$, as indicated by reduced line thicknesses (see legend). In all $9$ sets of simulations, we observe excellent agreement with the analytical predictions \eqref{eqn:linearisedresultantsScenarioII} for times $\hat{T}/(\epsilon L)^2\gtrsim 10^{-3}$, with the numerical curves collapsing onto the theoretical curve. However, at earlier times, the numerical curves deviate significantly from the diffusive behaviour ($\propto\hat{T}^{1/2}$) predicted by the equivalent-rod model. We attribute this to the large variations in angular velocity at early times (Fig.~\ref{fig:ScenarioIISol}b), consistent with the observation that the disagreement in Figs.~\ref{fig:ScenarioIIForceMoment}a--b is larger for simulations with a smaller length $L$ (i.e., a fewer total number of wavelengths).

\section{Discussion and conclusions}
\label{sec:conclusion}

\subsection{Summary of findings}
\label{sec:conclusionfindings}
In this paper, we have studied slender, helical rods undergoing unsteady deformations in the presence of distributed forces and moments. Focussing on the case when the helix axis remains straight, we have derived a (straight) equivalent-rod theory via an analytical reduction of the Kirchhoff rod equations. This analytical reduction is asymptotically valid in the limit of a highly-coiled filament, i.e., when the helical wavelength is much smaller than the typical deformation lengthscale. The (dimensionless) equivalent-rod equations comprise two coupled PDEs, Eqs.~\eqref{eqn:effectiveforcebalance}--\eqref{eqn:effectivemomentbalance} (Eqs.~\eqref{eqn:effectiveforcebalancelinear}--\eqref{eqn:effectivemomentbalancelinear} in the small-deformation limit $\epsilon L \ll 1$), which correspond to wavelength-averaged force and moment balances about the helix axis; together with constraints on the external loading needed for a straight helix axis. Equations \eqref{eqn:effectiveforcebalance}--\eqref{eqn:effectivemomentbalance} can further be written as a quasi-linear system of equations, in terms of two independent variables that uniquely characterise the locally-helical shape. We focussed on two such pairs of variables in \S\ref{sec:equivalent-rod}: the pitch angle and contour wavelength, and the wavelength-averaged longitudinal and rotational displacement.


The equivalent-rod equations provide a simplified modelling framework, applicable to a wide variety of physical and biological settings, that allows for a great deal of analytical progress or rapid numerical solution. In particular, the equations account for unsteady displacements and rod inertia. In the absence of distributed loads, we demonstrated that the linearised equations reduce to the coupled wave equations previously proposed to describe the free vibrations of helical coil springs \citep{phillips1972,jiang1989,jiang1991}. Our analysis therefore provides a rigorous justification that the linearised stiffness coefficients can be applied locally (i.e., for each infinitesimal element) in situations involving unsteady deformations and distributed loads, provided that the loading is consistent with the assumption of a straight helix axis. In addition to the free vibrations of helical coil springs, we illustrated the applicability of our theory with two physical scenarios: (I) the compression/extension of helices under gravity (\S\ref{sec:scenario1}), and (II) the over-damped dynamics of helical rods twirling in viscous fluid  (\S\ref{sec:scenario2}). In both scenarios, we obtained excellent agreement with solutions of the full Kirchhoff rod equations, even beyond the formal limit of validity of the highly-coiled assumption ($\epsilon \ll 1$).

\subsection{Discussion and outlook}
\label{sec:conclusiondiscussion}
Our equivalent-rod description is distinct from classic perturbative approaches, which consider small deformations from a known base state, usually taken to be the undeformed helical shape \citep{goriely1997a,goriely1997b,goriely1997c,takano2003,kim2005,katsamba2019}. The main difference here is that we consider slowly-varying changes to an \emph{unknown} leading-order shape. The relevant small parameter in our analysis is thus the gradient of the deformation along the arclength --- there is no restriction on the global size of the displacements, provided that the (local) strains remain small. Our analysis is therefore similar to dimension reduction methods for slender structures (rods, plates, shells), which are based on the assumption that variations in the strains occur on lengthscales much larger than the cross-section dimensions. These methods systematically eliminate the dependence of field variables over the cross-section, to obtain a lower dimensional model describing an effective centreline or mid-surface \citep{lestringant2020a}; examples include prismatic solids undergoing tensile necking \citep{audoly2016}, hyperelastic cylindrical membranes \citep{lestringant2018,yu2023}, and elastic ribbons \citep{audoly2021b,kumar2023,gomez2023}. For the highly-coiled helical rods considered here, the effective centreline is simply the helix axis.

In addition, the multiple-scales analysis developed in \S\ref{sec:multiscales} can be viewed as a homogenisation procedure in which the helical wavelength plays the role of a periodic unit cell over which the governing equations are averaged. In this sense, our analysis is similar to the work of \cite{kehrbaum2000} and \cite{rey2000} on straight rods with high intrinsic twist. However, in contrast to these studies, we did not pursue a Hamiltonian formulation of the Kirchhoff rod equations, but instead we worked directly with the equations of force and moment balance. This allowed us to readily incorporate general external forces and moments, including those that are non-conservative such as the hydrodynamic loading considered in \S\ref{sec:scenario2}. While simple hydrodynamic models such as resistive-force theory could be incorporated into a Hamiltonian formalism without much difficulty, we expect that the framework introduced here may more readily be extended to incorporate other, more complex, fluid frameworks, such as slender-body theory. 

The basis of our asymptotic method is the helical solution of the inextensible Kirchhoff rod equations, in the case of a constant wrench aligned with the helix axis \citep{love1944}; the assumptions of a highly-coiled helix and a straight helix axis then guarantee that this solution holds locally when distributed loads are present. Thus, we expect that our analysis can be extended to other situations, provided that there is sufficient symmetry such that a locally-helical solution persists. One important example is contact forces, either due to external radial confinement or self contact: under longitudinal compression, the helical symmetry guarantees that the net force arising from self-contact is directed towards the helix axis, i.e.~along $-\be_r$, for which a locally-helical solution to the Kirchhoff rod equations still holds --- see \cite{chouaieb2006}. Another example is the case of helical rods whose cross-section is rectangular (i.e., ribbons), which permit a helical solution in some cases \citep{goriely2001}. 

\paragraph{Physical significance of singularities}
The analysis in \S\ref{sec:effectivejacobian} indicates that the steady equivalent-rod equations become singular when the matrix of stiffness coefficients has zero determinant. In particular, in the $(\alpha,\Lambda)$ formulation of the dimensionless equivalent-rod equations, these singularities can be visualised as branches of critical points on the $(\alpha,\Lambda)$ phase-plane; recall Fig.~\ref{fig:determinantroots}. As the system approaches one of these critical points, the magnitude of the gradient vector $(\partial\alpha/\partial\hS,\partial\Lambda/\partial\hS)$ tends to infinity --- we observed such behaviour for compressive gravitational loading in \S\ref{sec:scenario1} (see Fig.~\ref{fig:ScenarioIPhasePlanes}). However, as soon as the helix geometry varies on a lengthscale comparable to the helical wavelength, $S = O(1)$, the multiple-scales analysis presented in \S\ref{sec:multiscales} will no longer be asymptotically valid. Nevertheless, we might expect that such critical points signal an underlying, physically-relevant instability, whose description would require a detailed analysis of the rod equations on the wavelength lengthscale.  

The behaviour of critical points as the undeformed pitch angle $\aref$ varies (Fig.~\ref{fig:determinantroots}) has implications for their physical relevance. For small $\aref$, the branches of critical points in Fig.~\ref{fig:determinantroots} evidently lie at values of $\alpha$ much larger than the undeformed value $\aref$ (drawn as circles in Fig.~\ref{fig:determinantroots}). Hence, in the vicinity of the critical points, we expect that the filament will be under a large amount of longitudinal compression (i.e., the force resultant $F_Z < 0$) as it must be compressed to a pitch angle $\alpha > \aref$. For given external loads, as $\hS$ varies and the system moves through the $(\alpha,\Lambda)$ phase plane, we therefore expect that the helix axis buckles before the system is able to approach the vicinity of any critical point (neglecting dynamic effects); we discussed this point in the context of gravitational loading in \S\ref{sec:scenario1}. However, for larger values of $\aref$, the branch closest to the undeformed point $(\aref,1)$ lies at values $\alpha < \aref$ where the filament is under longitudinal tension ($F_Z > 0$).  It is therefore conceivable that the system remains stable as it approaches such critical points, which may correspond to tensile instabilities that can be observed experimentally. 

\begin{figure}[t]
\centering
\includegraphics[width=0.9\textwidth]{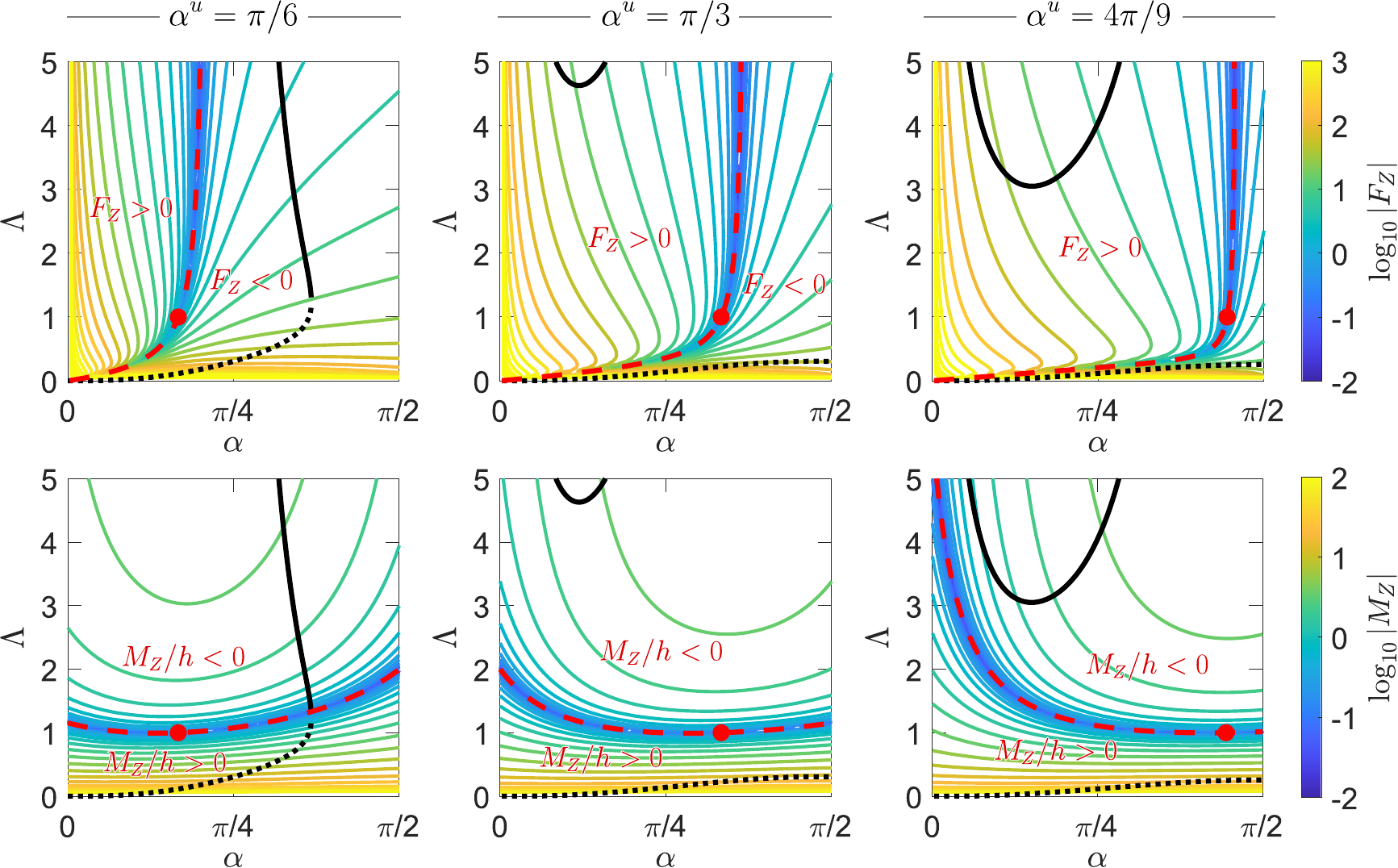}
\caption{Contour plots of the dimensionless resultant force $F_Z$ (top panels) and moment $M_Z$ (bottom panels) on the $(\alpha,\Lambda)$-plane, coloured according to the logarithm of their magnitude (see colourbars; here $\nu = 1/3$). In each plot, the zero contour separating regions of positive and negative force/moment resultant is shown (red dashed curves); this contour passes through the undeformed point $(\alpha,\Lambda) = (\aref,1)$ (red circles). Also shown are the roots $\Lambda_{-}$ and $\Lambda_{+}$ (black solid and dotted curves, respectively) from Eq.~\eqref{eqn:defnLambdapm}, at which the matrix of stiffness coefficients is singular.}
\label{fig:forcecontours}
\end{figure}

To further illustrate this behaviour, we show in Fig.~\ref{fig:forcecontours} contour plots of the dimensionless force resultant $F_Z$ (top panels) and moment resultant $M_Z$ (bottom panels) for three different values of $\aref$ in the physical range $0 < \alpha < \pi/2$ (these resultants are evaluated using Eq.~\eqref{eqn:resultantsleadZcomponentshelixgeometry}). For the smallest value $\aref = \pi/6$ (left column), the critical branches $\Lambda_{\pm}$ (black curves) both lie to the right of the zero-force contour $F_Z = 0$ (red dashed curve), where $F_Z < 0$. For $\aref \gtrsim \pi/3$, however, the branch closest to the undeformed point $(\aref,1)$ instead lies in the region $F_Z > 0$; see the middle and right columns in Fig.~\ref{fig:forcecontours}. We also observe that the critical branch lies in the region $M_Z/h < 0$. 

The above discussion indicates that tensile instabilities are a generic phenomenon when helical rods are subject to \emph{both} distributed forces and moments. Because these instabilities may be observed for isotropic rods with circular cross-section, they are qualitatively different to the tensile instabilities previously described for inextensible helical ribbons and anisotropic helical rods; see \cite{starostin2008} and references therein. Indeed, instabilities characterised by large stretching deformations (with approximately straight helix axis) have been observed in several systems involving distributed forces and moments. Examples include the abrupt, localised uncoiling of microscopic helical ribbons under directed fluid flow~\citep{pham2015}; and helices composed of magneto-rheological elastomers (MREs), which suddenly extend from a highly-collapsed state (attained due to long range dipole-dipole interactions) as the gradient of the magnetic field exceeds a critical value~\citep{sano2022}. It remains an intriguing possibility if these instabilities can be predicted using the analytical framework developed here.

\paragraph{Assumption of a straight helix axis}
Our analysis rests on a number of assumptions (summarised earlier in \S\ref{sec:papersummary}), the most severe of which is that the helix axis remains straight. This assumption places a strong restriction on the external forces and moments, since these must satisfy the off-axis solvability conditions obtained in \S\ref{sec:multiscales}. As might be expected, a straight helix axis is only possible if the external force exactly balances the off-axis components of the filament acceleration, when averaged over the helical wavelength. Surprisingly, we found that additional  constraints on the external loading --- which take the form of wavelength-averaged moment balances in the off-axis directions --- arise \emph{only} in the case $\epsilon L\ll 1$, corresponding to small deformations. Indeed, in Scenario II, these additional constraints are not satisfied and we observed oscillations in the strain components for $\epsilon L\ll 1$ resulting from axis bending. These oscillations diminish as $\epsilon$ increases up to the point where $\epsilon L=\ord(1)$, so that, counter-intuitively, the agreement between the simulation results and the linearised equivalent-rod theory \emph{improves}, even up to $\epsilon = 1$. Nevertheless, in other loading scenarios, the loading path may be more complicated, with the possibility of multiple stable states. In such scenarios, if significant axis bending occurs for $\epsilon L\ll 1$, the filament will not necessarily straighten as $\epsilon$ increases and so the (straight) equivalent-rod theory will not apply.

If the helix axis is not assumed straight, it is expected to bend over the deformation lengthscale $[s] = (B/[f])^{1/3}$, which arises from a balance between the external force density $[f]$ and the typical force (per unit length) required to bend the filament; recall the discussion in \S\ref{sec:highlycoiledassum}. In dimensionless variables, this introduces a bending lengthscale, $\aS$, defined by $S= \epsilon^{-1/3}\aS$ (using $S = s/\lamref$ and $\epsilon = (\lamref/[s])^3$). It is then necessary to generalise the locally-helical solution in \S\ref{sec:multiscaleskinematics} to incorporate a slowly-varying isometry $Q = Q(\aS,T)$ applied to the leading-order directors. The resulting bending strains, resulting from $\aS$-derivatives of $Q$, would contribute $\ord(\epsilon^{1/3})$ terms to the strain components, meaning that a regular asymptotic expansion in $\epsilon$ (like that in \S\ref{sec:perturbationscheme}) is no longer appropriate. This additional complexity is the topic of future work, which aims to obtain an effective-beam model describing the geometrically-nonlinear displacements of a helical rod in three dimensions. Such a model would have immediate applications in a number of contexts. For example, when combined with mechanical modelling of polymorphic shapes~\citep{srigiriraju2005}, it would allow the full dynamics of flagellar filaments on swimming bacteria to be modelled~\citep{turner2000}. Other potential biomechanical applications include the dynamics of flicking for bacteria~\citep{xie2011,son2013}, bacterial pumps~\citep{gao2015} and artificial flagella~\citep{zhang2010,huang2019}.

\section*{Acknowledgements}
M.G. gratefully acknowledges financial support from Peterhouse, University of Cambridge during early stages of this work. This project has received funding from the European Research Council (ERC) under the European Union's Horizon 2020 research and innovation programme (grant agreement 682754 to E.L.).

\appendix

\gdef\thesection{\appendixname \Alph{section}} 
\makeatletter

\section{Leading-order centreline and winding angle associated with the locally-helical ansatz}
\label{sec:appendixcentreline}
\setcounter{figure}{0}

In this Appendix we calculate the leading-order centreline, $\bRlead$, and winding angle, $\Psi$, associated with our ansatz of a locally-helical shape. From Eqs.~\eqref{eqn:FSframe} and \eqref{eqn:directorslead}, the unit tangent vector $\bdthreelead = h\sin\alpha\,\be_{\theta}+\cos\alpha\,\be_z$. Substituting into the inextensibility constraint \eqref{eqn:inextensibility} and expanding the derivative using the chain rule \eqref{eqn:chainrule}, we obtain
\beq
\pdS{\bRlead} + \epsilon\pdhS{\bRlead} = h\sin\alpha\,\be_{\theta}+\cos\alpha\,\be_z. \label{eqn:inextensibilityleading}
\eeq
We express $\bRlead$ in terms of components perpendicular and parallel to the helix axis:
\beq
\bRlead(S,\hS,T) = \bRleadperp(S,\hS,T) + \Zlead(S,\hS,T)\,\be_z \qquad \mathrm{where} \qquad \be_z\cdot\bRleadperp = 0. \label{eqn:defnbRleadperp}
\eeq
Inserting into Eq.~\eqref{eqn:inextensibilityleading} gives
\beq
\pdS{\bRleadperp} + \epsilon \pdhS{\bRleadperp} = h\sin\alpha\,\be_\theta, \quad \pdS{\Zlead} + \epsilon \pdhS{\Zlead} = \cos\alpha. \label{eqn:inextensibilityleadingsplit}
\eeq

Due to both $S$ and $\hS$ derivatives, it is not immediately clear how to integrate the equations in Eq.~\eqref{eqn:inextensibilityleadingsplit}. Consider first the equation for the longitudinal coordinate, $\Zlead$. Physically, we expect that $\Zlead$ consists of its average value over the wavelength centred at $\hS$, of size comparable to the total contour length, $L = O(\epsilon^{-1})$, together with an $O(1)$ part that has zero average. Seeking a solution in his form, we obtain
\beq
\Zlead(S,\hS,T) \sim \epsilon^{-1}\hZ(\hS,T) + \Delta S\cos\alpha(\hS,T) \qquad \mathrm{where} \qquad \pdhS{\hZ} = \cos\alpha. \label{eqn:deformedavgZ} 
\eeq
It may be verified that this satisfies Eq.~\eqref{eqn:inextensibilityleadingsplit} to leading order, i.e., up to terms of size $O(\epsilon)$, using
\beqn
\Delta S = S - \epsilon^{-1}\hS = O(1), \quad \pdS{(\Delta S)} = 1, \quad \pdhS{(\Delta S)} = -\epsilon^{-1},
\eeqn

For the transverse displacement, we seek a solution in the form
\beqn
\bRleadperp(S,\hS,T) = \hR(\hS,T)\be_r(S,\hS,T),
\eeqn
where $\hR = \Lambda\sin\alpha/(2\pi)$ is the slowly-varying helical radius. Using $\pdSt{\be_r} = \pdSt{\Psi}\be_{\theta}$ and $\pdhS{\be_r} = \pdhSt{\Psi}\be_{\theta}$ (which follow from Eq.~\eqref{eqn:defneRetheta}), Eq.~\eqref{eqn:inextensibilityleadingsplit} becomes, upon neglecting terms of $O(\epsilon)$,
\beqn
\pdS{\Psi} + \epsilon \pdhS{\Psi} \sim \frac{2\pi h}{\Lambda}.
\eeqn
Similar to the equation for $\Zlead$, the solution is
\beq
\Psi(S,\hS,T) \sim \epsilon^{-1}\hPsi(\hS,T) + \frac{2\pi h \Delta S}{\Lambda(\hS,T)}  \qquad \mathrm{where} \qquad \pdhS{\hPsi} = \frac{2\pi h}{\Lambda}. \label{eqn:deformedavgPsi}
\eeq

Physically, $\hZ$ and $\hPsi$ correspond to the wavelength-averaged longitudinal coordinate and winding angle, respectively; their slow derivatives, $\pdhSinline{\hZ}$ and  $\pdhSinline{\hPsi}$, can be interpreted as extensional and twist strains of the equivalent column. The boundary condition at the filament base,  Eq.~\eqref{eqn:BCs}, implies that  $\hZ(0,T) = \hPsi(0,T) = 0$; from Eqs.~\eqref{eqn:deformedavgZ}--\eqref{eqn:deformedavgPsi}, we then obtain expressions for $\hZ$ and $\hPsi$ in terms $\alpha$ and $\Lambda$:
\beqn
\hZ(\hS,T) = \int_0^{\hS}\cos\alpha(\xi,T)\,\id\xi, \quad \hPsi(\hS,T) = 2\pi h \int_0^{\hS}\frac{\id\xi}{\Lambda(\xi,T)}.
\eeqn
Combining with the above solutions for $\Zlead$ and $ \bRleadperp$ yields the expressions reported in the main text (Eqs.~\eqref{eqn:solnbRlead}--\eqref{eqn:solnshRavgZPsi}).


\section{Solvability conditions for the first-order problem: direct approach}
\label{sec:appendixsolvability}
In this Appendix, we present an alternative derivation of the solvability conditions for the first-order problem consisting of Eqs.~\eqref{eqn:Phideriv} and \eqref{eqn:forcebalancefirstexpanded}--\eqref{eqn:clawfirst}. We focus on the case $\epsilon L = \ord(1)$; the case $\epsilon L \ll 1$ is similar. We show that the same solvability conditions obtained via the Fredholm Alternative Theorem in \S\ref{sec:multiscalessolvability} can be derived by directly averaging the first-order equations.

From the periodicity of $\Fifirst$ and $\Mifirst$, it follows that
\beqn
\intwave{\pdS{\Fifirst}} = 0, \quad \intwave{\pdS{\Mifirst}} = 0,
\eeqn
From Eqs.~\eqref{eqn:FSframe} and \eqref{eqn:directorslead}, the leading-order directors $\bdilead$ depend on the fast variable $S$ only via the unit vectors $\be_r$ and $\be_{\theta}$; the coefficients of $\be_r$, $\be_{\theta}$ and $\be_z$ depend only on the slow variable, $\hS$. Hence,
\beq
\intwave{\be_r\cdot\left(\sum_{i = 1}^{3}\pdS{\Fifirst}\bdilead\right)} = \intwave{\be_{\theta}\cdot\left(\sum_{i = 1}^{3}\pdS{\Fifirst}\bdilead\right)} = \intwave{\be_z\cdot\left(\sum_{i = 1}^{3}\pdS{\Fifirst}\bdilead\right)} = 0. \label{eqn:zeroavgsforces}
\eeq
Identical expressions hold with $\Fifirst$ replaced by $\Mifirst$. 

Dotting the force balance \eqref{eqn:forcebalancefirstexpanded} by $\be_z$ and integrating over the helical wave, all unknown terms on the left-hand side vanish according to Eq.~\eqref{eqn:zeroavgsforces} and the fact that $\bUlead$ and $\bFlead$ are both parallel to $\be_z$ (recall Eqs.~\eqref{eqn:strainspinlead} and \eqref{eqn:resultantsleadpolars}). We are left with the solvability condition
\beq
\intwave{\be_z\cdot\left(\sum_{i = 1}^{3}\pdhS{\Filead}\bdilead + \bFelead - \epsilon\pdd{\bRlead}{T}\right)} = 0. \label{eqn:solvability1direct}
\eeq
If we instead dot Eq.~\eqref{eqn:forcebalancefirstexpanded} by $\be_z\times\bRlead$ ($=\hR\be_\theta$), dot the moment balance \eqref{eqn:momentbalancefirstexpanded} by $\be_z$ and add the resulting equations, the terms in $\Fifirst$ and $\Uifirst$ cancel. Integrating over the helical wave and making use of Eq.~\eqref{eqn:zeroavgsforces}, all remaining terms on the left-hand side again vanish, yielding
\beq
\intwave{\be_z\cdot\left[\sum_{i = 1}^{3}\pdhS{\Milead}\bdilead + \delta\bMlead + \bRlead\times\left(\sum_{i = 1}^{3}\pdhS{\Filead}\bdilead + \bFelead - \epsilon\pdd{\bRlead}{T}\right)\right]} = 0.  \label{eqn:solvability2direct}
\eeq
In component form, the solvability conditions \eqref{eqn:solvability1direct}--\eqref{eqn:solvability2direct} are equivalent to Eqs.~\eqref{eqn:solvability1component}--\eqref{eqn:solvability2component} in the main text.

The remaining solvability conditions can be formulated by noting from the definition of $\bTheta$ (Eq.~\eqref{eqn:defnTheta}) that $\be_x\cdot\bTheta$ and $\be_y\cdot\bTheta$ are $\Lambda$-periodic. Because the components of $\bPhi$ are assumed to be periodic, we also have that $\pdSinline{\bPhi}$ is $\Lambda$-periodic (since the $\bdilead$ are periodic). Equation \eqref{eqn:Phideriv} then gives
\beq
\intwave{\be_x\cdot\left(\sum_{i = 1}^{3}\Uifirst\bdilead\right)} = \intwave{\be_y\cdot\left(\sum_{i = 1}^{3}\Uifirst\bdilead\right)} = 0.
\label{eqn:zeroavgsstrains}
\eeq
We dot Eq.~\eqref{eqn:forcebalancefirstexpanded} in turn by $\be_x$ and $\be_y$ and integrate over the helical wave. The first two terms in Eq.~\eqref{eqn:forcebalancefirstexpanded} can be written as the single derivative $\pdSinline{(\sum_{i = 1}^{3}\Fifirst\bdilead)}$; since $\be_x$ and $\be_y$ are constant vectors, this derivative (after taking the dot product with $\be_x$ and $\be_y$) still vanishes upon integrating. The third term also vanishes from Eq.~\eqref{eqn:zeroavgsstrains}. We obtain 
\beqn
\intwave{\be_x\cdot\left(\sum_{i = 1}^{3}\pdhS{\Filead}\bdilead + \bFelead - \epsilon\pdd{\bRlead}{T}\right)} = \intwave{\be_y\cdot\left(\sum_{i = 1}^{3}\pdhS{\Filead}\bdilead + \bFelead - \epsilon\pdd{\bRlead}{T}\right)} = 0,
\eeqn
which, in component form, are precisely the solvability conditions \eqref{eqn:solvability3,4component} in the main text (the $\ssTheta\times\ssFlead$ terms in Eq.~\eqref{eqn:solvability3,4component} integrate to zero).

\section{Equivalent-rod equations in terms of Frenet curvature and torsion}
\label{sec:appendixFrenetformulation}
From Eq.~\eqref{eqn:deformedDarboux}, we have $\Lambda = 2\pi/\sqrt{\hK^2+\hT^2}$, $\cos\alpha = h\hT/\sqrt{\hK^2+\hT^2}$ and $\sin\alpha = \hK/\sqrt{\hK^2+\hT^2}$. The leading-order moment and force resultants, $M_Z$ and $F_Z$ (defined in Eq.~\eqref{eqn:resultantsleadZcomponents}), and the slowly-varying helical radius, $\hR$ (defined in Eq.~\eqref{eqn:solnshRavgZPsi}), can then be written in terms of $(\hK,\hT)$ alone:
\beqn
M_Z = \frac{h}{\sqrt{\hK^2 + \hT^2}}\left[\hK\left(\hK-\hKref\right) + \hT\frac{\hT-\hTref}{1+\nu}\right], \quad
F_Z = h\sqrt{\hK^2 + \hT^2}\left[\frac{\hT-\hTref}{1+\nu}-\hT\frac{\hK-\hKref}{\hK}\right], \quad \hR = \frac{\hK}{\hK^2 + \hT^2}.
\eeqn
The equivalent-rod equations \eqref{eqn:effectiveforcebalance}--\eqref{eqn:effectivemomentbalance} yield the first-order system for $\hK(\hS,T)$ and $\hT(\hS,T)$:
\begin{align*}
& \hC_1 \pd{\hT}{\hS} + \hC_2\pd{\hK}{\hS} + \overline{\be_z\cdot\bFelead} - h\pdd{}{T}\left(\int_0^{\hS}\frac{\hT}{\sqrt{\hK^2+\hT^2}}\Bigg\lvert_{\hS=\xi}\id\xi\right) = 0, \\
& \hC_3 \pd{\hT}{\hS} + \hC_4\pd{\hK}{\hS} + \delta\overline{\be_z\cdot\bMelead} + \frac{\hK}{\hK^2 + \hT^2}\overline{\be_\theta\cdot\bFelead} - h\pd{}{T}\left[\frac{\hK^2}{\left(\hK^2+\hT^2\right)^2}\pd{}{T}\left(\int_0^{\hS}\sqrt{\hK^2+\hT^2}\bigg\lvert_{\hS=\xi}\id\xi\right)\right]= 0,
\end{align*}
where the dimensionless stiffness coefficients $\hC_i = \hC_i(\hK,\hT)$ ($i=1,2,3,4$) are 
\begin{align*}
\hC_1 & = \pd{F_Z}{\hT} = h\frac{(1+\nu)\hKref\left(\hK^2 + 2\hT^2\right) - \nu\hK^3 - \hK\hT\left(\hTref + 2\nu\hT\right)}{(1+\nu)\hK\sqrt{\hK^2+\hT^2}}, \\
\hC_2 & = \pd{F_Z}{\hK} = -h\frac{(1+\nu)\hKref\hT^3 + \hK^3\left(\hTref+\nu\hT\right)}{(1+\nu)\hK^2\sqrt{\hK^2+\hT^2}}, \\
\hC_3 & = \pd{M_Z}{\hT} = h\frac{\hT^3 -\hK^2\left[\hTref - (1 - \nu)\hT\right] + (1+\nu)\hKref\hK\hT}{(1+\nu)\left(\hK^2+\hT^2\right)^{3/2}}, \\
\hC_4 & = \pd{M_Z}{\hK} = h\frac{(1+\nu)\left(\hK^3 -\hKref\hT^2\right) + \hK\hT\left[\hTref + (1+2\nu)\hT\right]}{(1+\nu)\left(\hK^2+\hT^2\right)^{3/2}}.
\end{align*}
The boundary conditions \eqref{eqn:effectiveBCs} are equivalent to
\beqn
\hK(\epsilon L,T) = \hKref, \quad \hT(\epsilon L,T) = \hTref.
\eeqn
Similar to the $(\alpha,\Lambda)$-formulation, the external forces and moments can (in principle) be expressed in terms of $\hK$ and $\hT$ using the expressions in \S\ref{sec:multiscaleskinematics} together with Eq.~\eqref{eqn:deformedDarboux}. 

\section{Buckling threshold of a vertical helical filament under self-weight}
\label{appendix:gravitybuckling}
Here we estimate the buckling threshold of a helical filament under gravity using an effective-beam approximation. We set the effective-beam length to be the linear length along the helix axis, $l_{\mathrm{eff}} = l\cos\aref$, and the effective density from equating the total mass, $\rho_{\mathrm{eff}} = \rho_s\sec\aref$. The effective bending stiffness, $B_{\mathrm{eff}}$, can be determined by calculating the elastic energy associated with uniform bending due to small moment applied normal to the helix axis, and comparing this with the corresponding energy for a naturally-straight beam \citep{vogel2012}. The result is
\beqn
B_{\mathrm{eff}} = \frac{B\cos\aref}{1+\left(\nu/2\right)\sin^2\aref}.
\eeqn
The classical result for a straight, vertical column \citep{wang1986} predicts buckling occurs under self-weight when
\beqn
\frac{\rho_{\mathrm{eff}}A|g|l_{\mathrm{eff}}^3}{B_{\mathrm{eff}}} \approx 7.84 \qquad \mathrm{(buckling\ threshold)}.
\eeqn
Substituting for the effective parameters, re-arranging and making use of the expression \eqref{eqn:scenarioIdefnepsilon} for $\epsilon$, we obtain Eq.~\eqref{eqn:approxbucklethresh} in the main text.



\end{document}